# Geological flows

Yu. N. Bratkov

In this paper geology and planetology are considered using new conceptual basis of high-speed flow dynamics. Recent photo technics allow to see all details of a flow, 'cause the flow is static during very short time interval. On the other hand, maps and images of many planets are accessible. Identity of geological flows and high-speed gas dynamics is demonstrated. There is another time scale, and no more. All results, as far as the concept, are new and belong to the author. No formulae, pictures only.

CONTENTS

# 1. CENTRAL AMERICA IS A KARMAN VORTEX STREET BEHIND BALLISTIC SOUTH AMERICA.

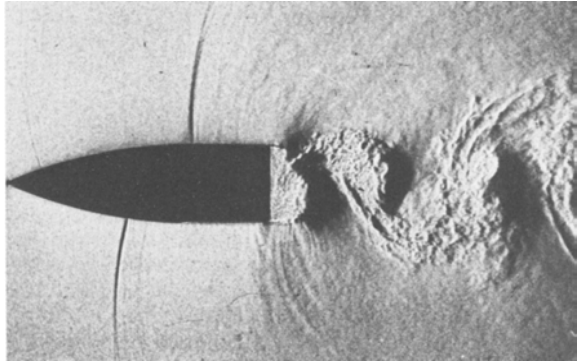

**Fig. 1.** Oscillating trace (a Karman vortex street) in air behind a moving body. Mach = 0.6, Reynolds = 220 000. High-speed photo [1, ph. 67].

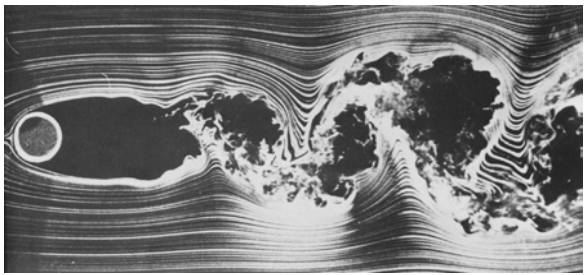

**Fig. 2.** Oscillating trace in liquid behind streamlined cylinder [1, ph. 48]. Re = 10 000.

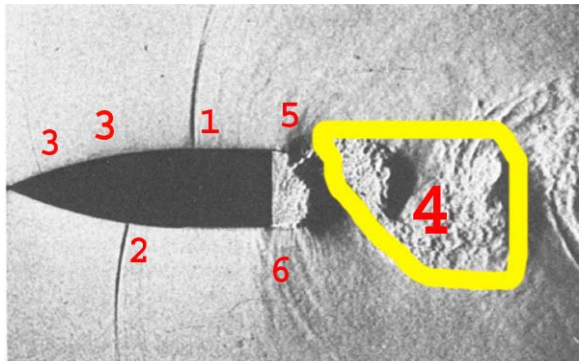

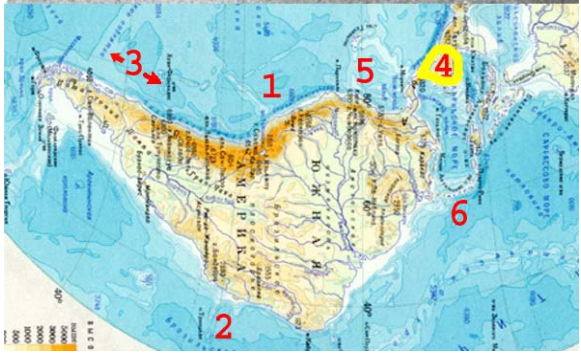

**Fig. 3.** Comparing two flows. See Fig. 1. 1 – shock wave, 2 – pair of shock waves, 3 – two weak shock waves.

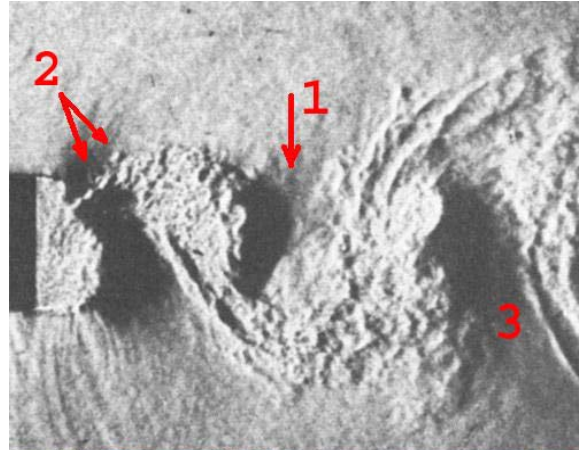

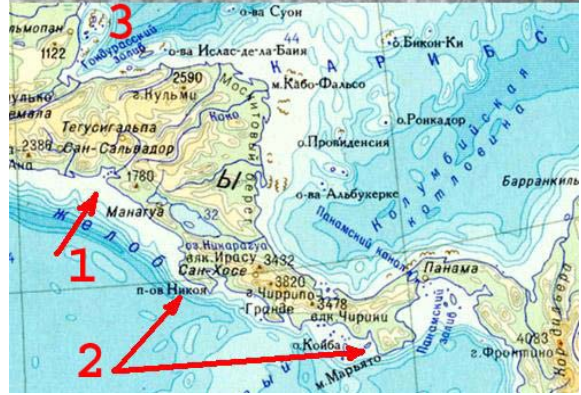

**Fig. 4.** Central America. See Fig. 1.

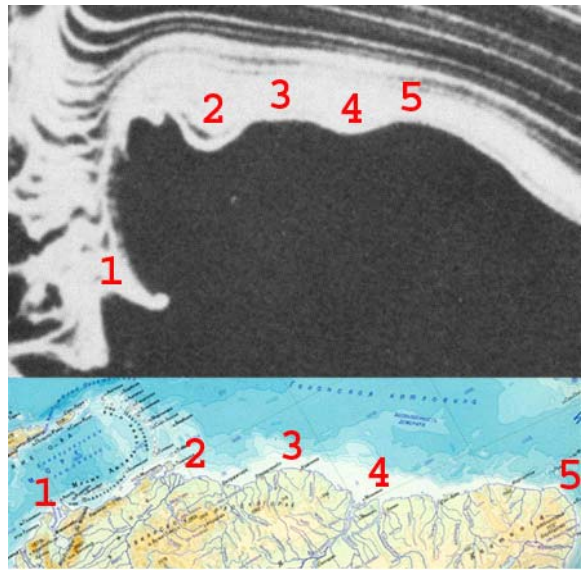

**Fig. 5.** South America. See Fig. 2. 1 – the Gulf of Venezuela.

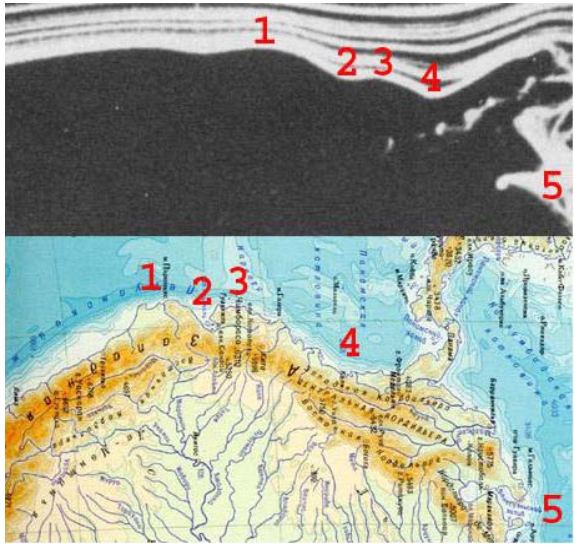

**Fig. 6.** South America. See Fig. 2.

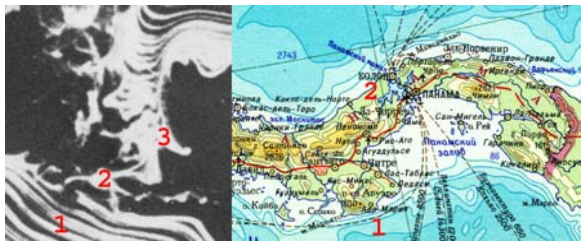

**Fig. 7.** The Panama Canal. See Fig. 2. 2 – Gatun Lake, 3 – the Gulf of Venezuela.

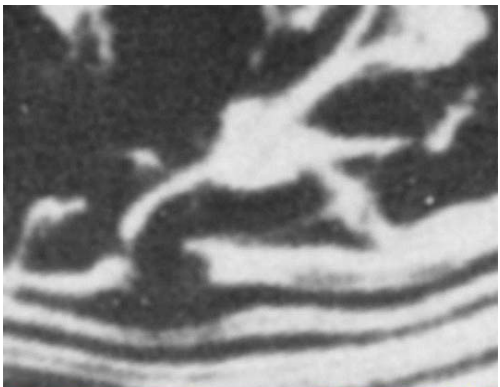

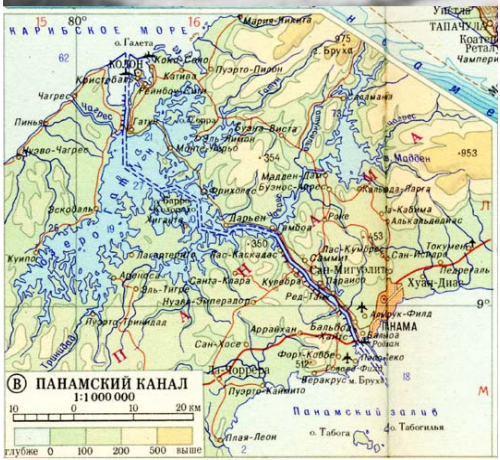

**Fig. 8.** Panama Canal's Gatun Lake. The upper photo was made *after* making the canal. If before?

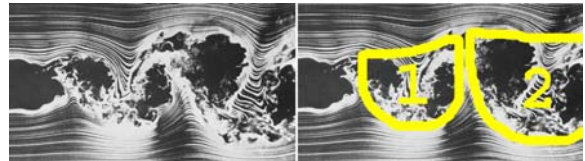

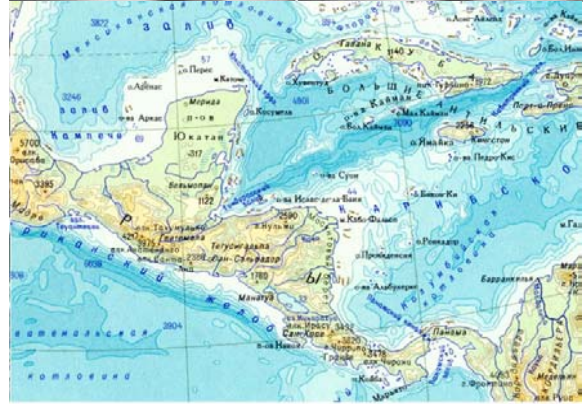

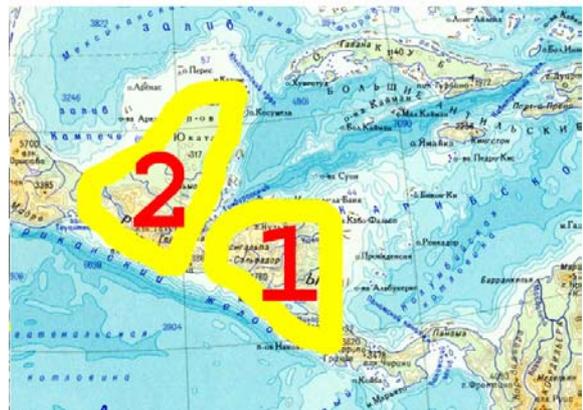

**Fig. 9.** Central America. See Fig. 2.

Maps from [2] were used here and further.

**Conclusions:** 1) There are many examples of global and local exact structures in relief [3, 4]. We see that such structures are elements of flows, so they will be replaced by another exact structures during flow moving. Thus a turbulent flow is an exact structure. Also note that a sphere or a ball is a compact body, so there could be strong dependence between elements of fows. Such flows could be very structurized and interdependent.

2) Exact relief structures exist in some "another space" (in topos) [5], see examples in [4]. Thus pre-images of slow planetary flows exist in the same "another space". Some nontrivial confirmation of this concept is given in [6].

3) The Panama Canal problem is an interesting one (Fig. 8). This is a problem of interdependence of so-called "real world" and "space of shapes". We'll give here some examples of natural Panama Canals, which canals are initially non-through. So this construction is stable.

## 2. NORTH AMERICA ON TITAN.

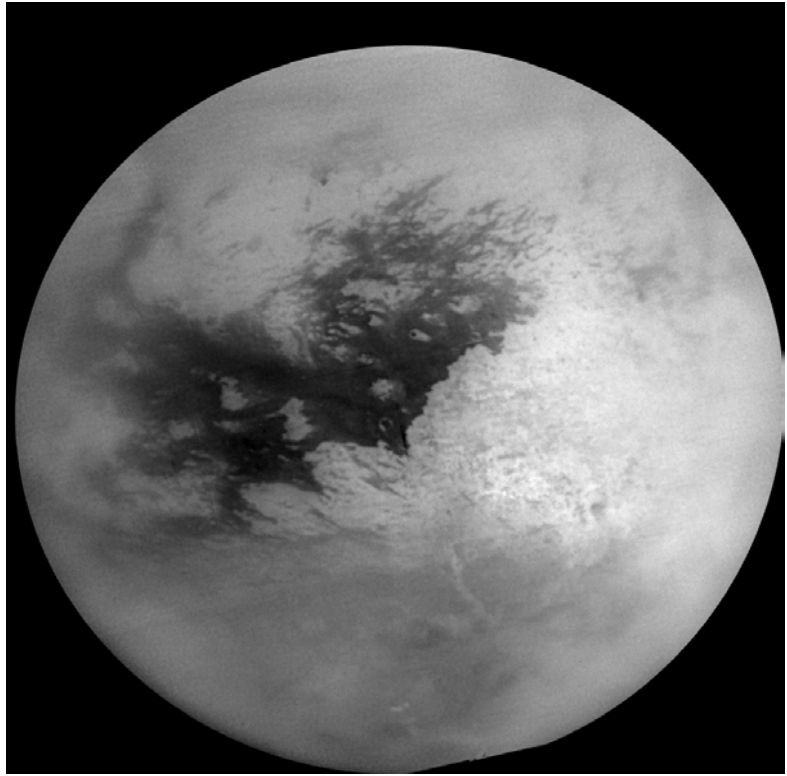

**Fig. 1.** PIA06185. Titan: Shangri-La (dark area) and Xanadu (bright area at center right).

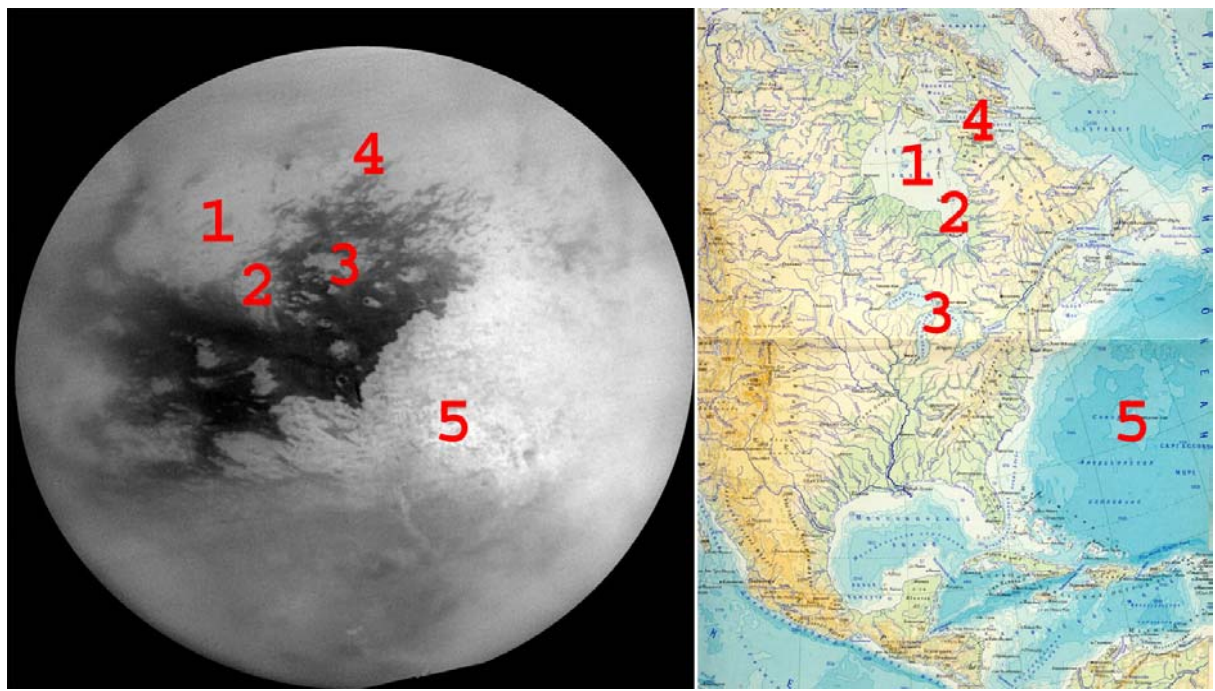

**Fig. 2.** 1 – Hudson Bay, 2 – James Bay, 3 – Great Lakes, 4 – the Hudson Strait, 5 – the Bermudas.

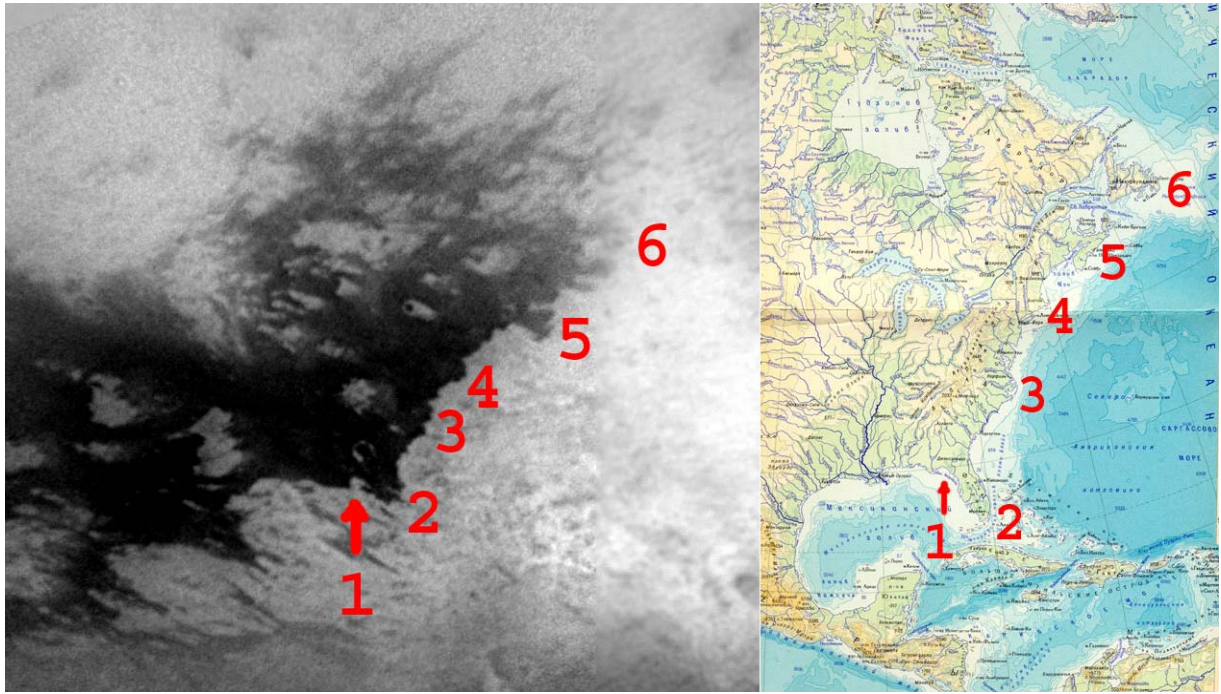

**Fig. 3.** 1 – Apalachee Bay, 2 – Florida, 6 – Newfoundland.

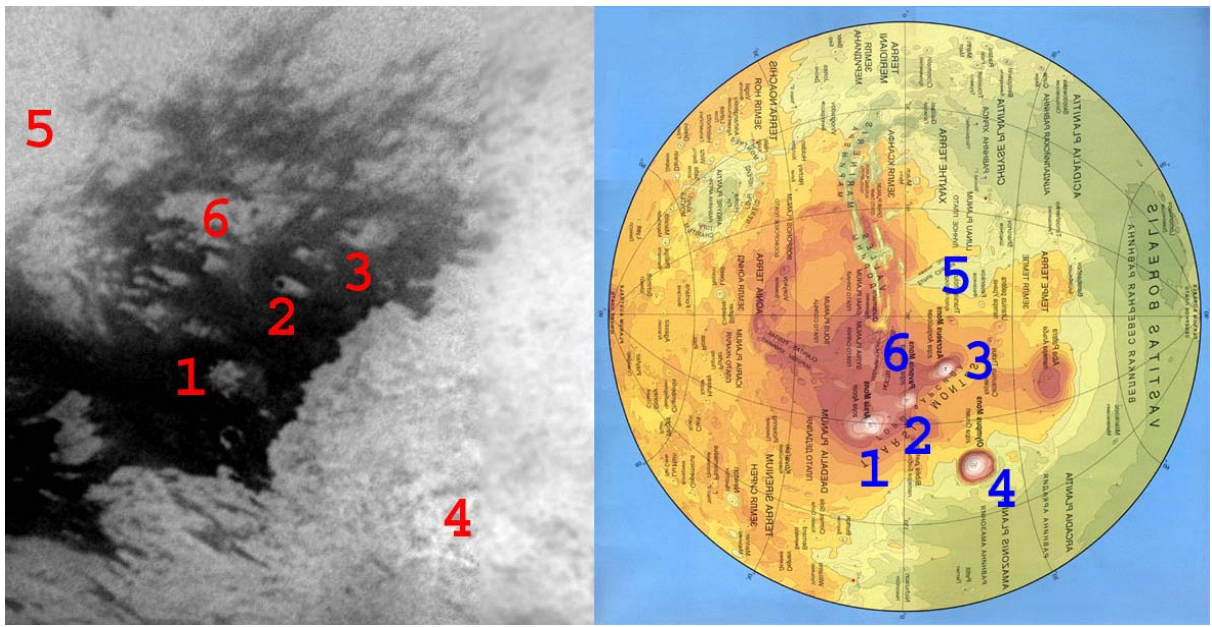

**Fig. 4.** Comparing with North America on Mars. 1 – Arsia, 2 – Pavonis, 3 – Ascraeus, 4 – Olympus/Bermudas, 5 – Hudson Bay, 6 – Great Lakes. **Right:** Map of Mars [7], mirror image, rotated cw $90^0$.

The map of Titan has two colors: black and white. Black areas are dark for location, and white areas are bright. So dark areas seem to be convex down (concave), and bright areas maybe are convex up (heights, mountains, etc.). At Fig. 4 (left) objects 1, 2, 3 could become deepenings, i.e. wholes in dark matter. Thus North America on Titan (concave continent) maybe is antisymmetric to convex North Americae on the Earth and on Mars.

## 3. THE SEA OF AZOV IS A CONCAVE SOUTH AMERICA.

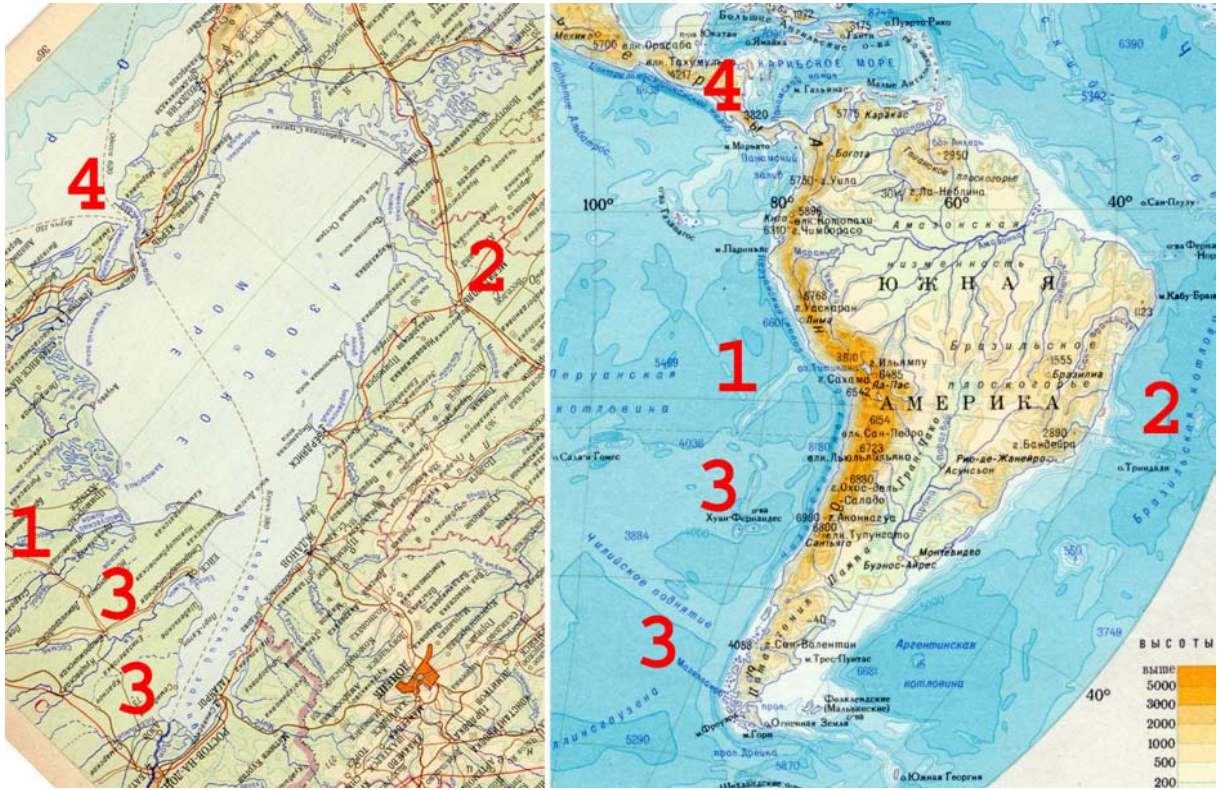

**Fig. 1.** Comparing flows. The Sea of Azov and South America. 1 – a shock wave, 2 – a pair of shock waves, 3 – two weak shock waves, 4 – Panama. See the "Central America" section. Diameter (length) of the Sea of Azov (including Sivash Bay) is $(\pi - 3)/2$ radians, see the "$(\pi - 3)$-basins" section.

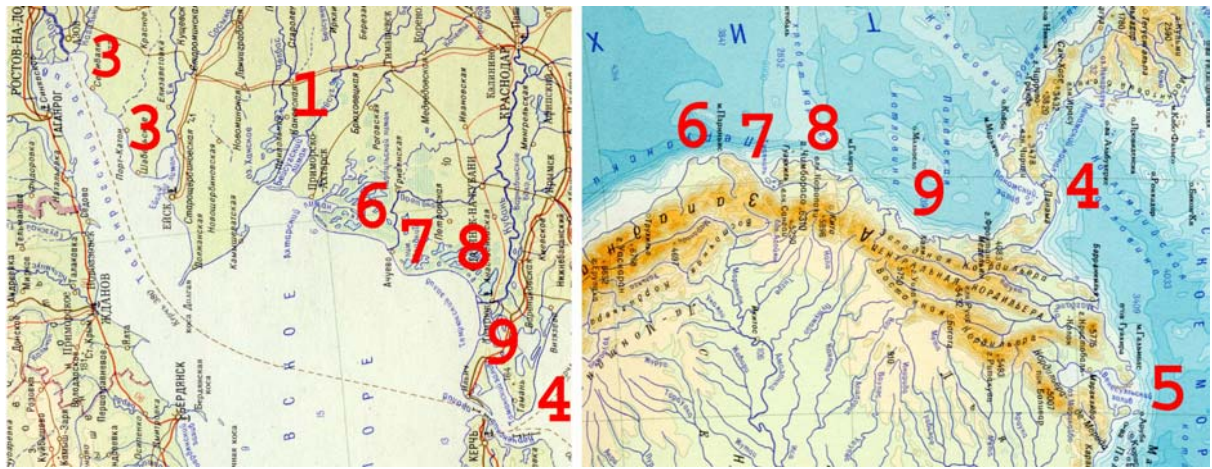

**Fig. 2.** Comparing flows. 1 – a shock wave, 3 – two weak shock waves, 4 – Panama, 5 – the Gulf of Venezuela, 6-7-8 – a bipole (a modon). (A modon is a strongly nonlinear 2-vortex, a 2-resonance object. See [8], p. 84 in Russian edition.) Bipoles are studying in [3].

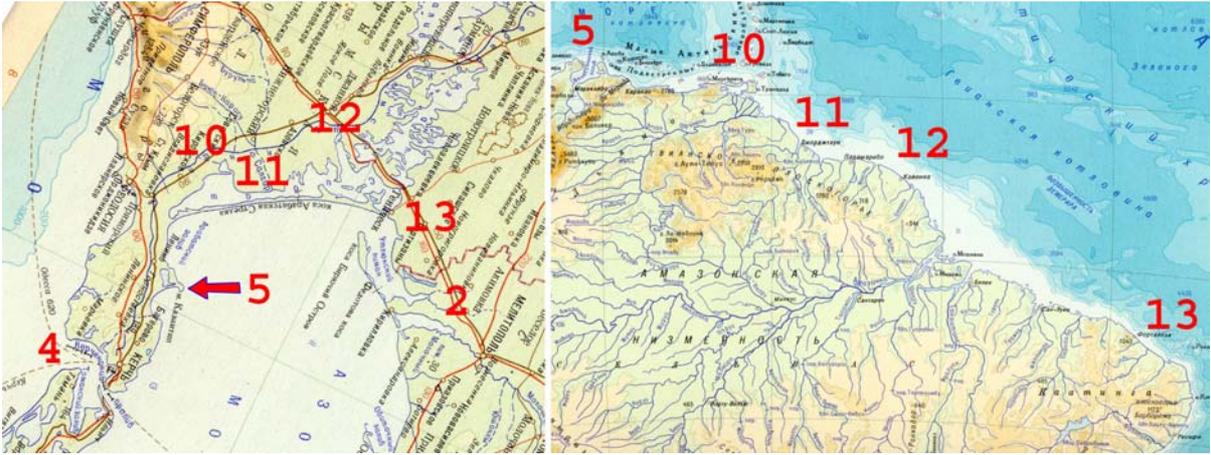

**Fig. 3.** 2 – a pair of shock waves, 4 – Panama, 5 – the Gulf of Venezuela, 10-11-12 – a bipole.

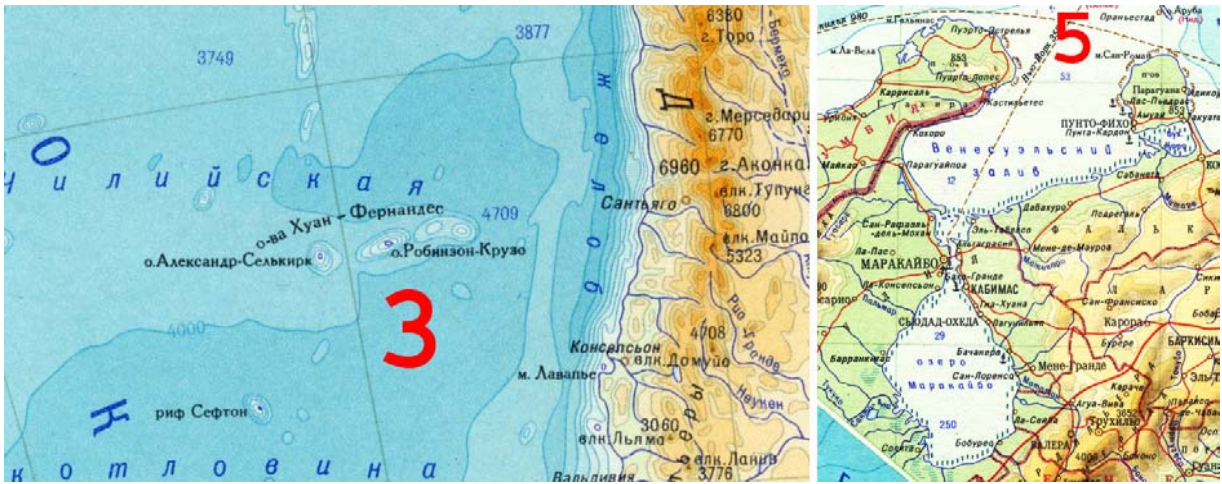

**Fig. 4.** South America: 3 – the Juan Fernandes Islands are the shock wave 3 (see Fig. 1); 5 – the Gulf of Venezuela.

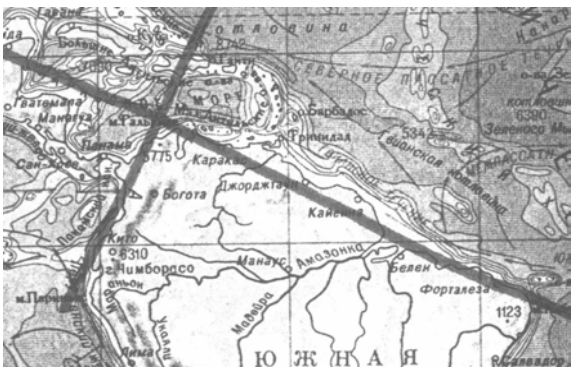

**Fig. 5.** [3, Fig. 145]. The Chomolungma–Perm straight line (vertical) and it's perpendicular (horizontal). See Fig. 3 left, the 10-11-12 object. There is some analogous horizontal straight line.

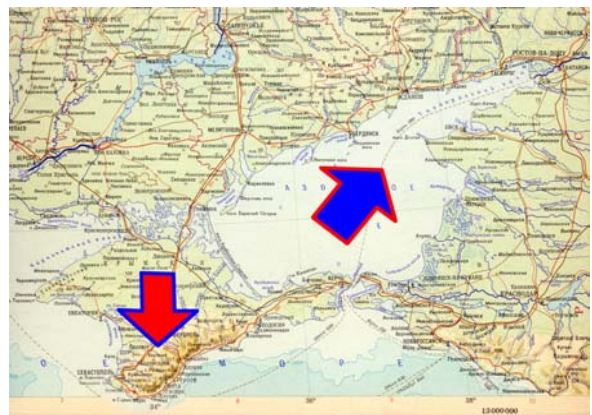

**Fig. 6.** The Crimea and the Sea of Azov are contrary flows.

## 4. $(\pi - 3)$-BASINS.

Define angular radius of a basin. If we have a circular structure on a planet, we get its radius (in kilometers) and divide it by the radius of the planet (in kilometers). We obtain angular radius of the ring in radians. A planet is supposed to be an exact sphere. A distance on a planet is a distance along the exact sphere.

In this section we draw circles at maps of planets. We set the center and the radius of the circle under consideration (radius is a distance along the surface of the sphere), and we calculate some points of the circle. These points are drawn at the map. Sometimes we connect the points by polygonal line.

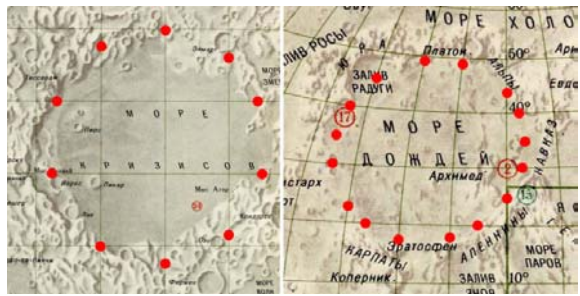

**Fig. 1.** The Moon [2]. **Left:** Mare Crisium. Radius of the circle is $(\pi - 3)$ radians. **Right:** Mare Imbrium. Radius of the circle is $2(\pi - 3)$ radians. See images of the Moon at the "Extraterrestrial Antarctidae" section.

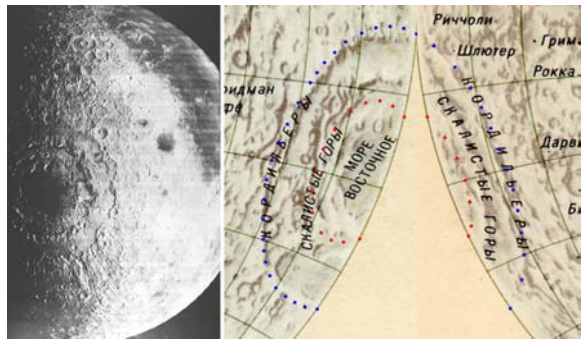

**Fig. 2.** The Moon, Mare Orientale. **Left:** Image of Lunar Orbiter 4. **Right:** Map of the Moon [2]. Radii of the circles are $(\pi - 3)$ and $2(\pi - 3)$ radians.

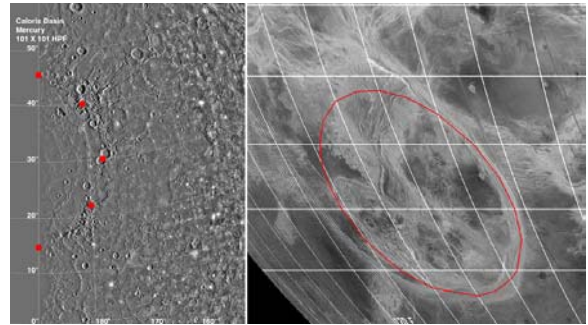

**Fig. 3.** **Left:** Mercury, Caloris basin. PIA02439. Radius of the circle is $2(\pi - 3)$ radians. **Right:** Venus, Artemis Corona. PIA00478 (fragment). Radius of the circle is $(\pi - 3)$ radians.

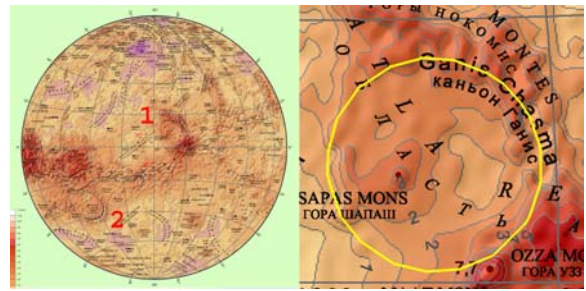

**Fig. 4.** **Left:** Venus [9]. 1 – the Maat-Sapas region, 2 – Artemis Corona. **Right:** Disk 1 at the Maat-Sapas region. Radius of the circle is $(\pi - 3)$ radians.

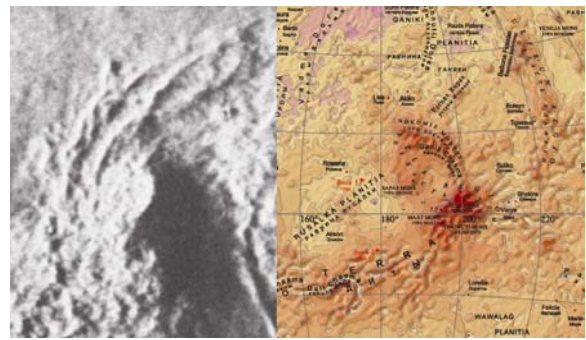

**Fig. 5.** Vortex discs at high-speed gas flow (left; see the "Central America" section, Fig. 1, 4) and on Venus at the Maat–Sapas region (see Fig. 4).

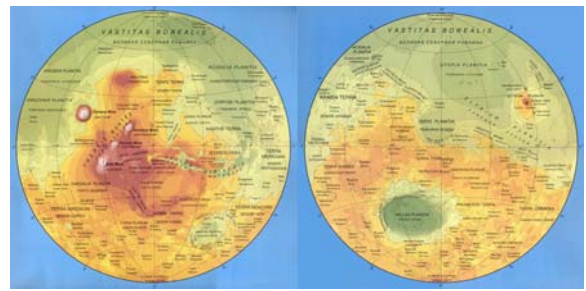

**Fig. 6.** Mars [7]. The main Martian basins are given at Fig. 7–9. Isidis: the center of the Eastern hemisphere. Elysium: the Eastern hemisphere, upper

right. Hellas: the Eastern hemisphere, lower left. Argyre: the Western hemisphere, lower right.

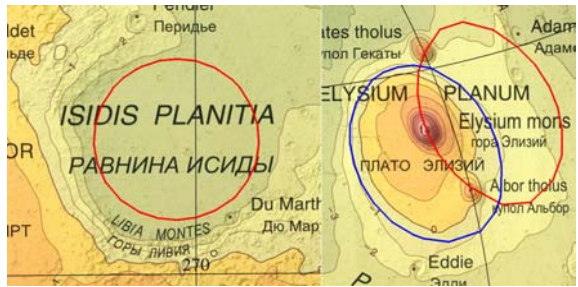

**Fig. 7.** Mars [7]. Isidis (left), Elysium (right). Radii of the circles are $(\pi - 3)$ radians.

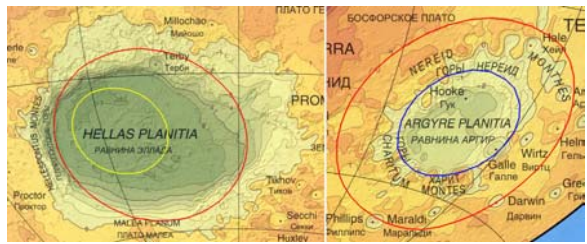

**Fig. 8.** Mars [7]. Hellas (left), Argyre (right). Radii of the circles are $(\pi - 3)$ and $2(\pi - 3)$ radians.

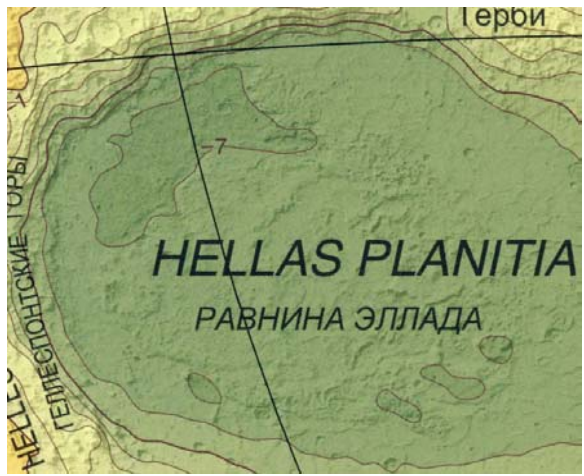

**Fig. 9.** Mars [7]. Hellas, the region of the yellow circle, Fig. 8.

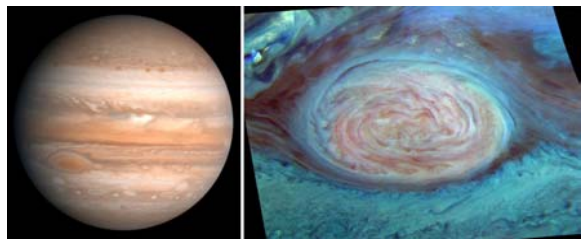

**Fig. 10.** Jupiter, the Great Red Spot. PIA00489. They say the size of the Spot was varying. However, in comments to PIA00488 (it is analogous to PIA00489) the size 20 000 km is given. So radius of the circumscribed circle is $(\pi - 3)$ radians. **Left:** PIA00343. **Right:** PIA00489. The Great Red Spot in false colors.

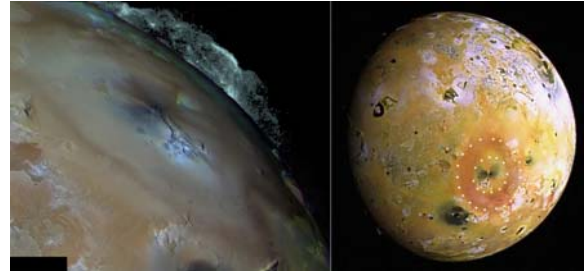

**Fig. 11.** Io, Jupiter's satellite. Pele, a ring structure. **Left:** PIA00323. **Right:** PIA00738. Pele is red, at the center. White points (two circles) are drawn by the author. Radii of the circles are $(\pi - 3)$ and $2(\pi - 3)$ radians.

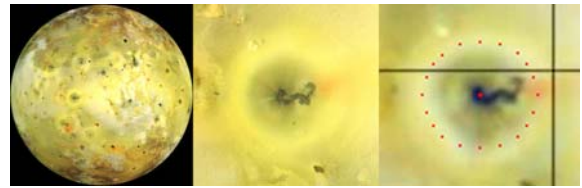

**Fig. 12.** Io, Jupiter's satellite. Prometheus, a volkano. **Left:** PIA02308, Prometheus is at the center left. **Center:** PIA02308 (fragment). **Right:** PIA00585 (fragment). Radius of the circle is $(\pi - 3)/2$ radians.

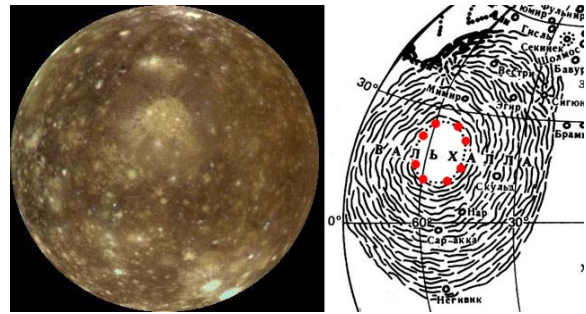

**Fig. 13.** Callisto, Jupiter's satellite. Valhalla, a ring structure. **Left:** PIA01298. Valhalla is near the center, a bit upper. **Right:** Map of Callisto [10]. Radius of the circle is $(\pi - 3)$ radians.

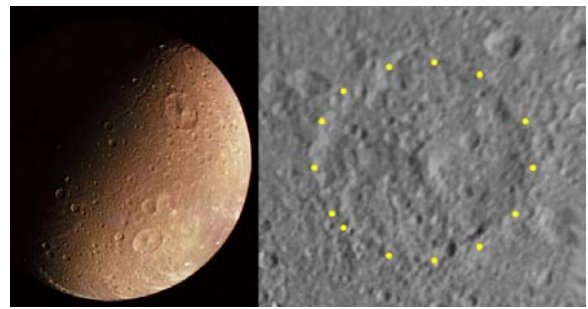

**Fig. 14.** Dione, Saturn's satellite. Aeneas, a crater. **Left:** PIA01482 (fragment), faulse colors. Aeneas is upper right. **Right:** PIA08341 (fragment). Radius of the circle is $(\pi - 3)$ radians.

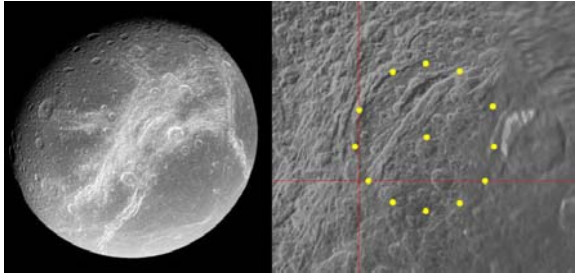

**Fig. 15.** Dione, Saturn's satellite. Some segment of a disk. **Left:** PIA08256. Object is at the center. **Right:** PIA08341 (fragment). Radius of the circle is $(\pi - 3)$ radians.

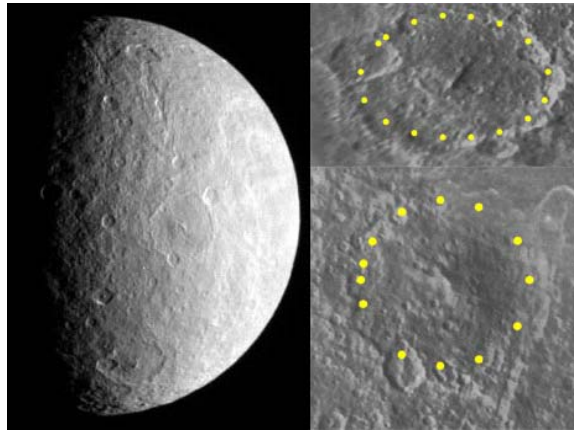

**Fig. 16.** Rhea, Saturn's satellite. Izanagi and its pair basin. **Left:** PIA09019. Izanagi is at the lower left, its pair is at the center. **Right:** PIA08343 (fragments). Izanagi is upper. Radii of the circles are $(\pi - 3)$ radians.

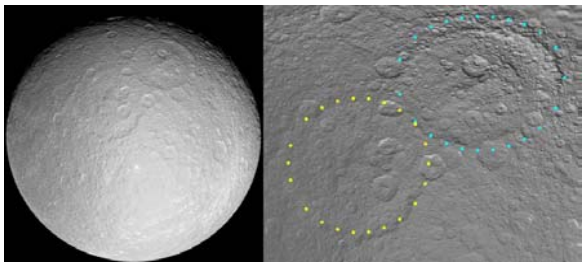

**Fig. 17.** Rhea, Saturn's satellite. Tirawa (upper right) and its pair basin. **Left:** PIA07763. **Right:** PIA08343 (fragment). Radii of the circles are $2(\pi - 3)$ radians.

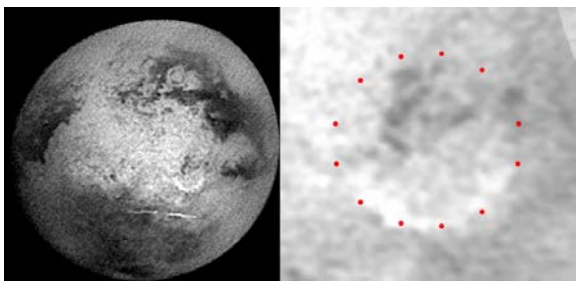

**Fig. 18.** Titan, Saturn's satellite. The Smile (Hotei), a ring structure. **Left:** PIA06154. The Smile is at the center right. **Right:** PIA08399 (fragment). Radius of the circle is $(\pi - 3)$ radians.

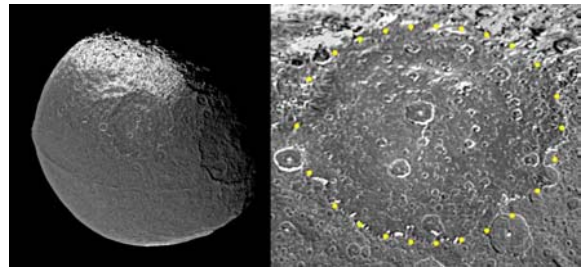

**Fig. 19.** Iapetus, Saturn's satellite. **Left:** PIA06166. **Right:** PIA08406 (fragment). Radius of the circle is $2(\pi - 3)$ radians.

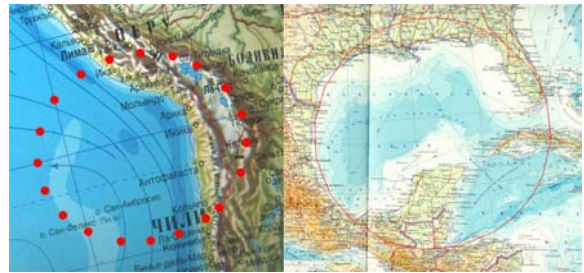

**Fig. 20. Left:** South America, the Andes. Globe. **Right:** The Gulf of Mexico. Radii of the circles are $(\pi - 3)$ radians.

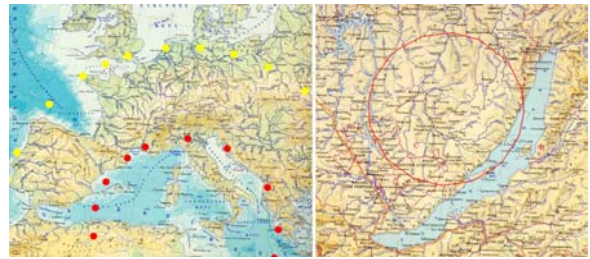

**Fig. 21. Left:** The Mediterranean Sea. Radii of the circles are $(\pi - 3)$ and $2(\pi - 3)$ radians. **Right:** Siberia, Baikal Lake. Radius of the circle is $(\pi - 3)/5$ radians.

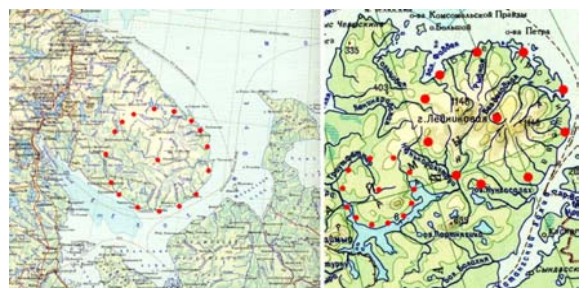

**Fig. 22. Left:** The Kola Peninsula and the White Sea. Radius of the circle is $(\pi - 3)/8$ radians. **Right:** The Taimyr Peninsula. Taimyr Lake is a small copy of the White Sea (left). Radii of the circles are $(\pi - 3)/8$ and $(\pi - 3)/16$ radians.

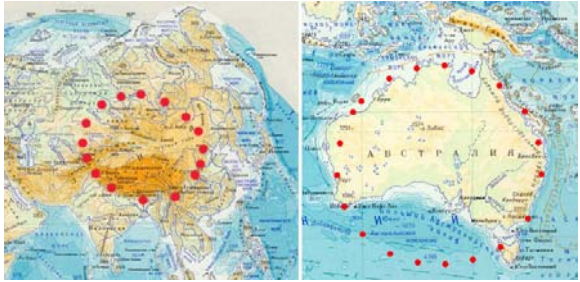

**Fig. 23.** Radii of the circles are $2(\pi - 3)$ radians.

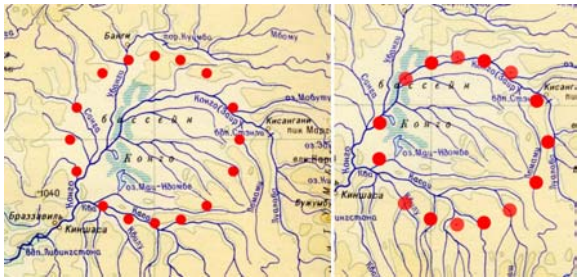

**Fig. 24.** Africa, Congo basin. Radii of the circles are $(\pi - 3)/2$ radians.

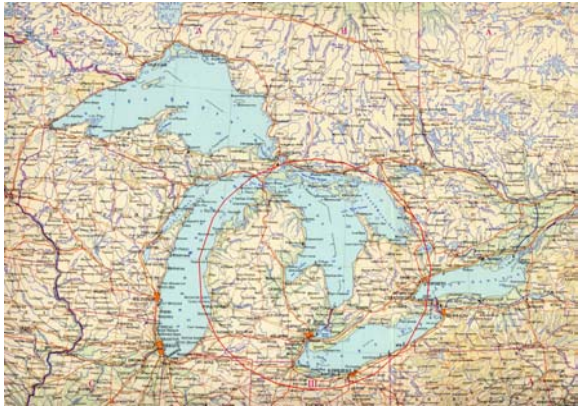

**Fig. 25.** North America, Great Lakes. Radius of the circle is $(\pi - 3)/3$ radians.

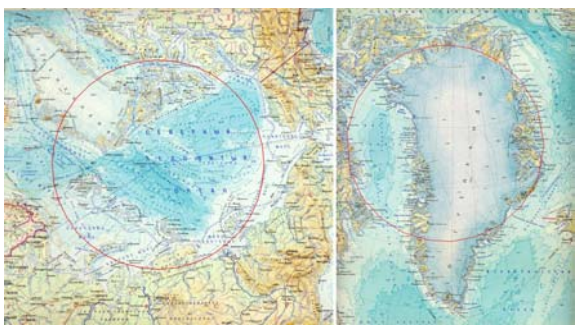

**Fig. 26. Left:** The Arctic. Radius of the circle is $2(\pi - 3)$ radians. **Right:** Greenland. Radius of the circle is $(\pi - 3)$ radians.

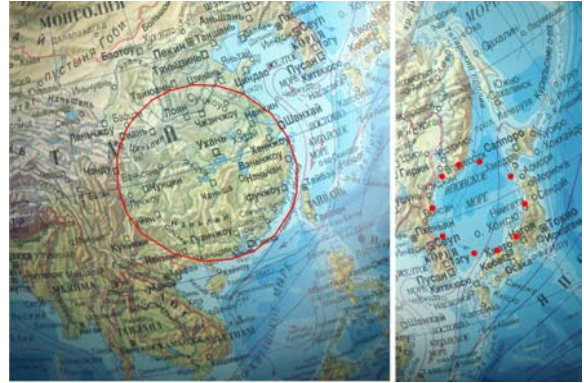

**Fig. 27.** SE Asia. Globe. **Left:** Eastern China. Radius of the circle is $(\pi - 3)$ radians. **Right:** The Sea of Japan. Radius of the circle is $(\pi - 3)/2$ radians.

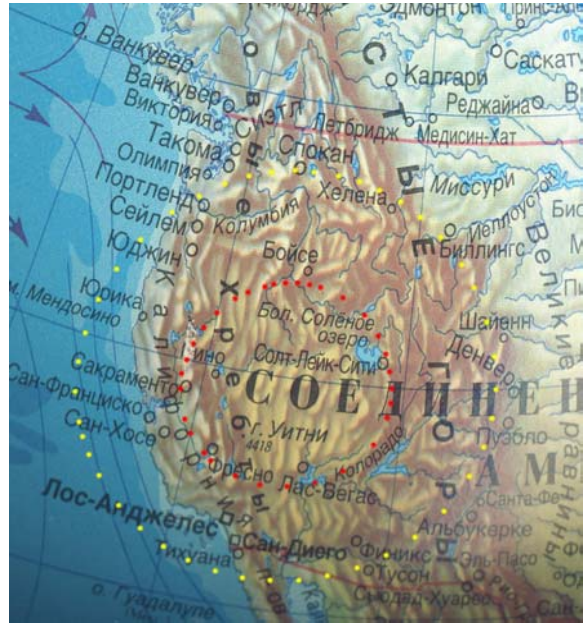

**Fig. 28.** North America, California, Big Basin. Globe. Radii of the circles are $(\pi - 3)/2$ and $(\pi - 3)$ radians.

**Conclusion:** $(\pi - 3)$ is some vortex eigenvalue for flows in a ball or on a sphere.

## 5. 0.5-NORTH-AMERICA IN ASIA.

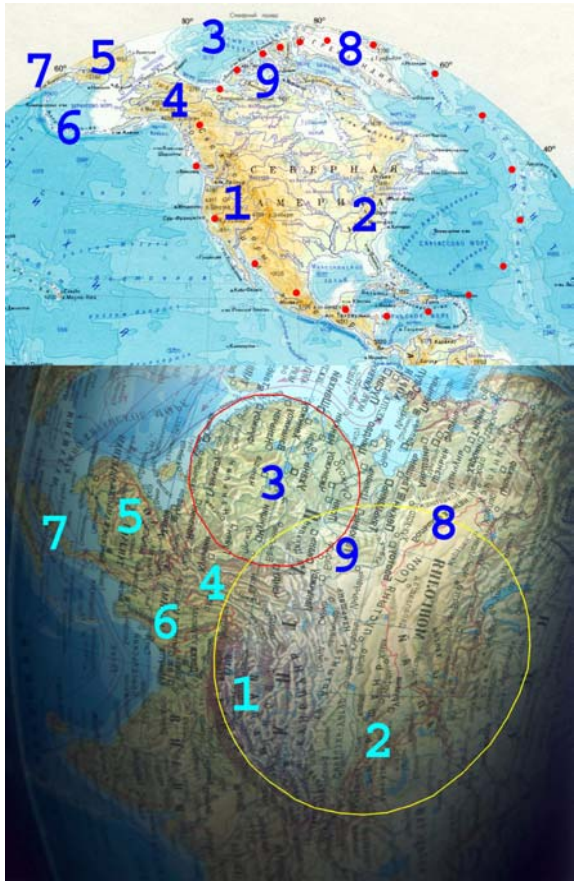

**Fig. 1.** North America and SE Asia. Asia (globe, mirror image) is 90⁰ cw rotated. Radii of circles are: the North America circle is of $4(\pi - 3)$ radians, yellow circle is of $2(\pi - 3)$ radians, red circle is of $(\pi - 3)$ radians.

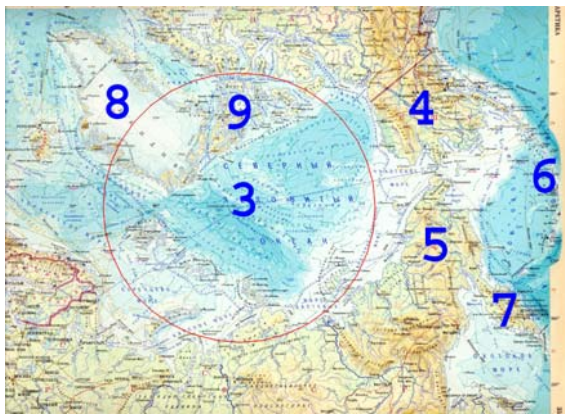

**Fig. 2.** Radius of Arctic basin is $2(\pi - 3)$ radians. This circle is identified with red $(\pi - 3)$-circle from Fig. 1.

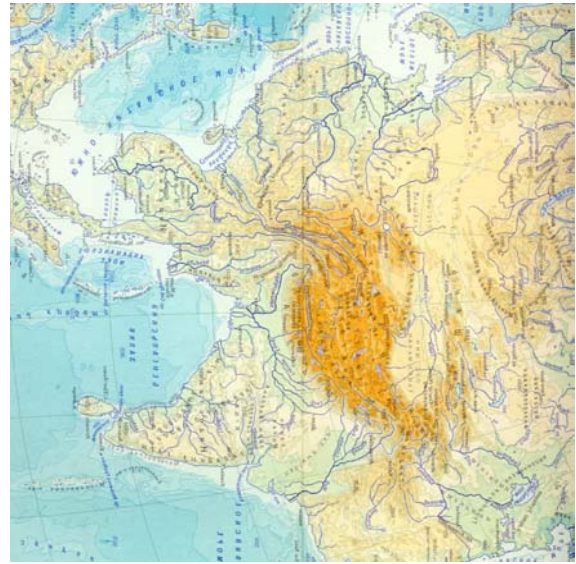

**Fig. 3.** South Asia (mirror image), 90⁰ cw rotated.

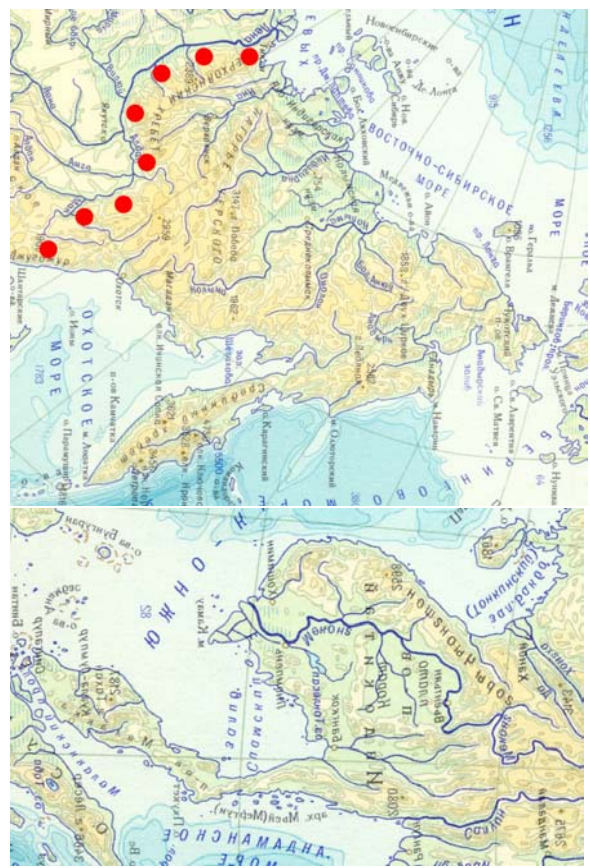

**Fig. 4.** Chukotcae (5, 7). Asian Chukotca (Indo-China) is a mirror image. The body 5 of Chukotca is between Kamchatka (7) and Arctic basin (3).

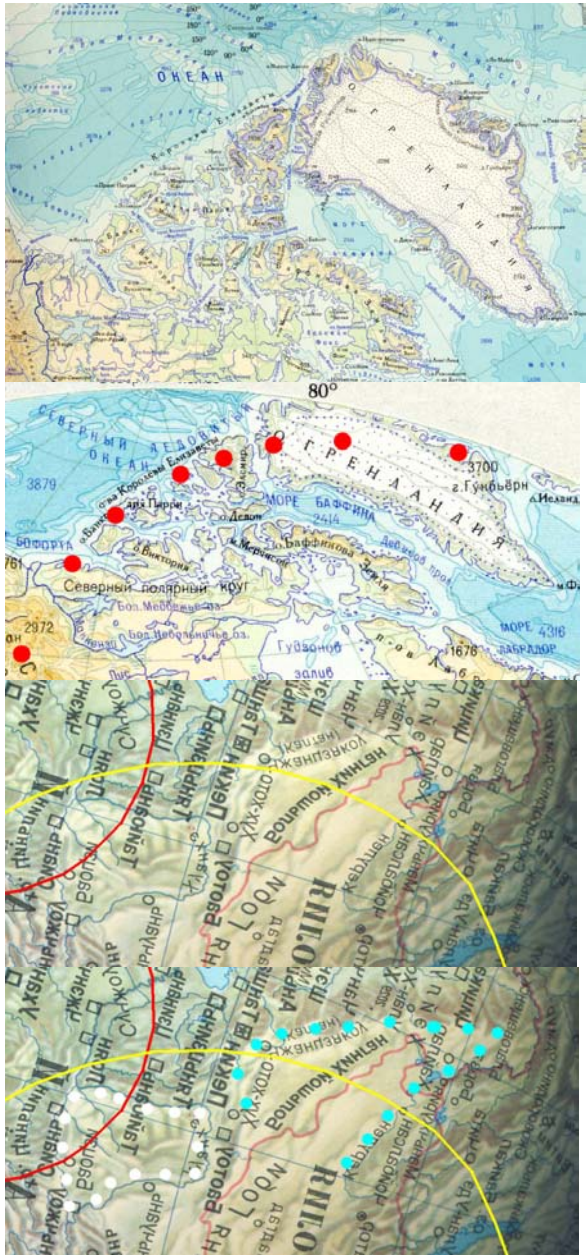

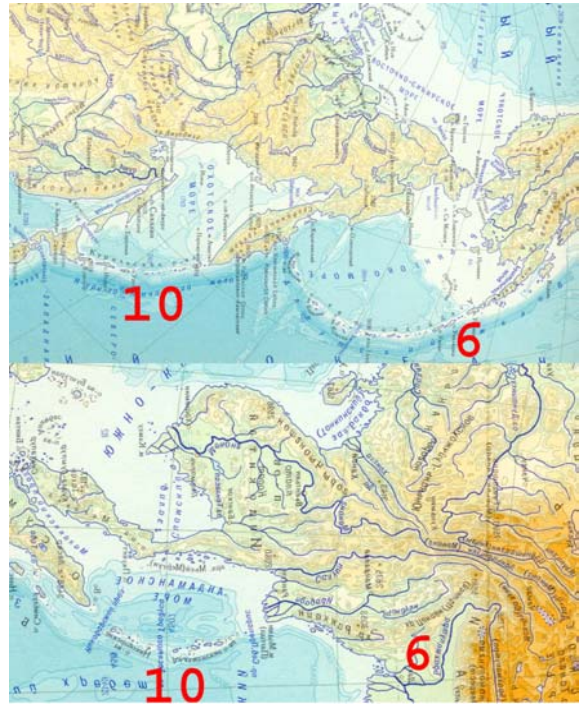

**Fig. 6.** Two arcs of the shock wave (6), (10).

**Fig. 5.** At the lower image (globe) yellow triangle (9) and yellow Greenland (8) (Fig. 1, 2) are recognizable. The triangle (9) partially is inside Arctic circle (3), and the triangle (9) is fully inside the North-America/Himalaya circle. Greenlandiae (8) are placed on the border of the North-America/Himalaya circle.

## 6. AFRICA AND THE PANAMA CANAL INSIDE ARTEMIS CORONA (VENUS).

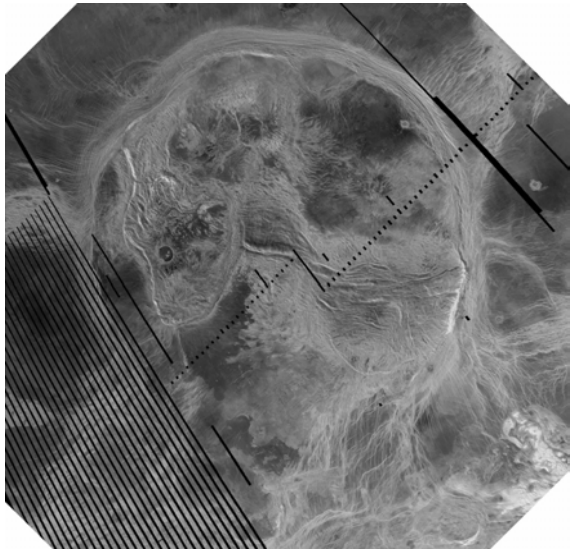

**Fig. 1.** Artemis Corona. PIA00101 after a) 135$^0$ ccw rotation, and b) mirror image with vertical axis of symmetry. There is some Africa inside.

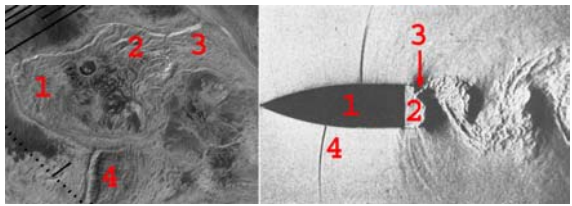

**Fig. 2. Left:** Artemis Africa from Fig. 1 is some kind of South America. See the "Central America" section. **Right:** Karman vortex street (oscillating trace) in air after moving object (1) [1, ph. 67]. High-speed photo. M = 0.6, Re = 220 000.

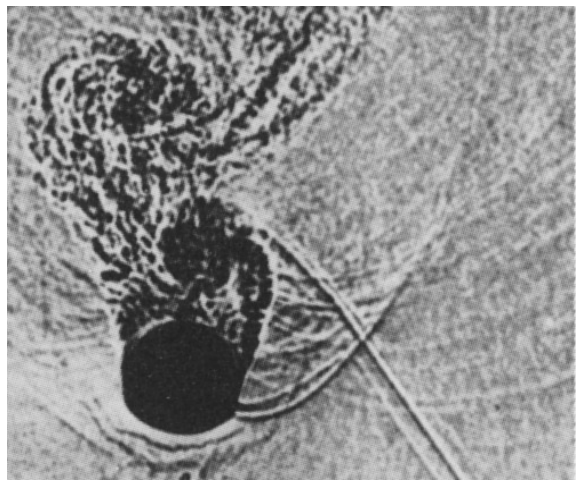

**Fig. 3.** Oscillating trace behind streamlined cylinder [1, ph. 221]. High-speed photo. M = 0.64, Re = 1.35·10$^6$.

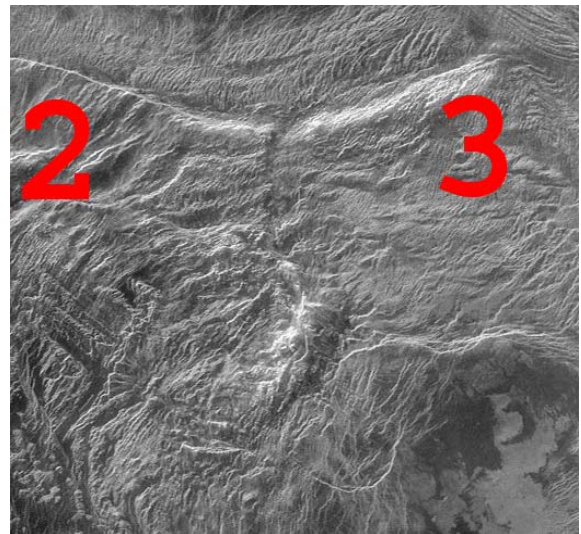

**Fig. 4.** The Panama Canal inside Artemis Corona. The lower side of the canal is ending before connecting the Atlantic.

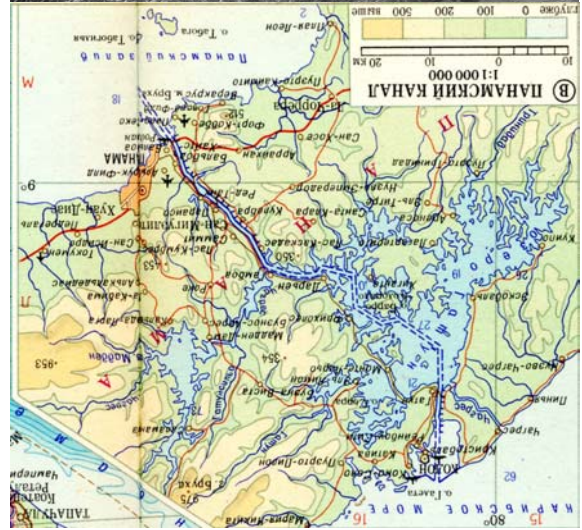

**Fig. 5.** Africae. 1 – nonoscillating body of the flow, 2 – beginning of oscillating trace, 3 – Panama, 4 – triangle of shock waves, 5 – Congo basin. The Mediterranean Sea is some version of the Panama Canal. There is another direction of Venus rotation, and the left picture is a mirror image of Venusian PIA00101 (see Fig. 1).

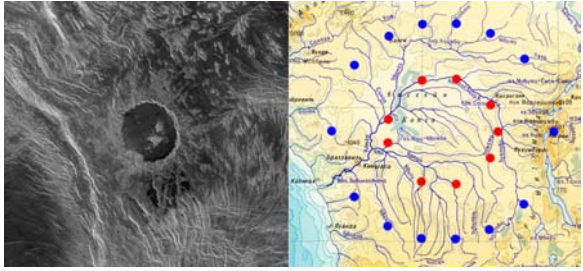

**Fig. 6.** Congo basin inside Artemis Corona. **Right:** Radii of the circles are $(\pi - 3)/2$ and $(\pi - 3)$ radians.

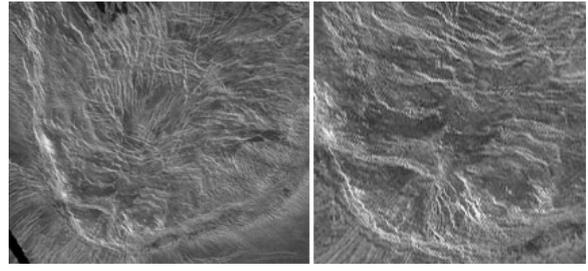

**Fig. 9.** Fragment of Fig. 1. **Right:** The initial vortex of the flow. Artemis Africa is a result of its moving.

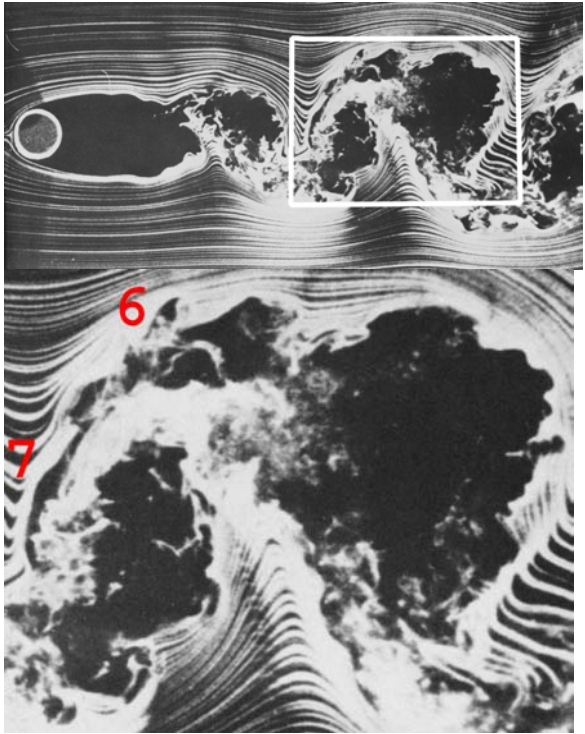

**Fig. 7.** Oscillating trace in liquid behind streamlined cylinder [1, ph. 48]. Re = 10 000. 6 – England, 7 – the border of Artemis Corona. There is some Arabia at the lower left corner.

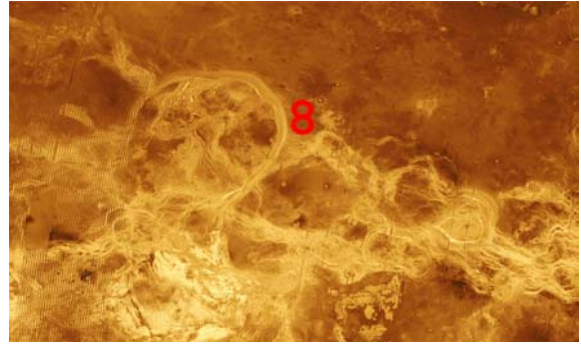

**Fig. 10.** PIA00256 (fragment), mirror image. Suppose 8 be Hindustan.

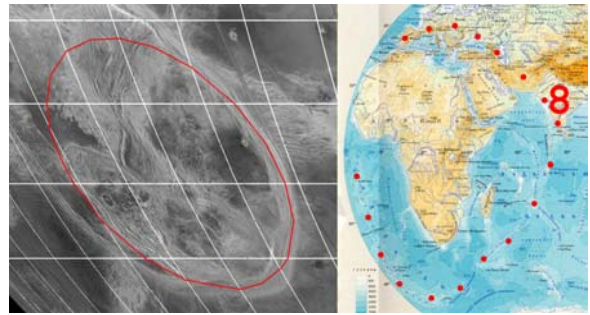

**Fig. 11.** Living areas of Africae. Radius of the left circle is $(\pi - 3)$ radians. The initial point of the left flow (Fig. 9) is being at the border of the $(\pi - 3)$-area.

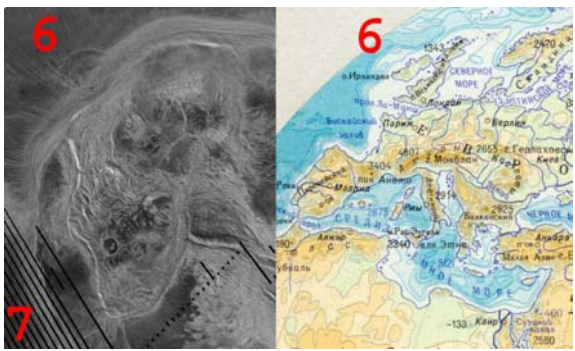

**Fig. 8.** Venusian Europe. 6 – England.

## 7. SUPERSONIC AUSTRALIA.

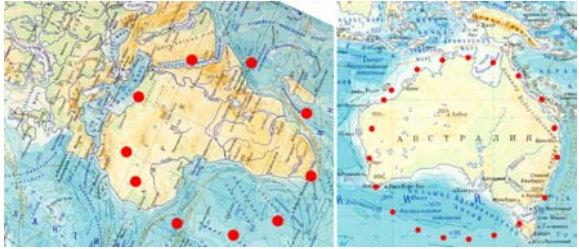

**Fig. 1.** Radii of circles are $4(\pi - 3)$ and $2(\pi - 3)$ radians. The direction of concavity is defined by the direction of planet's rotation (see the "Africa in Artemis Corona" section).

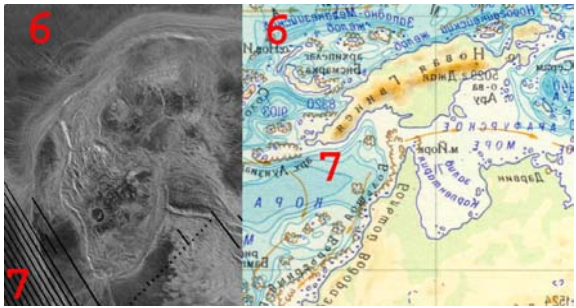

**Fig. 2. Left:** Artemis Corona (Venus). 6 – England, 7 – the border of Artemis Corona. See the "Africa in Artemis Corona" section. **Right:** Australian Europe (mirror image). 6 – England. Names of the islands: New Britain, New Ireland.

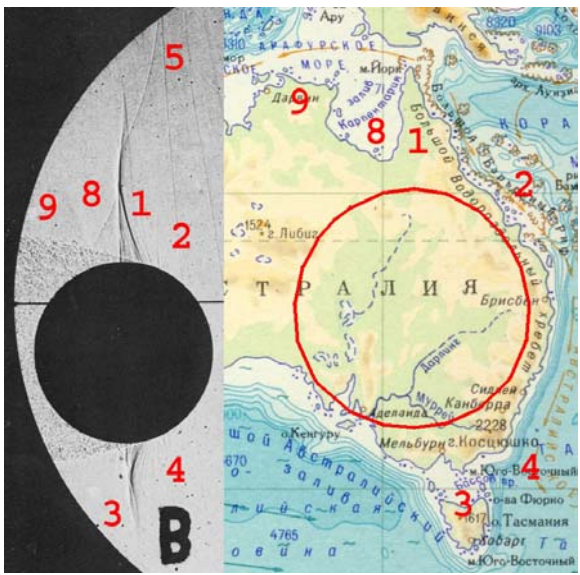

**Fig. 3. Left:** [1, ph. 220]. The ball is moving in air. Re = 920 000. 1–5 – shock waves, 8 – Hudson Bay (Gudzonov zaliv). **Right:** Australia. Radius of the circle is $(\pi - 3)$ radians. Shock waves 1–4 show the direction of moving of the $(\pi - 3)$-basin (the 1-st

Australian moving). The 1-st moving is subsonic, but nearly transonic ($\approx 1$ M), 'cause the shock wave 5 is very long (see Fig. 5, 9).

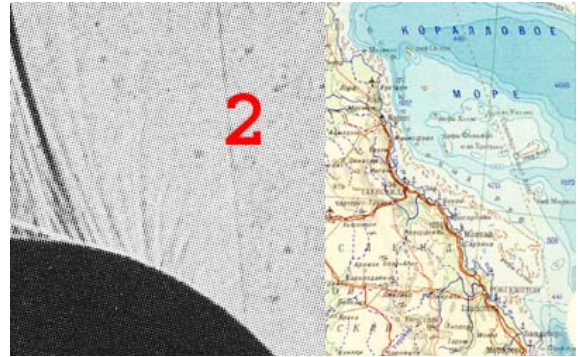

**Fig. 4. Left:** Fragment of Fig. 3 (left). **Right:** Shock waves 2 near Australian body.

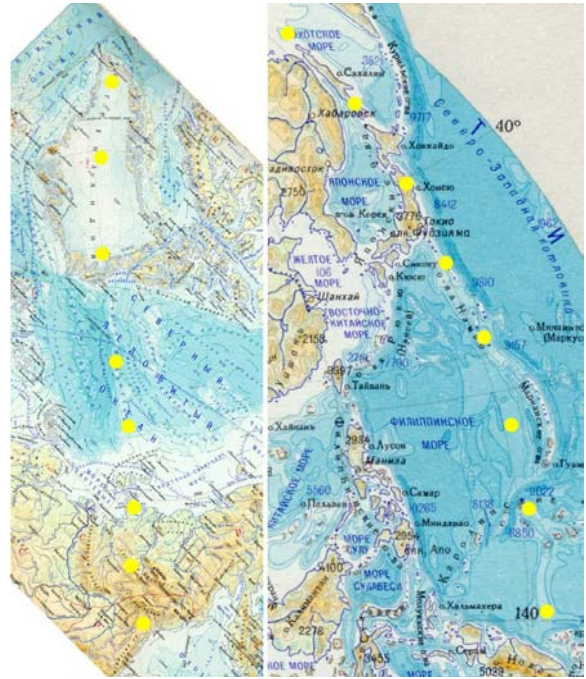

**Fig. 5.** Shock wave 5 (Fig. 3 left) of the 1-st moving at the Arctic and at the Pacific.

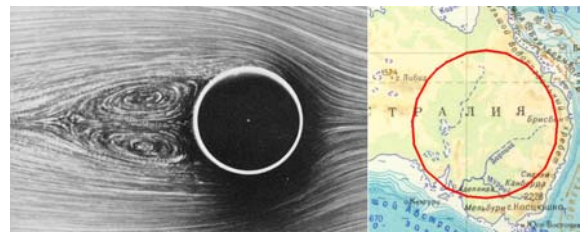

**Fig. 6. Left:** [1, ph. 42]. Two vortices behind streamlined cylinder. Re = 26. **Right:** Two mountain ranges behind moving $(\pi - 3)$-basin.

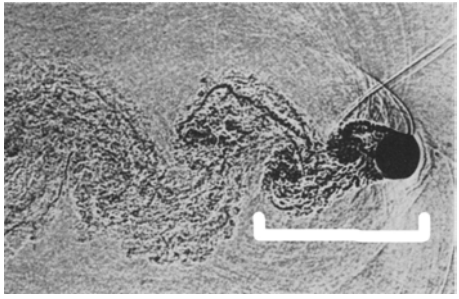

**Fig. 7.** Oscillating trace behind streamlined cylinder [1, ph. 221]. M = 0.64, Re = 1.35·10⁶. Body of Australia is shown. There isn't any Karman vortex street behind Australian body (at the direction of the 1-st moving). However, some another trace exists (the 2-nd Australian moving). See Fig. 9, 11. Maybe the 1-st trace is suppressed by the 2-nd trace, or maybe the 1-st moving is short in time.

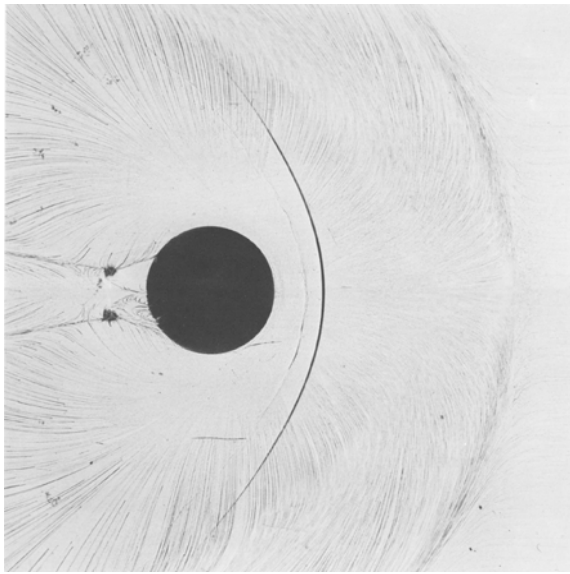

**Fig. 8.** Supersonic flow around cylinder [1, Introduction]. M = 2.5, Re = 735 000. There are shock waves before the cylinder.

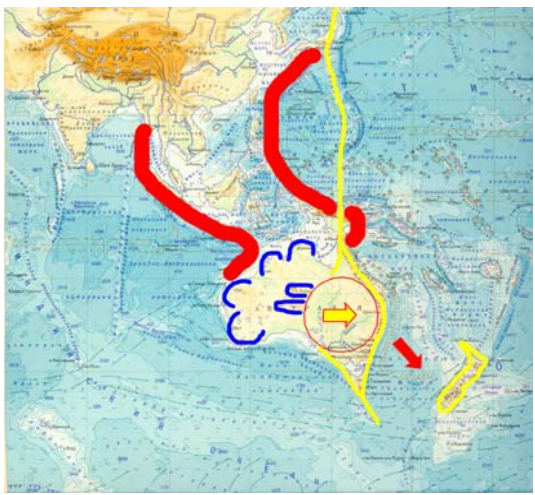

**Fig. 9.** Oscillating trace (red) is a trace of the 2(π − 3)-body. New Zealand is a shock wave of the 2-nd

moving. Shock waves are before moving body, so moving is supersonic. Oscillating of the trace is small, and the trace is very turbulent (isn't continuous).

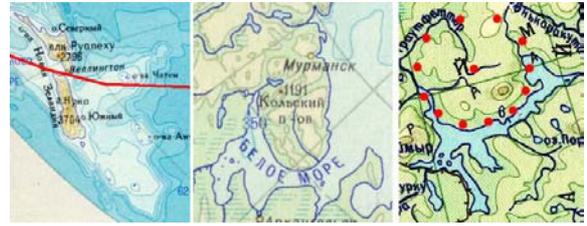

**Fig. 10.** White Seae. **Left:** New Zealand, fragment of Fig. 11. **Center:** The White Sea. **Right:** Taimyr Lake at the Taimyr Peninsula. Radius of the circle is (π − 3)/16 radians.

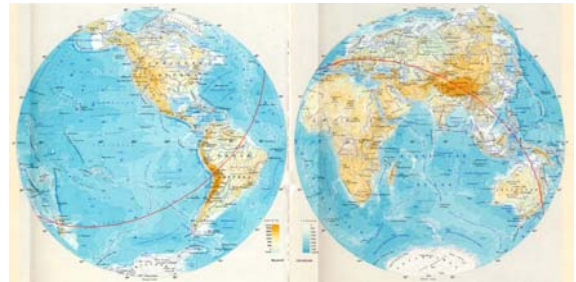

**Fig. 11.** Straight line contains the most Eastern point of Australia and the deepest point of the South America's Western concavity (the Andes, Peru). This line goes through: 1) the center of the 1-st shock wave (New Zealand, see Fig. 10 left), and 2) through the center of the 2-nd shock wave (the Pacific, see Fig. 12). Thus the Andes are the result of the action of the 3-rd shock wave. The 3-rd shock wave maybe is acting on all Western American coast. (We don't discuss here such question as: whether the shock waves are "perfect shock waves" (breaks) or "approximately shock waves". )

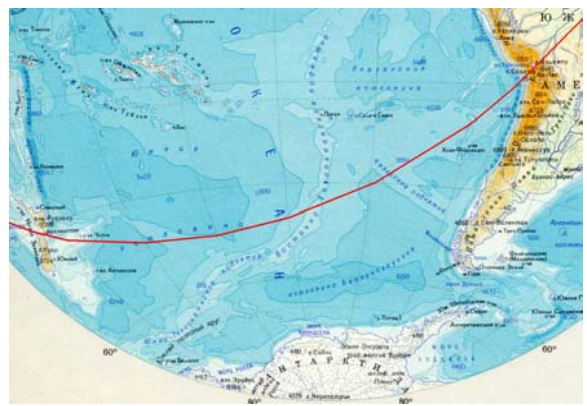

**Fig. 12.** Fragment of Fig. 11. The 1-st (left), the 2-nd (center), and the 3-rd (right) shock waves of the 2-nd Australian moving. See Fig. 8.

## 8. MOVING BASINS.

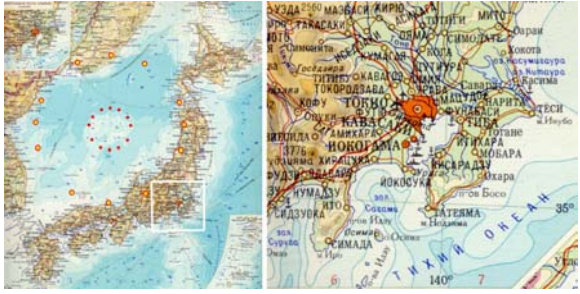

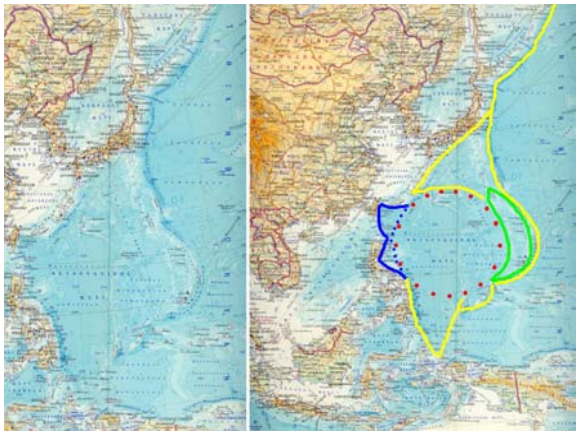

**Fig. 1.** The Sea of Japan. **Upper left:** Radii of the circles are ($\pi - 3$)/2 (big circle) and $(\pi - 3)^2$ (small circle) radians. **Upper right:** Fragment of the upper left. Breast of the moving basin. This element doesn't exist in [1]. Here breast is oriented by the Australian subsonic shock wave 5 (see the "Supersonic Australia" section). **Lower:** Scanned globe. Yellow: clusters of shock waves (fins); green: vortex area before moving body (foregoing area); orange: tits; blue: two back vortices (buttocks, or a fish tail). The initial basin is concave ($< 0$), so the foregoing area is convex ($> 0$). Shock waves (yellow) are very long, it means subsonic ($< 1$ M), but nearly transonic ($\approx 1$ M) moving.

**Fig. 2.** The Philippine Fish. Radius of the circle is ($\pi - 3$) radians. Fins (triangle clusters of shock waves, yellow) are shorter than at Fig. 1, so speed is less than at Fig. 1. The foregoing area is nearly convex ($> 0$), with islands. The frontal shock wave is placed on the secondary body, not foregoing. Subsonic moving? See, however, the "Chucotcae and Siberiae" section, Fig. 1. The upper shock wave (concave) is very long.

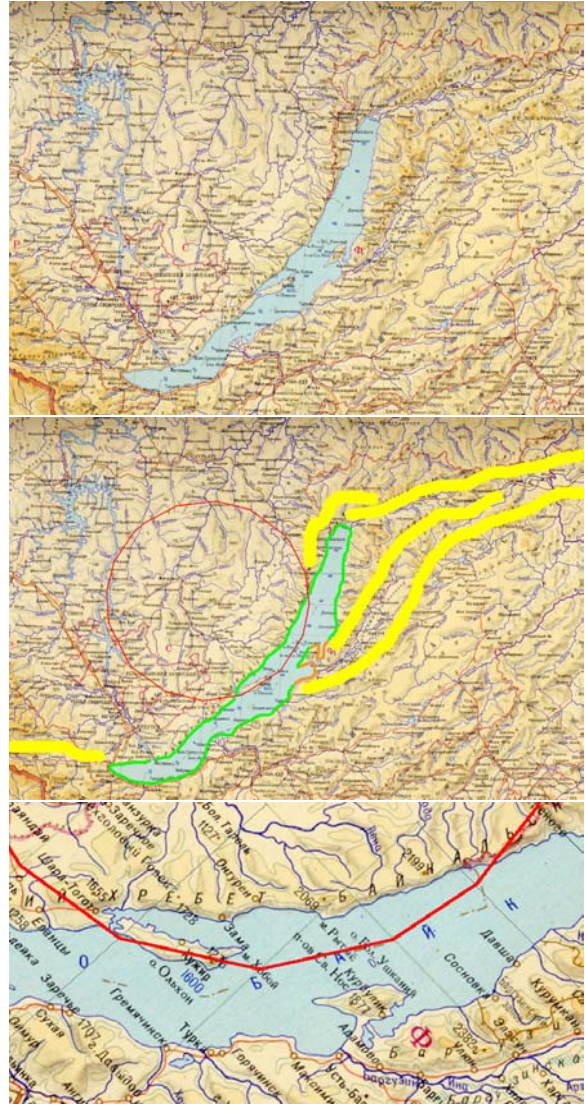

**Fig. 3.** Baikal Lake. Radius of the circle is ($\pi - 3$)/5 radians. 1) Here we see an explanation of flat fronts of some moving basins. 2) Back side seems to be triangle, not as at Fig. 1, 2. See Fig. 4.

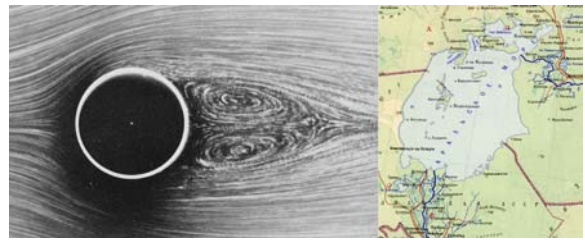

**Fig. 4.** Triangle tail of moving basin. **Left:** [1, ph. 42]. Triangular trace behind streamlined cylinder. Re = 26. **Right:** The Aral Sea. Triangle clusters of shock waves, flat front, triangle back. Radius of moving circle is ($\pi - 3$)/8 radians.

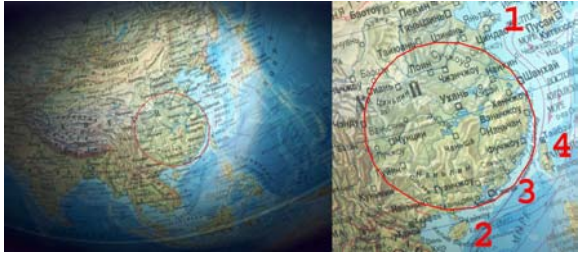

**Fig. 5.** SE Asia. Globe. See the "North America in Asia" section. 1, 2 – shock waves, 3 – flat front. Radius of the circle is $(\pi - 3)$ radians.

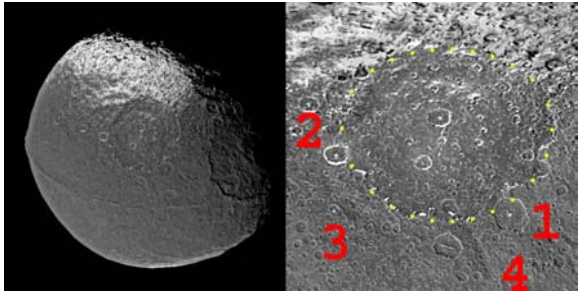

**Fig. 6.** Iapetus, Saturn's satellite. **Left:** PIA06166. **Right:** PIA08406 (fragment). Radius of the circle is $2(\pi - 3)$ radians. 1, 2 – shock waves, 3 – foregoing area, 4 – Taiwan, see Fig. 5.

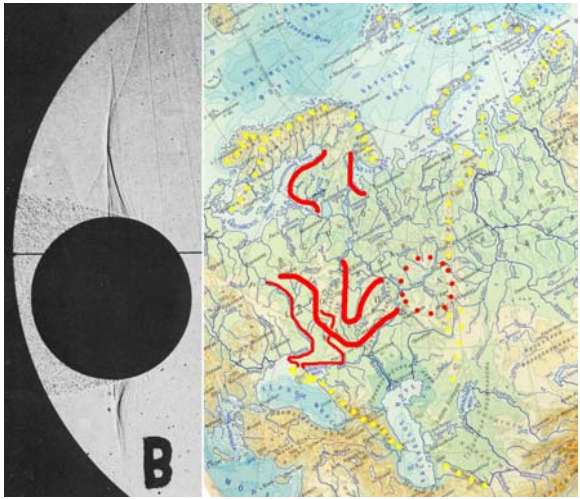

**Fig. 7. Left:** [1, ph. 220]. The ball is moving in air. Re = 920 000. **Right:** The Urals, Scandinavia, the Crimea. Radius of the Ural circle is $(\pi - 3)/3$ radians, the center is $(55.9911^0$ N, $54.439^0$ E). Distances from this center are: 1) to Yaroslavl (the 2-bridge over Volga): $(\pi - 3)$ radians exactly (for exactly spherical Earth); 2) to the center of the Eastern hemisphere $(0^0$ N, $70^0$ E): 1 radian, the difference is 11.2 km [3, 4, 6].

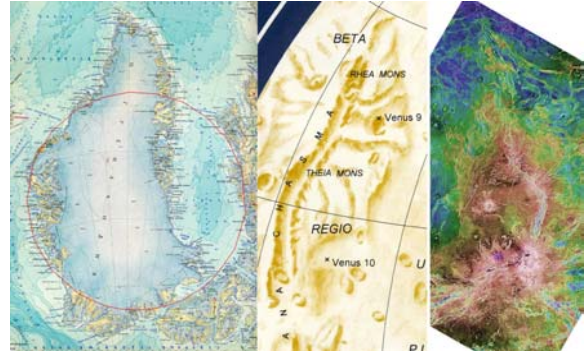

**Fig. 8.** Greenlandiae. See Fig. 7 left. See the "North America in Asia" section. **Left:** Greenland. Radius of the circle is $(\pi - 3)$ radians. **Center:** Venus, Beta [11]. **Right:** Venus, Maat. PIA00159 (fragment).

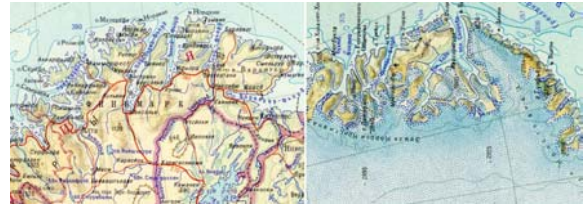

**Fig. 9.** Breast of Scandinavia (left) and of Greenland (right). See Fig. 1, 3.

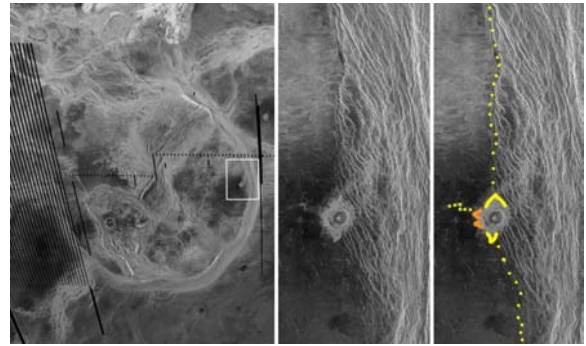

**Fig. 10.** PIA00101. Behn, a crater inside Artemis Corona (Venus), has triangle fins. It stands on the top of its trace. Shock waves (Fig. 7) are very long.

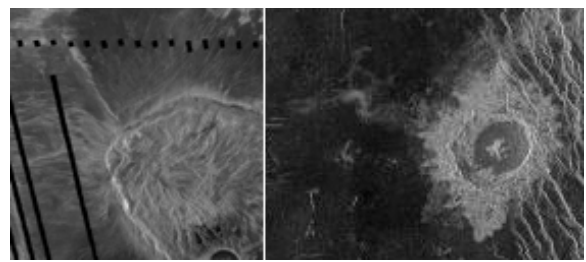

**Fig. 11.** PIA00101 (fragments, see Fig. 10). Shock waves and other features towards the direction of moving. See Taiwan 4 in Fig. 5, 6. **Left:** Artemis Africa, see the "Africa inside Artemis" section.

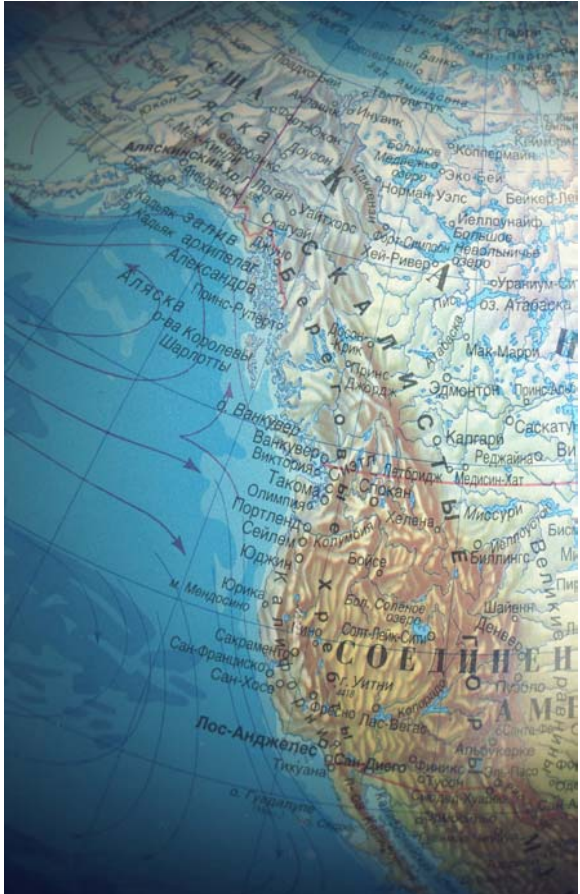

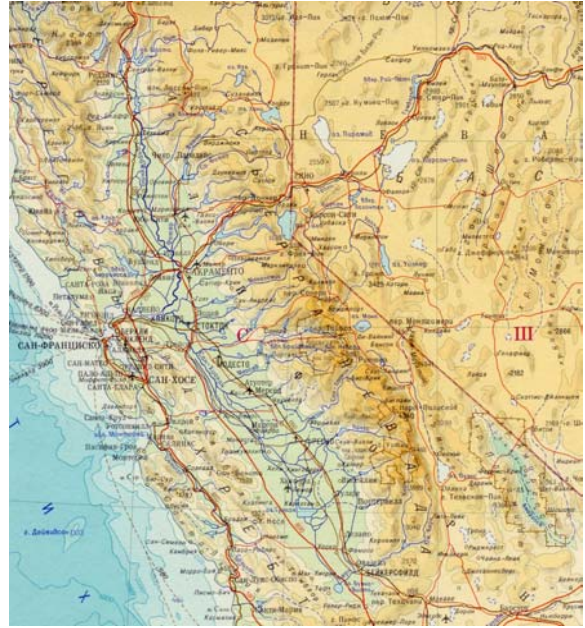

**Fig. 13.** North America, California, Big Basin. A folding before the $(\pi - 3)/2$-disc (<span style="color:red">red</span> circle at Fig. 12).

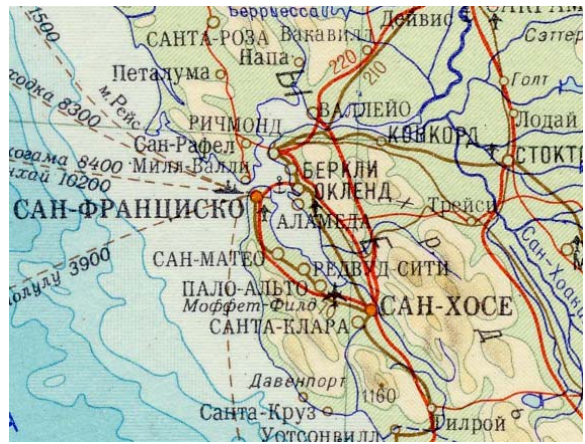

**Fig. 14.** California, fragment of Fig. 13. Breast of moving basin. Compare with Fig. 1, 3, 9.

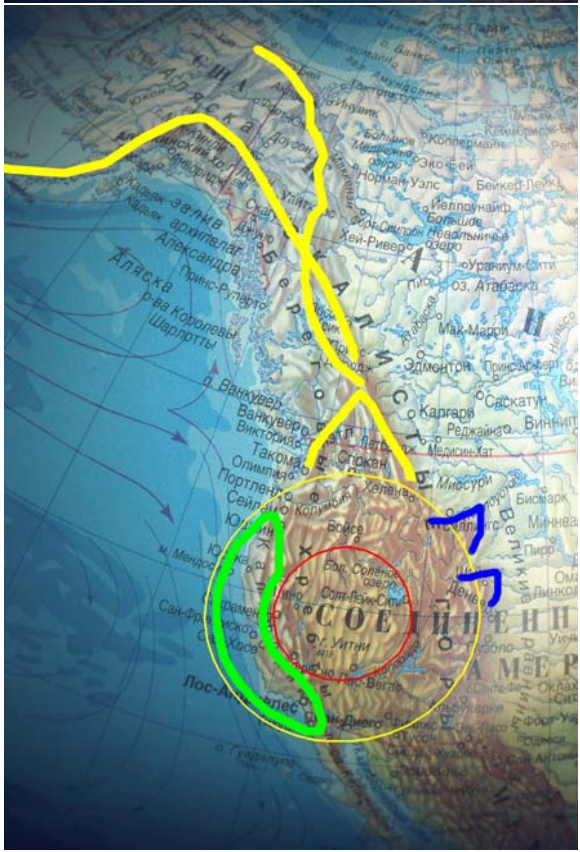

**Fig. 12.** North America, Big Basin. Globe. Radii of the circles are $(\pi - 3)/2$ and $(\pi - 3)$ radians.

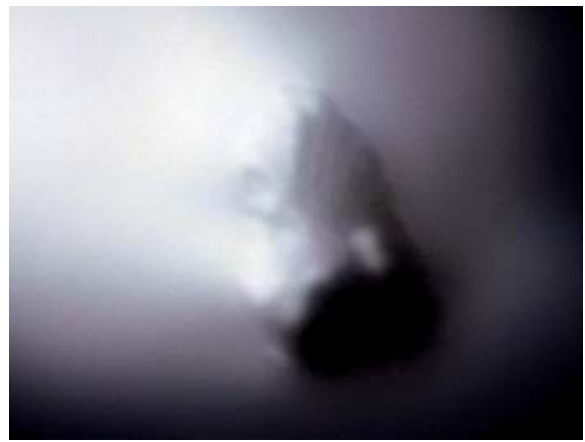

**Fig. 15.** The Halley comet nucleus. The image of "Giotto" 13 March 1986. Moving from left to right. Compare with Elysium, the "North America on Mars" section, Fig. 8.

## 9. NORTH AMERICA ON MARS.

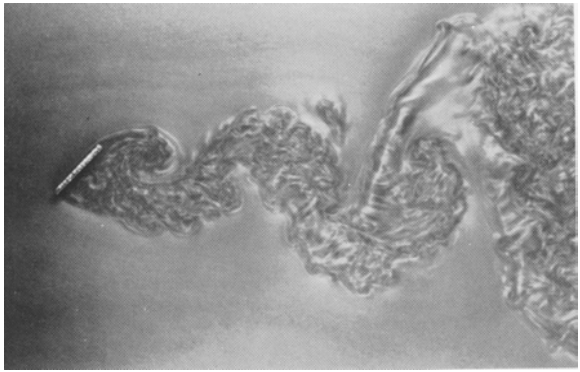

**Fig. 1.** Oscillating trace behind inclined planar plate [1, ph. 172]. The angle is $45^0$, Re = 4300.

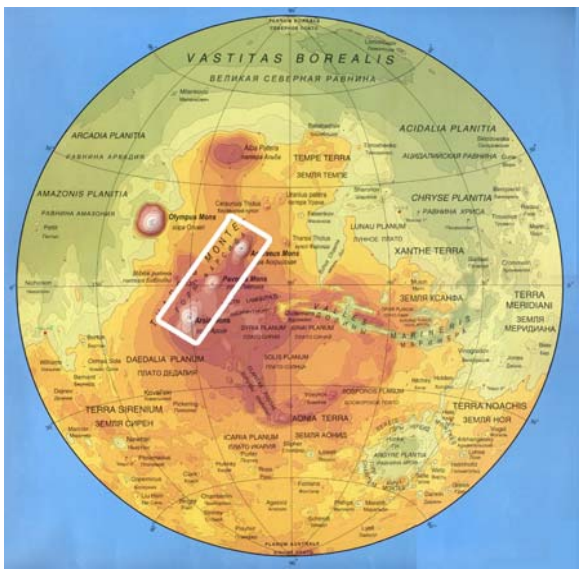

**Fig. 2.** Mars [7]. The Tharsis Montes are a moving inclined plate (Fig. 1). Valles Marineris are a vortex trace behind the plate.

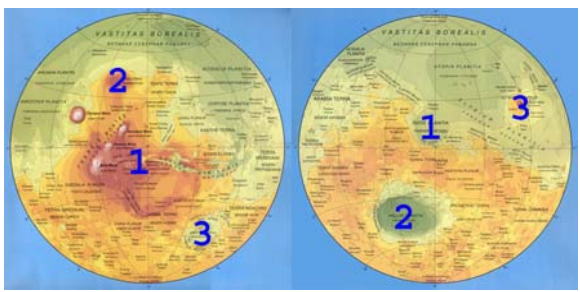

**Fig. 3.** Well-known antipodal objects. Distances are: 1–1: 3 radians, the difference $\Delta \approx 2$ mm on the map 1:26 000 000; 2–2: a) from Alba to the $2(\pi-3)$-center of Hellas $\approx \pi$ radians, $\Delta \approx 3.5$ mm; b) from Alba to the $(\pi-3)$-center of Hellas $3 + (\pi-3)/2$ radians, $\Delta \approx 0.4$ mm; 3–3: $\approx 2.7$ radians.

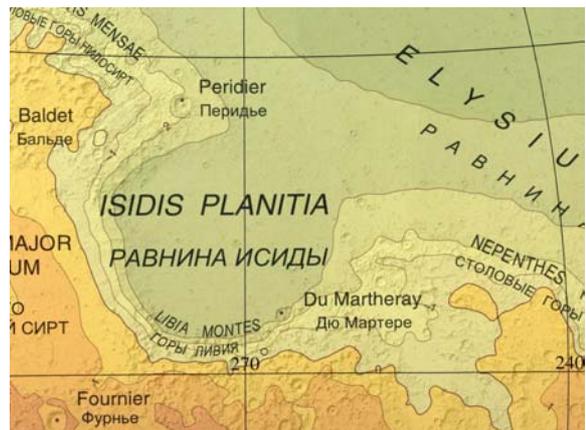

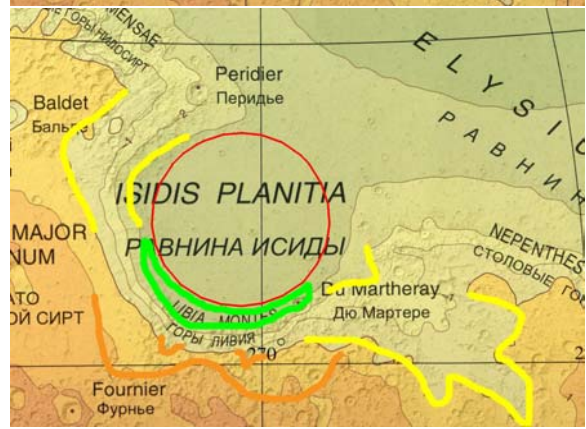

**Fig. 4.** Isidis is a moving basin. See the "Moving basins" section. The direction of moving is coordinated with moving of the antipodal vortice. Radius of the circle is $(\pi-3)$ radians.

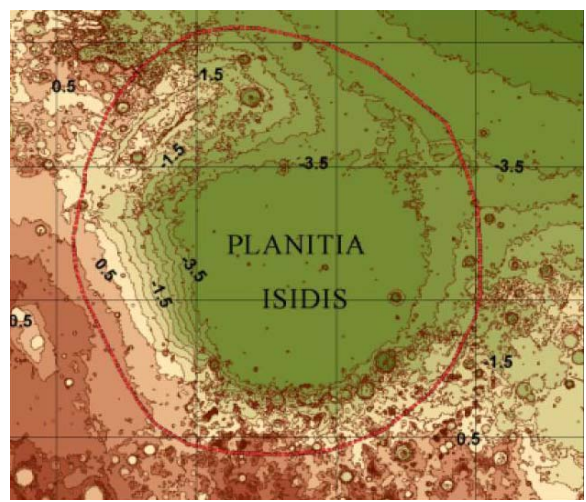

**Fig. 5.** Isidis [12]. The isthmus (Panama) of the oscillating trace (deep green). Isidis is a concave analog of the Crimea (see the "Antarctidae" section, Fig. 4).

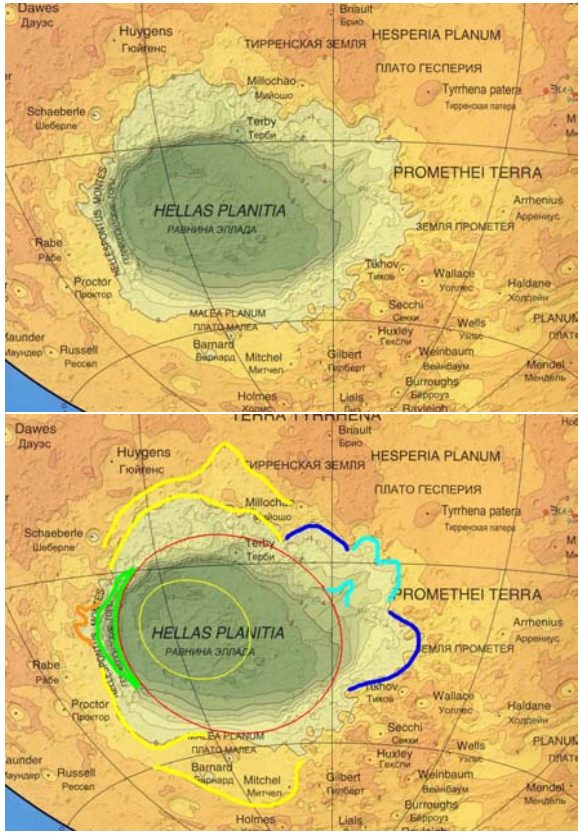

**Fig. 6.** Hellas is a moving basin. Radii of the circles are $(\pi - 3)$ and $2(\pi - 3)$ radians. The oscillating tail is being, see Fig. 7.

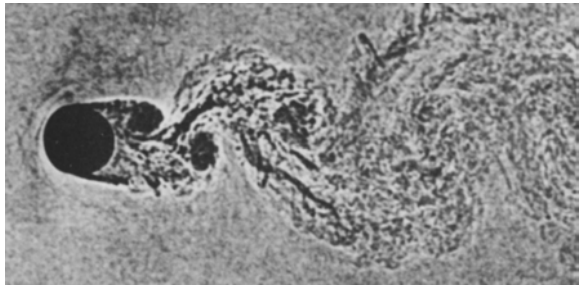

**Fig. 7.** [1, ph. 221]. M = 0.45, Re = 110 000.

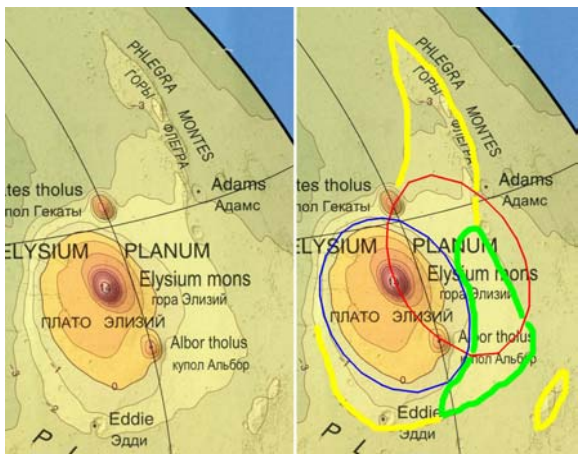

**Fig. 8.** Elysium is a moving basin. Radii of the circles are $(\pi - 3)$ radians.

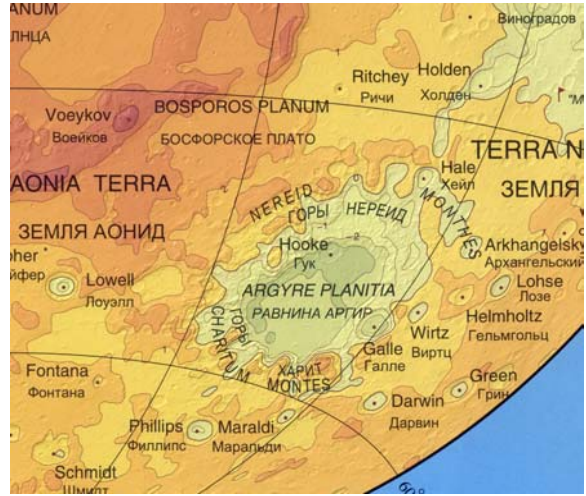

**Fig. 9.** Argyre is a moving basin: green shock wave (upper right) and flat front (green, right side) are being. The direction is coordinated with Elysium (an antipode). See the "$(\pi-3)$-basins" section, Fig. 8.

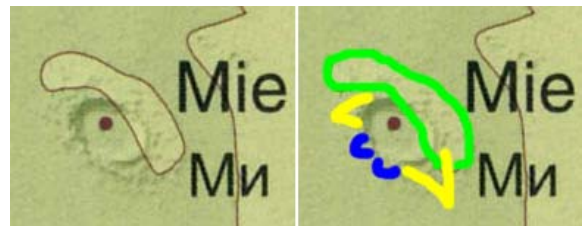

**Fig. 10.** Mie is an exact antipode of Argyre. Mie's direction of moving is coordinated with Argyre.

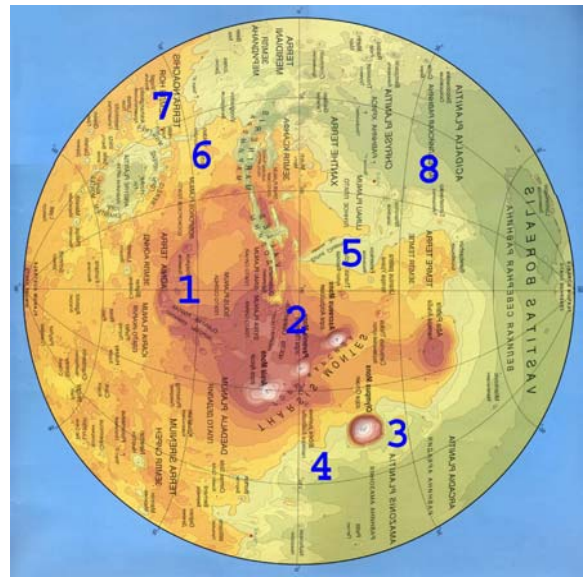

**Fig. 10.** Mars, $90^0$ ccw rotated mirror image. The Tharsis Montes are the Appalachians. 1 – California, 2 – Great Lakes, 3 – the Bermudas, 4 – Florida, 5 – Hudson Bay (Gudzonov zaliv), 6 – Alaska, 7 – Chukotca, 8 – Greenland.

## 10. ANTARCTIDAE.

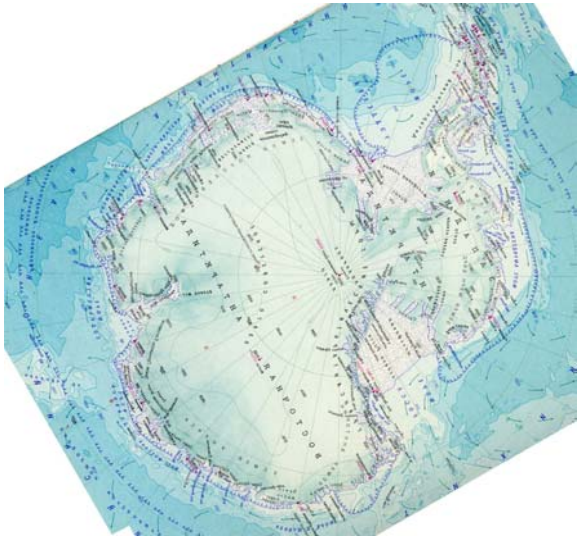

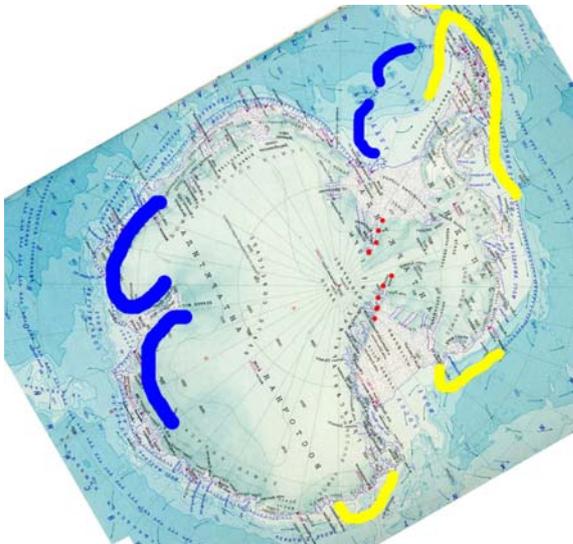

**Fig. 1.** Mirror image. Antarctida is a moving basin. Yellow are fins of shock waves, red is an isthmus (Panama) of an oscillating trace, blue are buttocks.

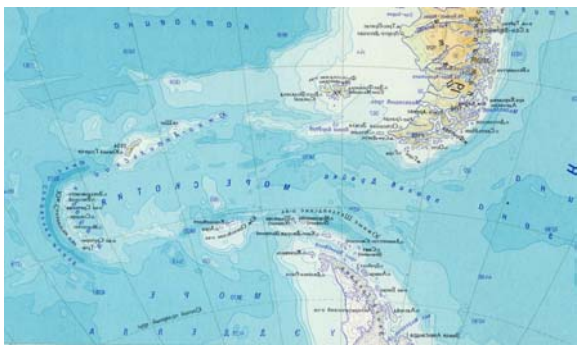

**Fig. 2.** Long shock wave of Antarctida. Mirror image.

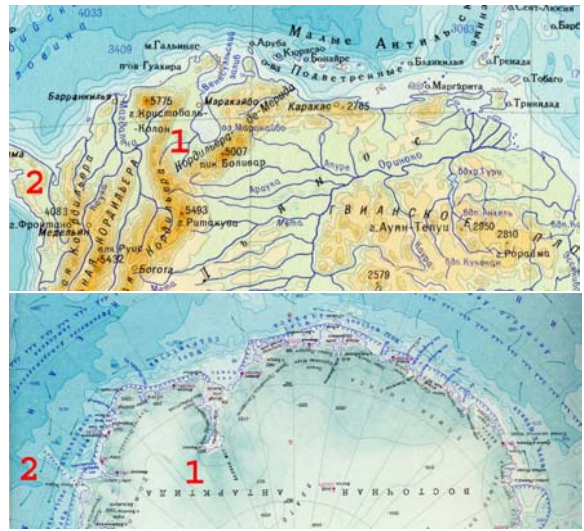

**Fig. 3.** The Gulf of Venezuela (1) and Panama (2) in South America and in (non-mirror) Antarctida. See the "Central America" section.

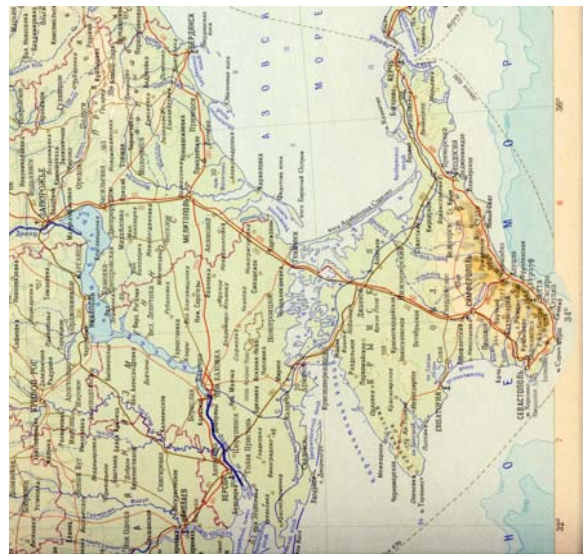

**Fig. 4.** The Crimea. There are analogous Crimean mountains in Antarctida. Also see the "Moving basins" section, Fig. 7.

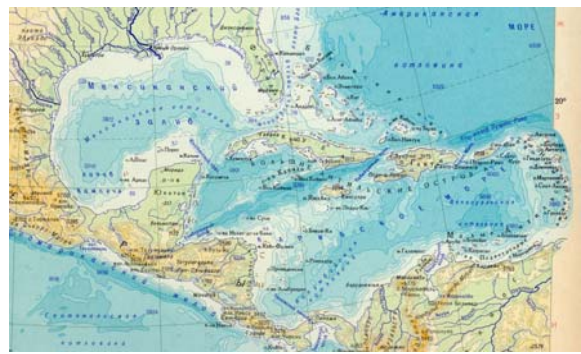

**Fig. 5.** Caribbean Antarctida (concave).

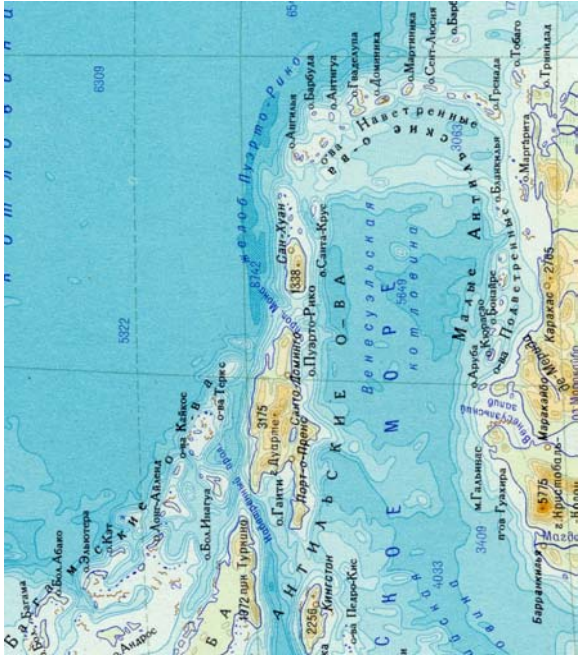

**Fig. 6.** Shock wave of Caribbean Antarctida (Fig. 2).

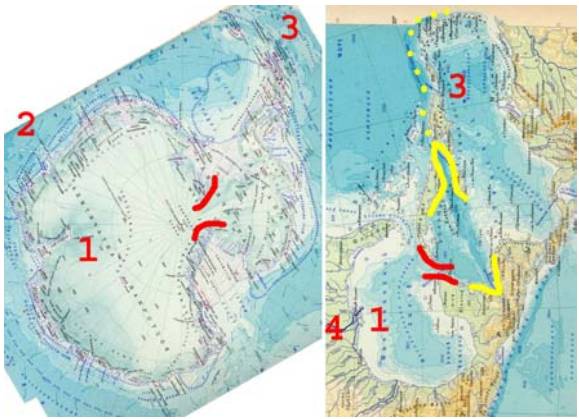

**Fig. 7.** Red lines are an isthmus of an oscillating trace. It is oscillating down in Antarctida, and it is oscillating up in the Caribbean Sea. 1 – the Gulf of Venezuela. 3 – shock wave of another sign, it is a deepening in Antarctida, and it is a chain of islands in the Caribbean Sea. Caribbean Antarctida has flat front, see the "Moving basins" section.

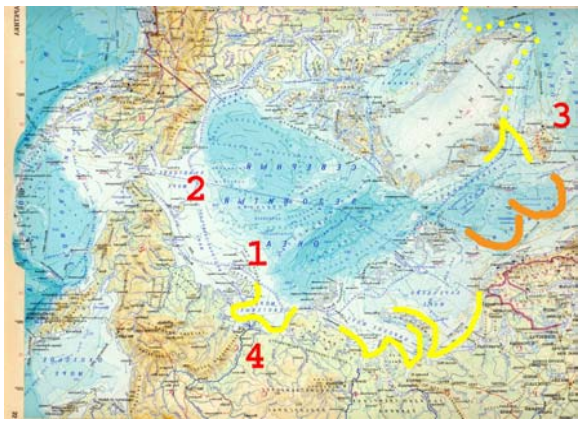

**Fig. 8.** Arctic Antarctida, mirror image.

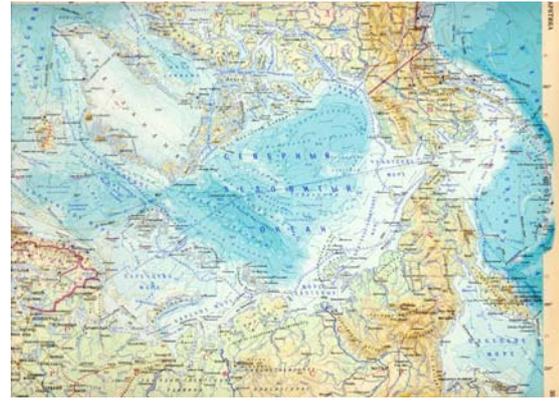

**Fig. 9.** Antarctida on North Pole. Non-mirror image.

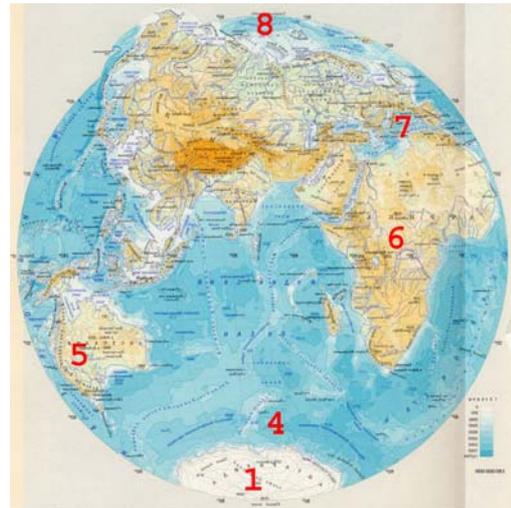

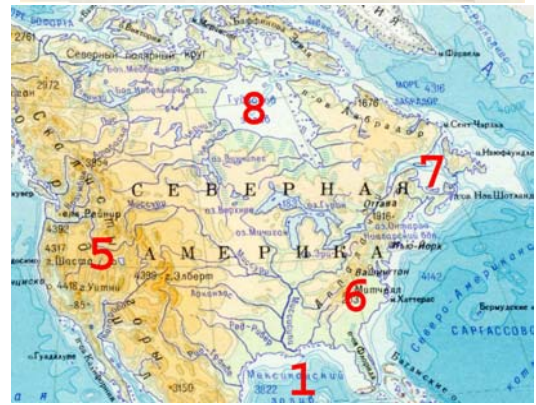

**Fig. 10.** Mississippi (Fig. 7) in the Indian Ocean.

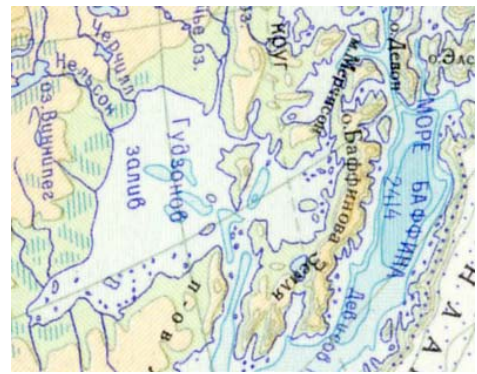

**Fig. 11.** Hudson Antarctida. See Fig. 9, 10.

## 11. MOSCOW ANTARCTIDA IN THE MILKY WAY.

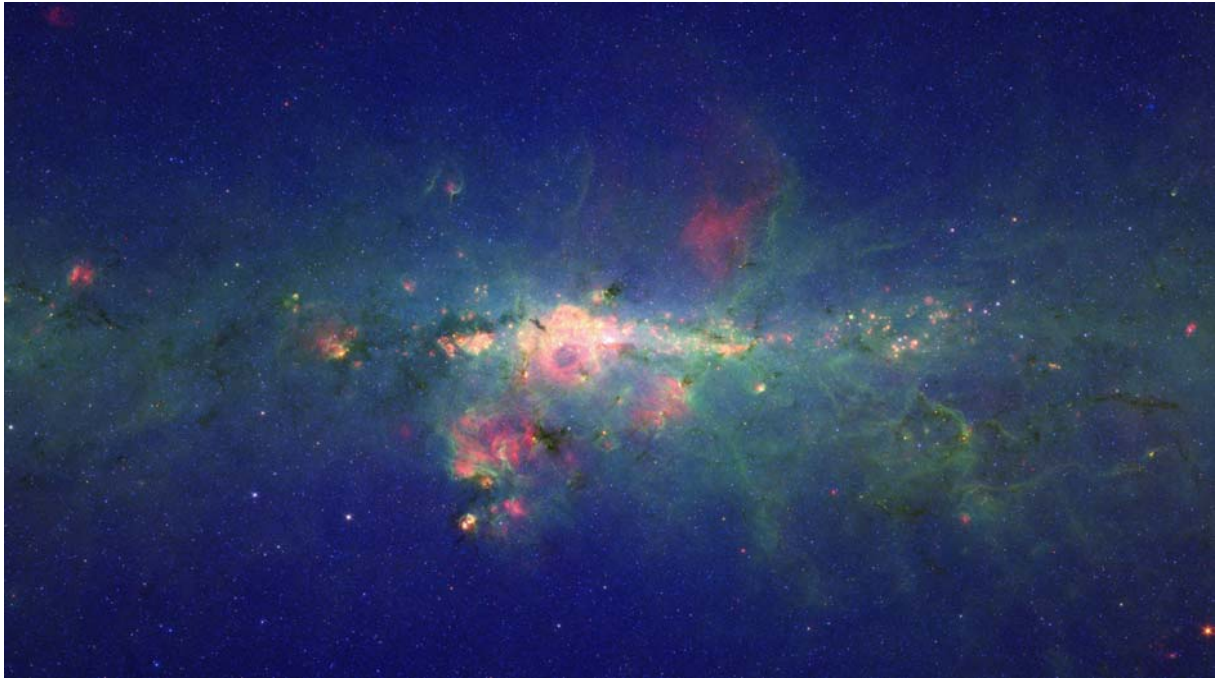

**Fig. 1.** PIA10955. The center of the Milky Way.

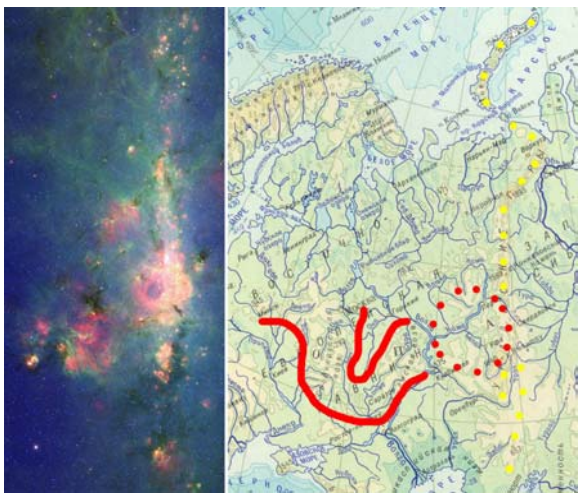

**Fig. 2.** Ural moving basin and (mirror image) the Milky Way. See the "Moving basins" section.

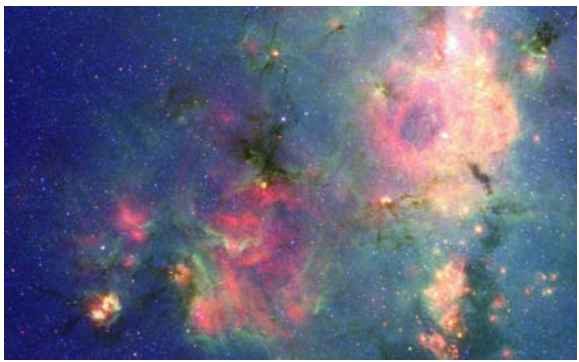

**Fig. 3.** Central Russia. Antarctida is left.

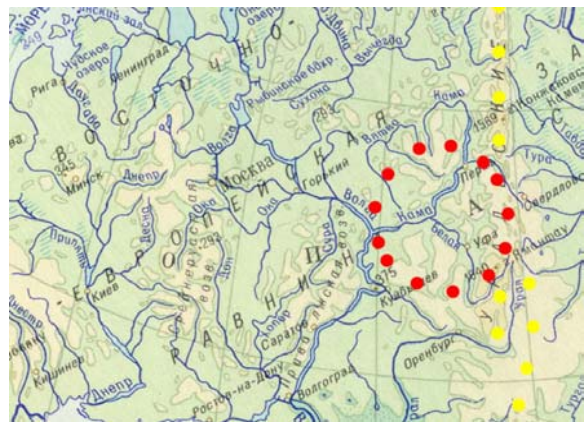

**Fig. 4.** Oscillating trace behind Ural moving basin.

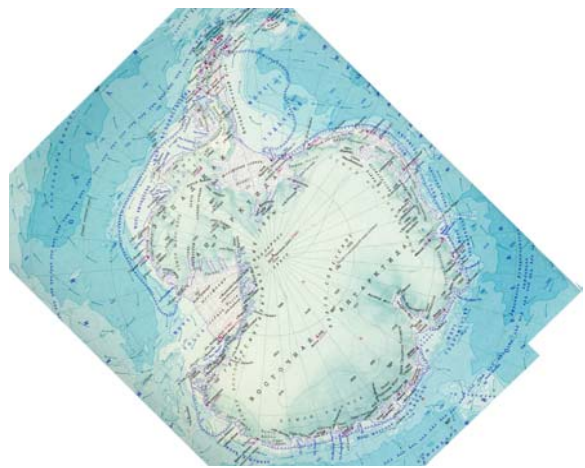

**Fig. 5.** Antarctida. See the "Antarctidae" section.

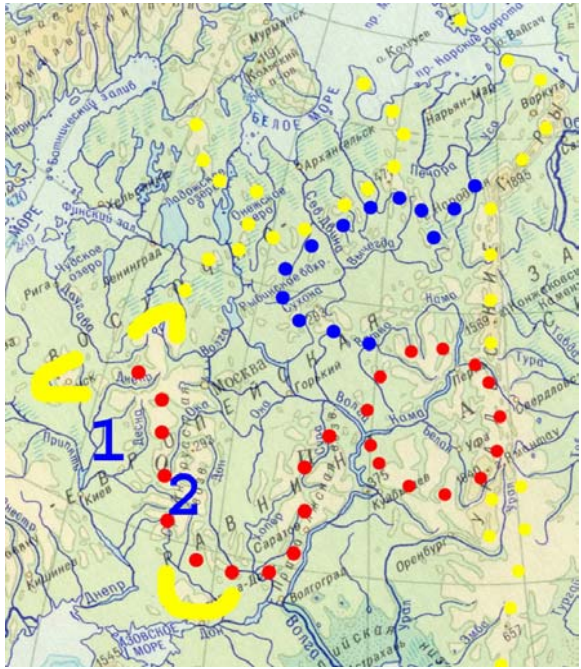

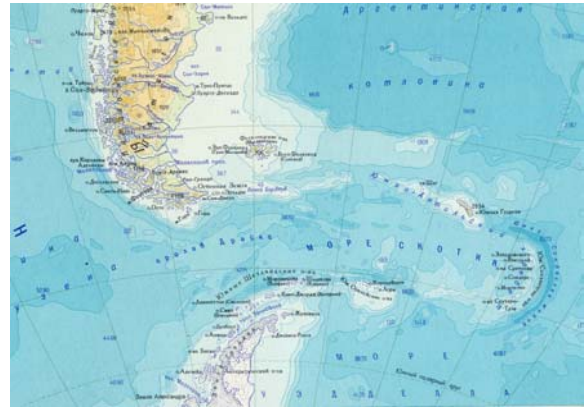

**Fig. 7.** Long shock wave connects Antarctida with a big continent (South America).

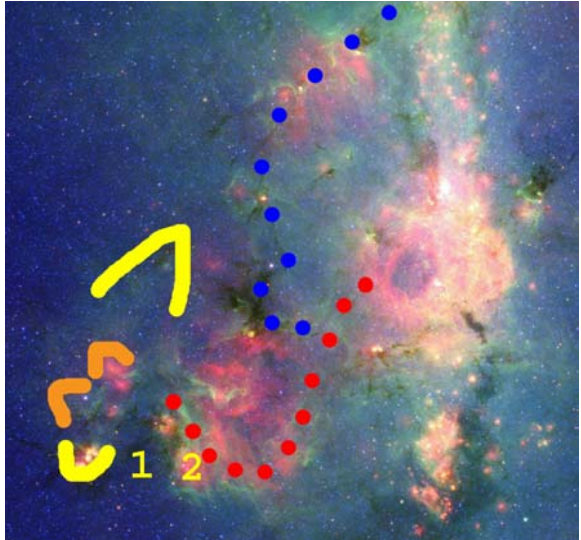

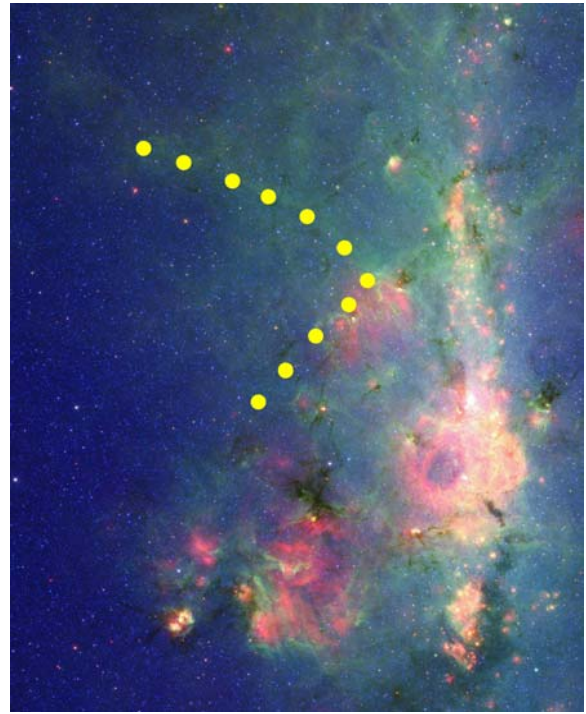

**Fig. 8.** By Fig. 6 this shock wave seems to be Antarctidean. See Fig. 7.

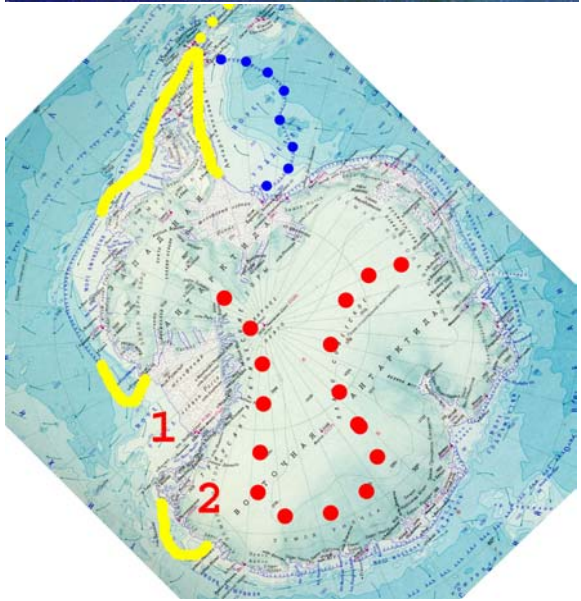

**Fig. 6.** Comparative planetology.

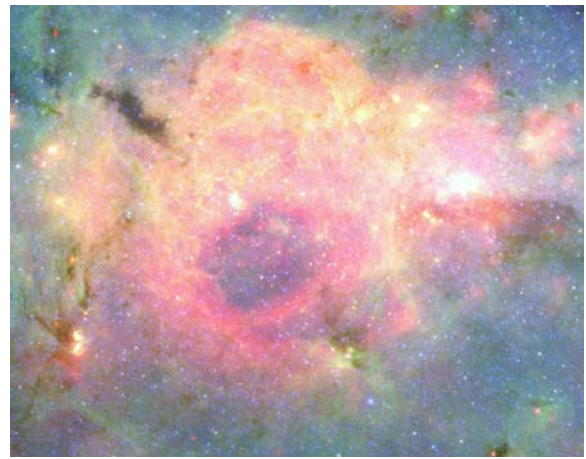

**Fig. 9.** Structure of the moving basin.

## 12. EXTRATERRESTRIAL ANTARCTIDAE.

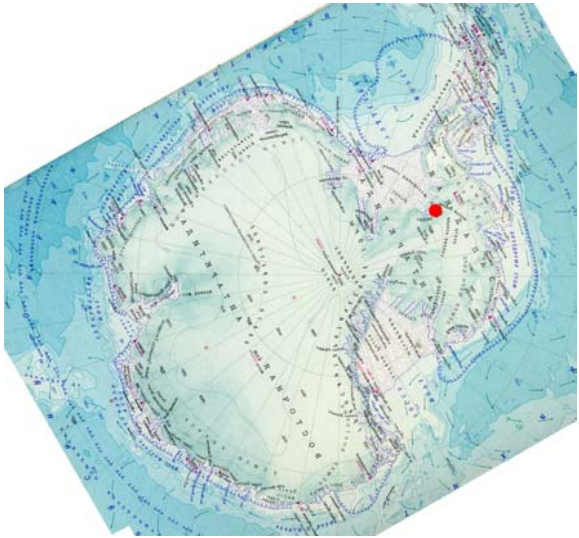

**Fig. 1.** Antarctida (mirror image) in its normal form (from left to right). Winson, the highest mount of Antarctida, is pointed.

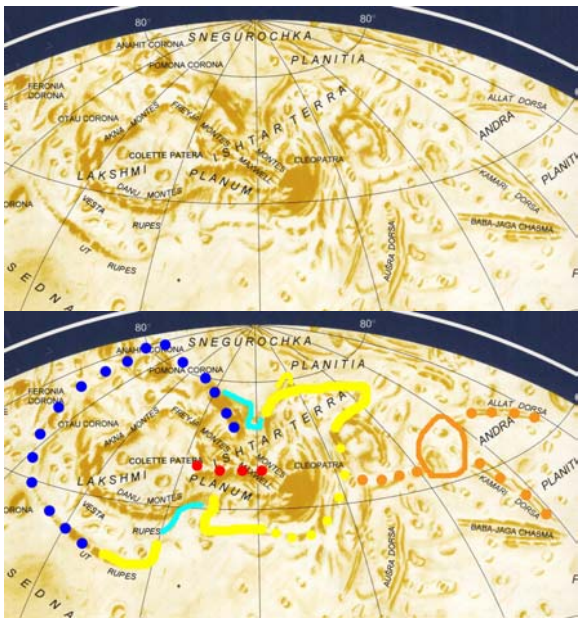

**Fig. 2.** Venus, Ishtar [11]. Isthmus of oscillating trace is red. Clusters of shock waves are yellow. Foregoing feature (orange) is an interesting one.

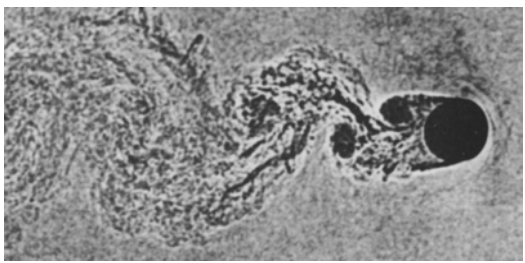

**Fig. 3.** [1, ph. 221]. M = 0.45, Re = 110 000.

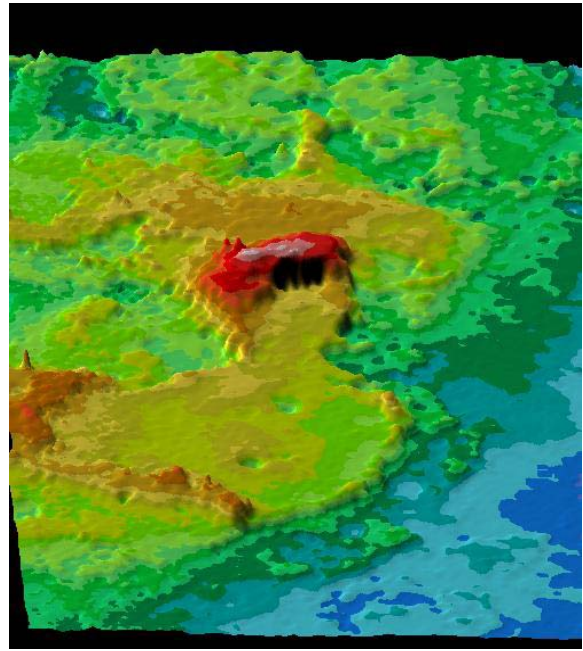

**Fig. 4.** PIA00093. Ishtar. Maxwell = Winson (Fig.1).

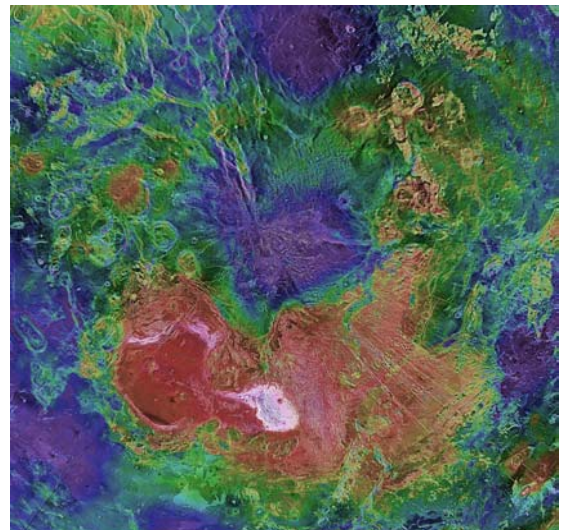

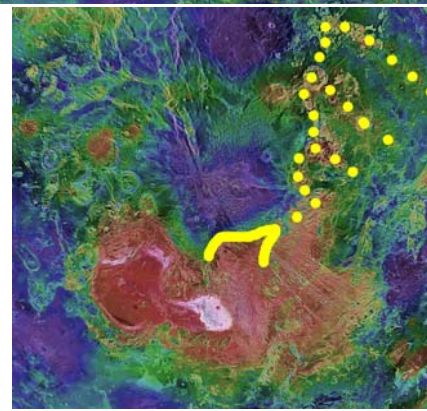

**Fig. 5.** PIA00007 (fragment). The upper shock wave of Ishtar Antarctida.

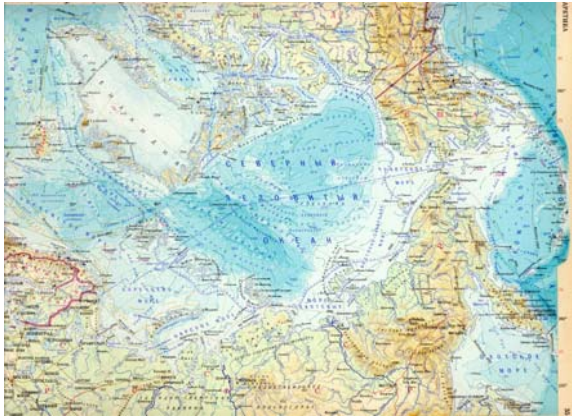

**Fig. 6.** Arctic Antarctida, see the "Antarctidae" section, Fig. 8, 9.

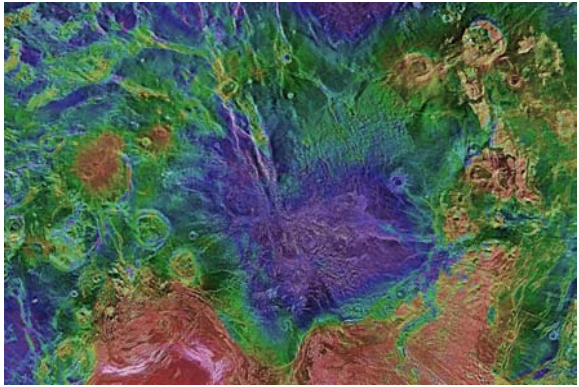

**Fig. 7.** PIA00007 (fragment). Venus, North Pole. Venusian Arctic Antarctida (blue, concave) is one-to-one to its analog on the Earth (Fig. 6).

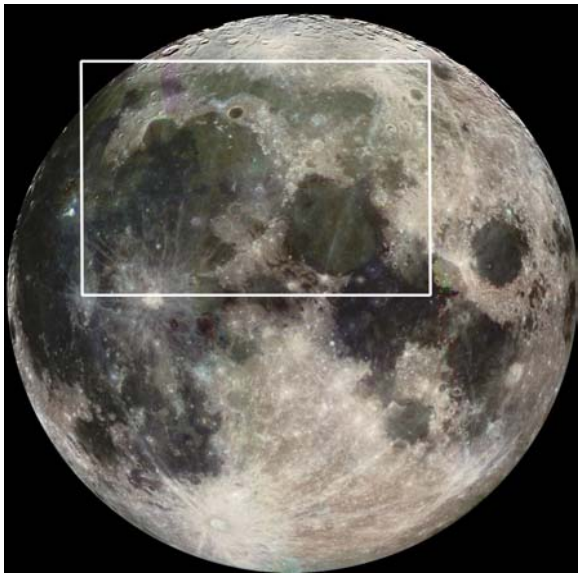

**Fig. 8.** PIA00405. There is very revealing Antarctida on the Moon. It is in normal form (from left to right). See Fig. 3, 10.

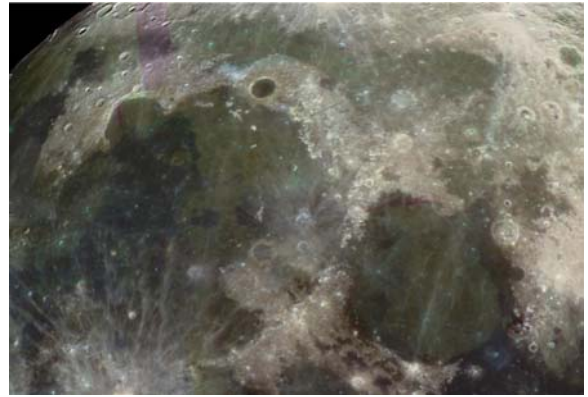

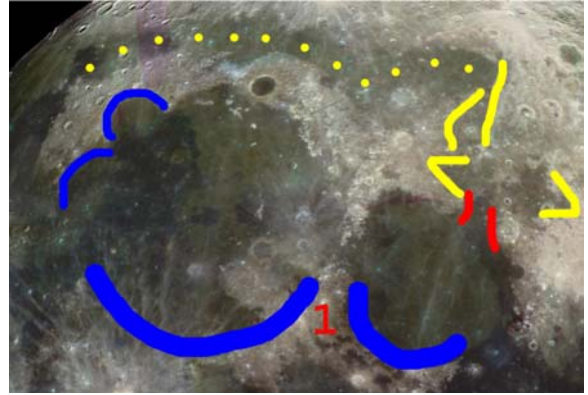

**Fig. 9.** Imbrium Antarctida (concave).

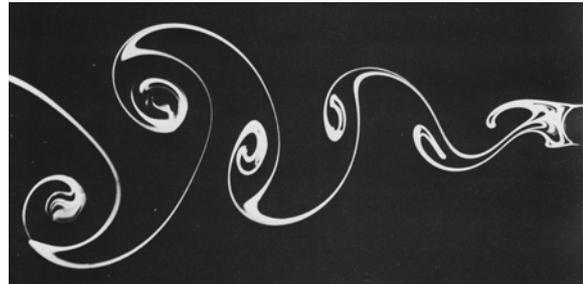

**Fig. 10.** [1, ph. 94], Re = 140. Oscillating trace in water behind streamlined cylinder explains Fig. 9.

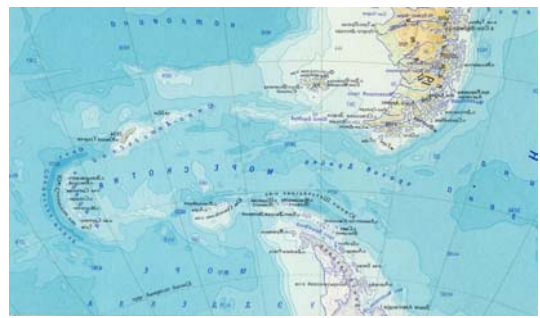

**Fig. 11.** Mirror image. Long shock wave connects Antarctida with some big continent (South America). Imbrium Antarctida is concave, and Procellarum is a concave continent.

# 13. AITKEN BASINS.

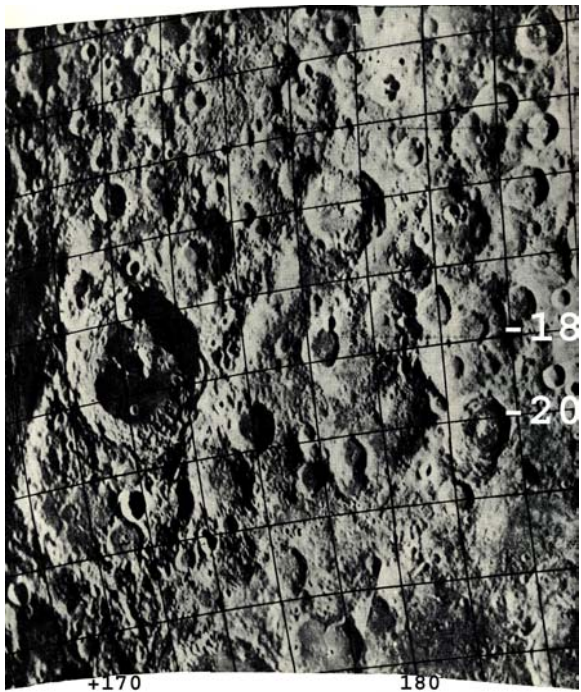

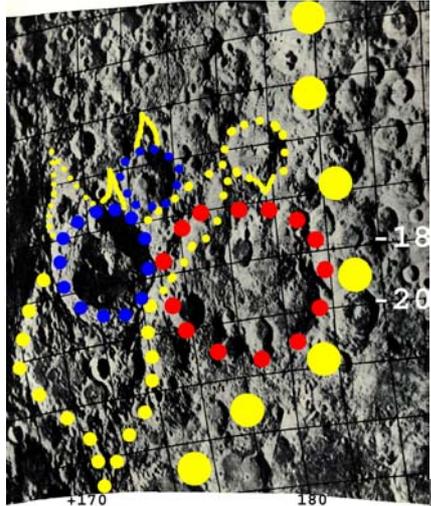

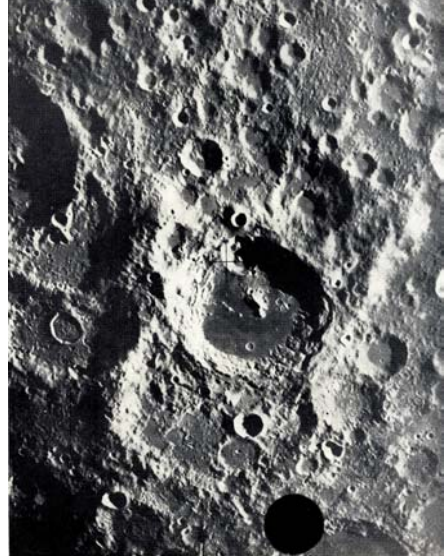

**Fig. 1.** [13], "Zond-8". The Moon. Aitken basin (big blue) is a back vortex of some unvisible moving basin (red).

**Fig. 2.** [13, p. 238]. Two back vortices (left) behind Aitken basin. Moving from left to right.

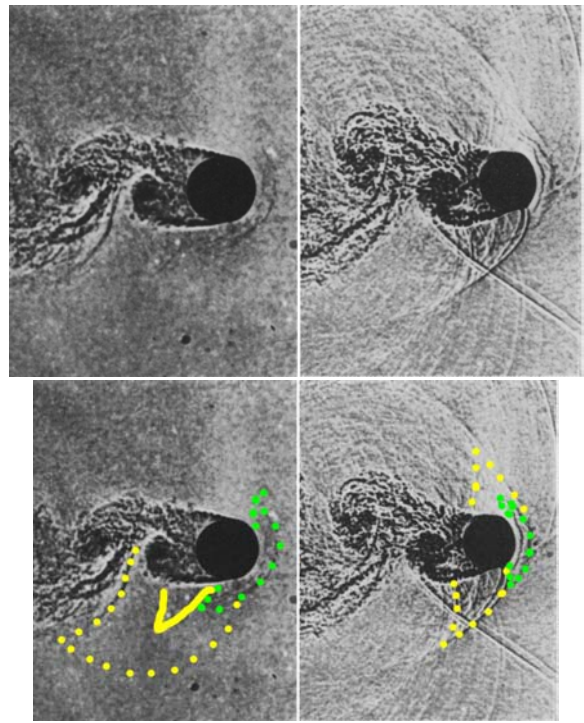

**Fig. 3.** Oscillating trace behind streamlined cylinder [1, ph. 221]. **Left:** M = 0.45, Re = 1.1·10⁵. **Right:** M = 0.64, Re = 1.35·10⁶.

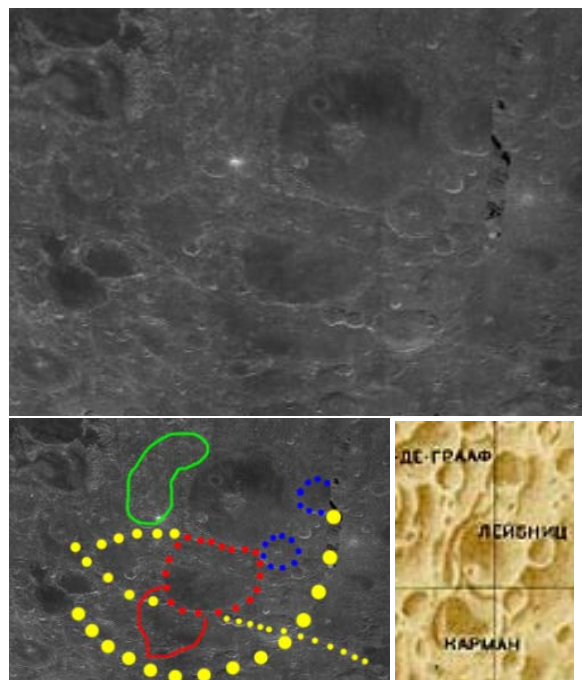

Fig. 4. PIA00304 (fragment), [14]. The Moon. Karman (lower basin, red points) and Leibniz (upper). Karman (an Aitken) is a cluster of shock waves of moving Leibniz. The right wall is a shock wave.

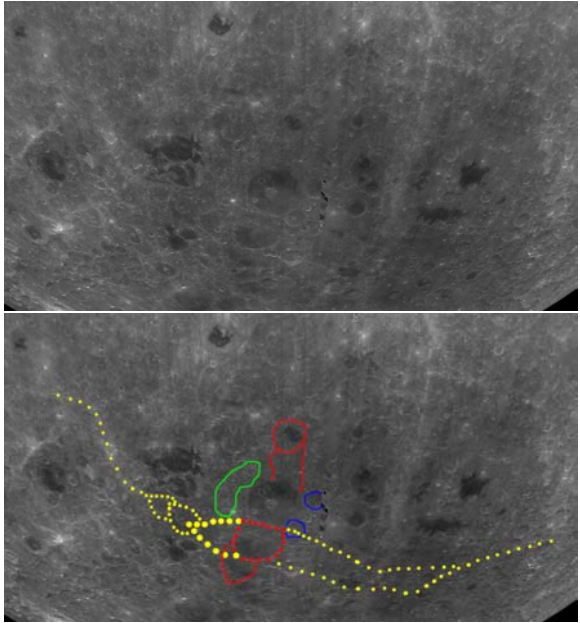

**Fig. 5.** PIA00304 (fragment). The Moon. Leibniz and Karman basins. See Fig. 4.

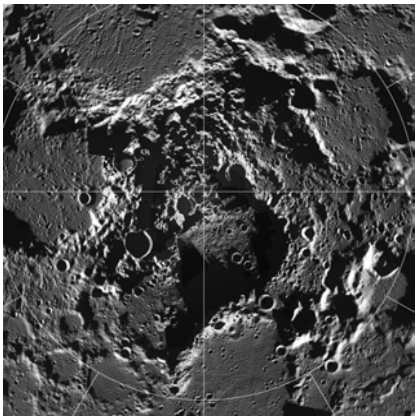

**Fig. 6.** PIA00002 (fragment). The Moon. Some Aitken basin on North Pole.

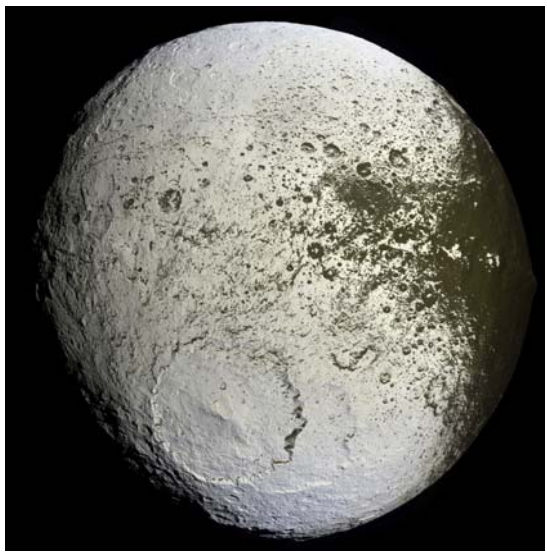

**Fig. 7.** PIA08384. Iapetus, Saturn's satellite. Here is some Aitken basin.

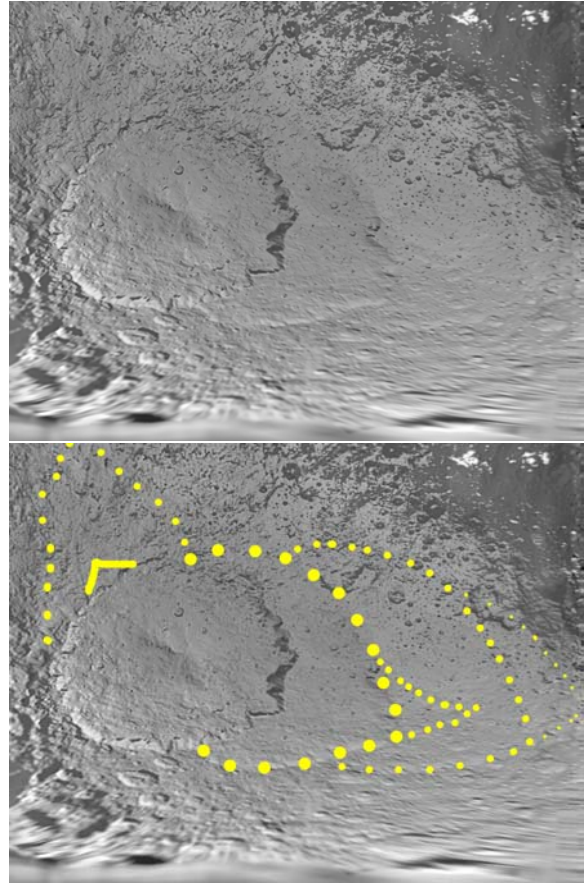

**Fig. 8.** PIA08406 (fragment). Iapetus, an Aitken basin. Two clusters on right side. One could try to see the 3-rd cluster at Fig. 7.

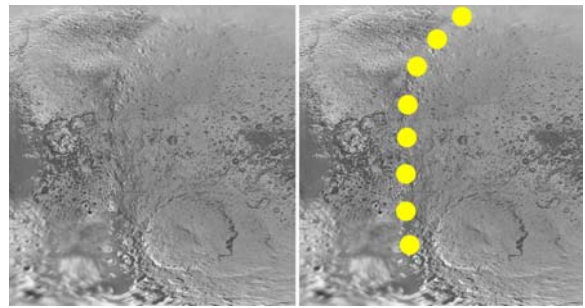

**Fig. 9.** PIA08406 (fragment). Iapetus, Saturn's satellite. This wave front (if it really exists) seems to be associated with the Aitken basin. See Fig. 1, 4.

## 14. ICELANDIAE.

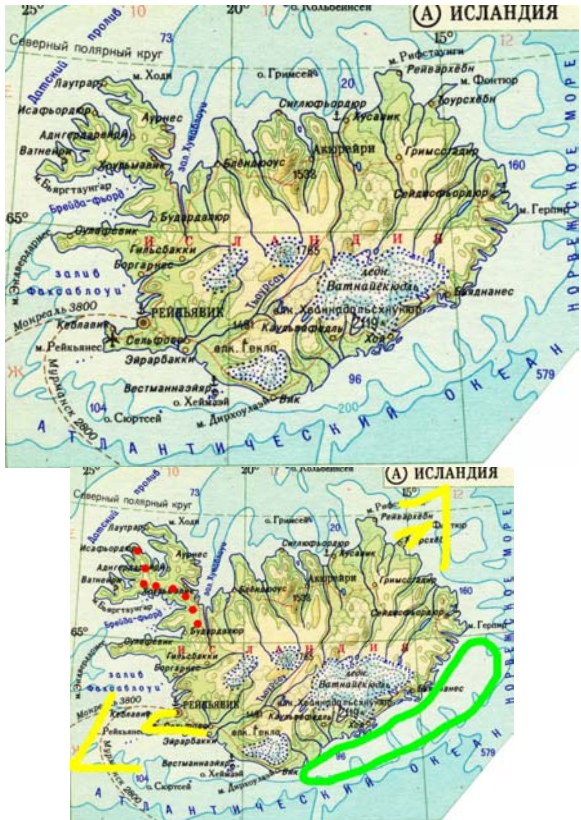

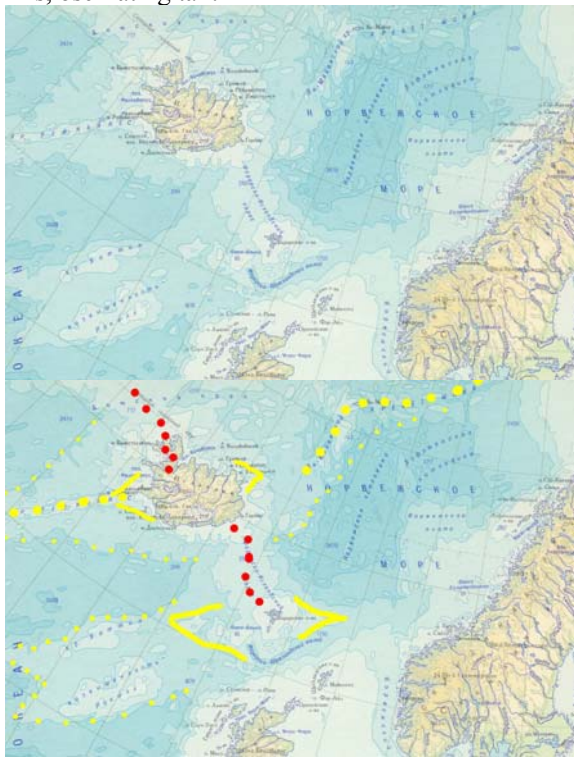

**Fig. 1.** Iceland is a moving basin: flat front, triangle fins, oscillating tail.

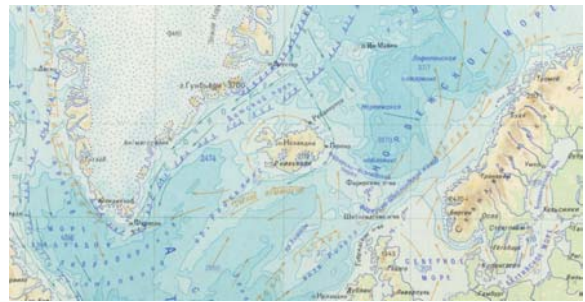

**Fig. 3.** Iceland's oscillating trace goes from Greenland. So the foregoing basin (Fig. 2) seems to be an Iceland of Iceland. See Fig. 2 at the "Extraterrestrial Antarctidae" section.

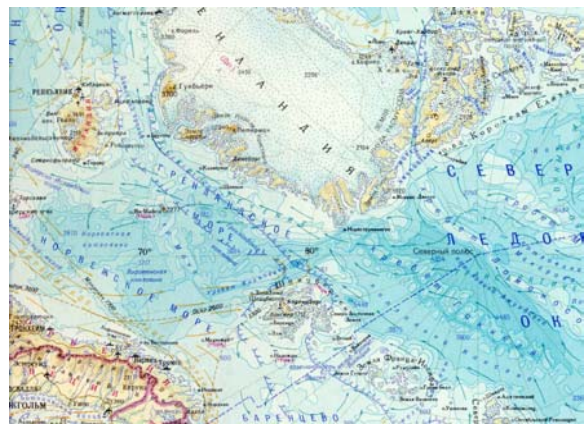

**Fig. 2.** There is some foregoing moving basin before Iceland. Oscillating trace, triangle fins.

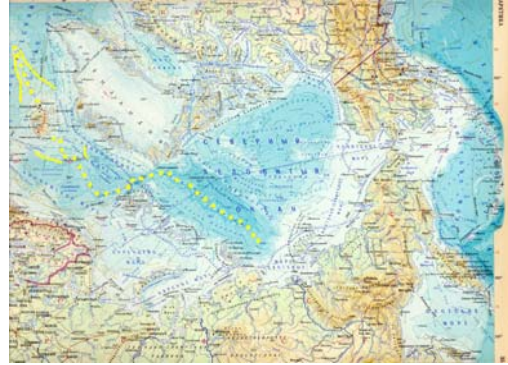

**Fig. 4.** Greenland's oscillating trace goes from North America. Long shock wave of Iceland in the Arctic.

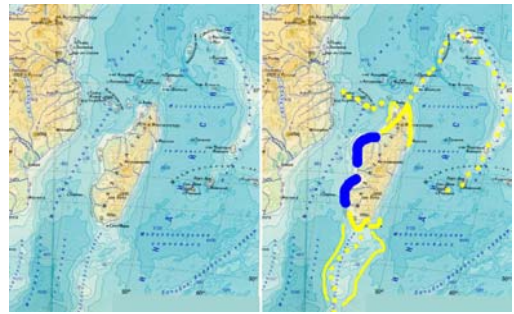

**Fig. 5.** Madagascar is a moving basin, an Iceland, and an Aitken basin (see the upper right big fin).

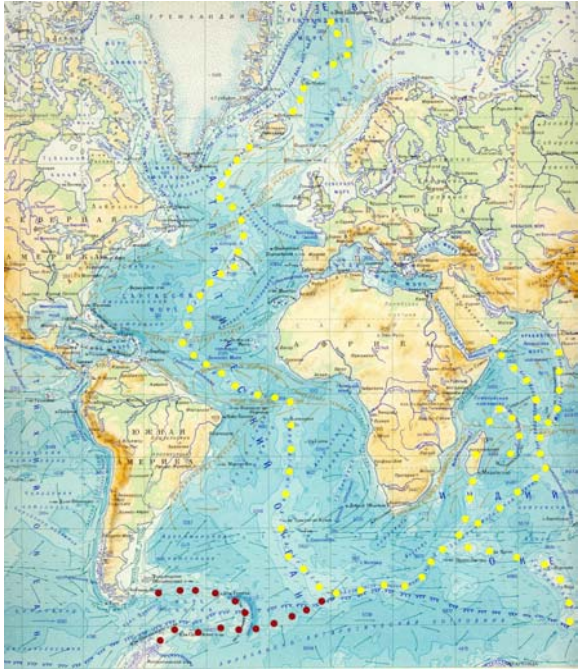

**Fig. 6.** Long shock waves of Iceland. The Middle Atlantic Range is a shock wave from Iceland.

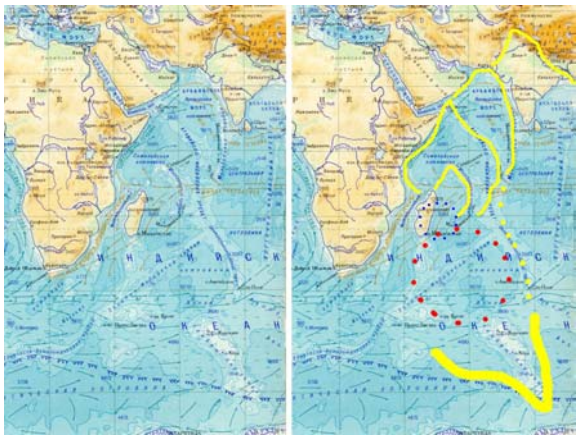

**Fig. 7.** Madagascar, chain of upper fins. Madagascar basin (blue) seems to be a satellite vortex of some big unvisualized basin (red). See the "Aitken basins" section.

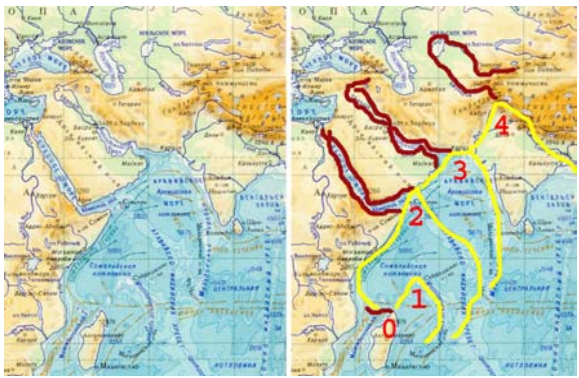

**Fig. 8.** Sequence of shock waves (dark red). The upper border of the 4-th shock wave goes up too, so it isn't the last element of the chain.

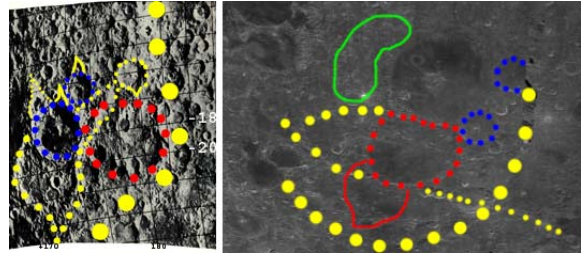

**Fig. 9.** Lunar Icelandiae. See the "Aitken basins" section. Here Icelandiae are craters on straight shock waves.

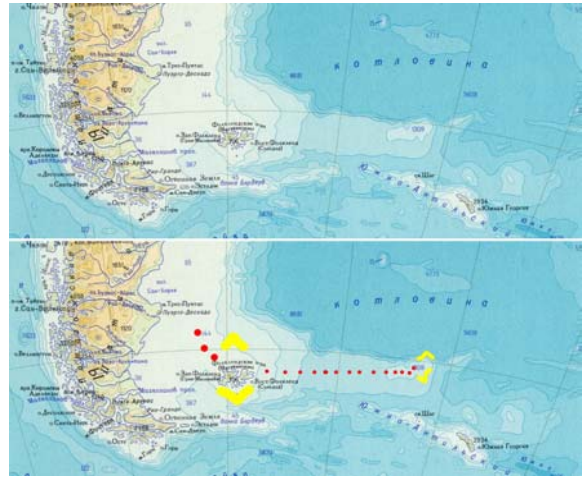

**Fig. 10.** South America, an Iceland and a sub-Iceland. The Falklands are an Isidis-like basin, see the "North America on Mars" section.

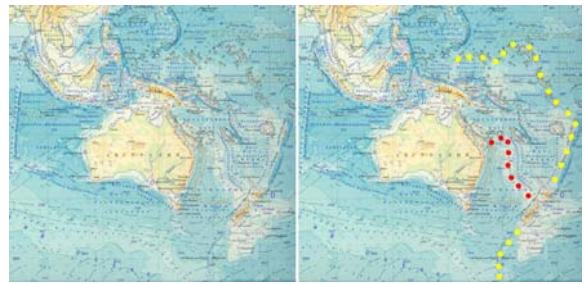

**Fig. 11.** Australia and New Zealand are an Iceland and a sub-Iceland. See the "Supersonic Australia" section. New Zealand's oscillating trace connects Australia at interesting point, see Fig. 9.

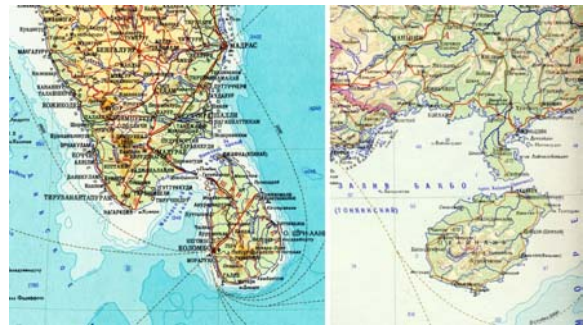

**Fig. 12.** Ceylon and Hainan are moving basins with oscillating traces.

## 15. CHUKOTCAE AND SIBERIAE.

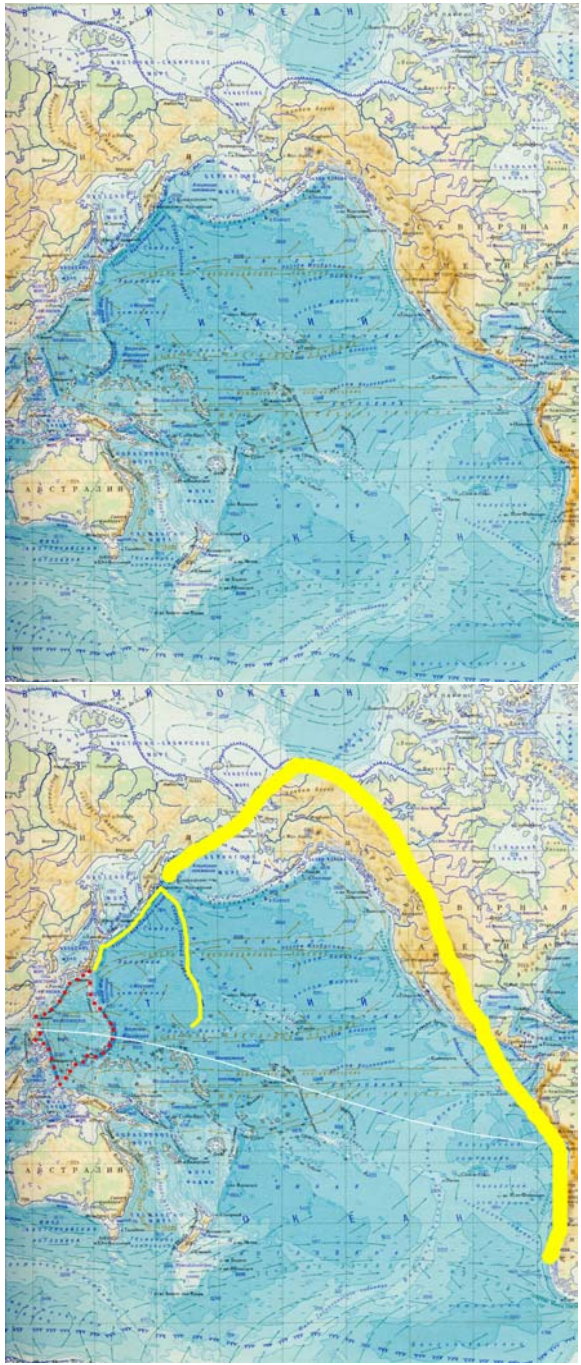

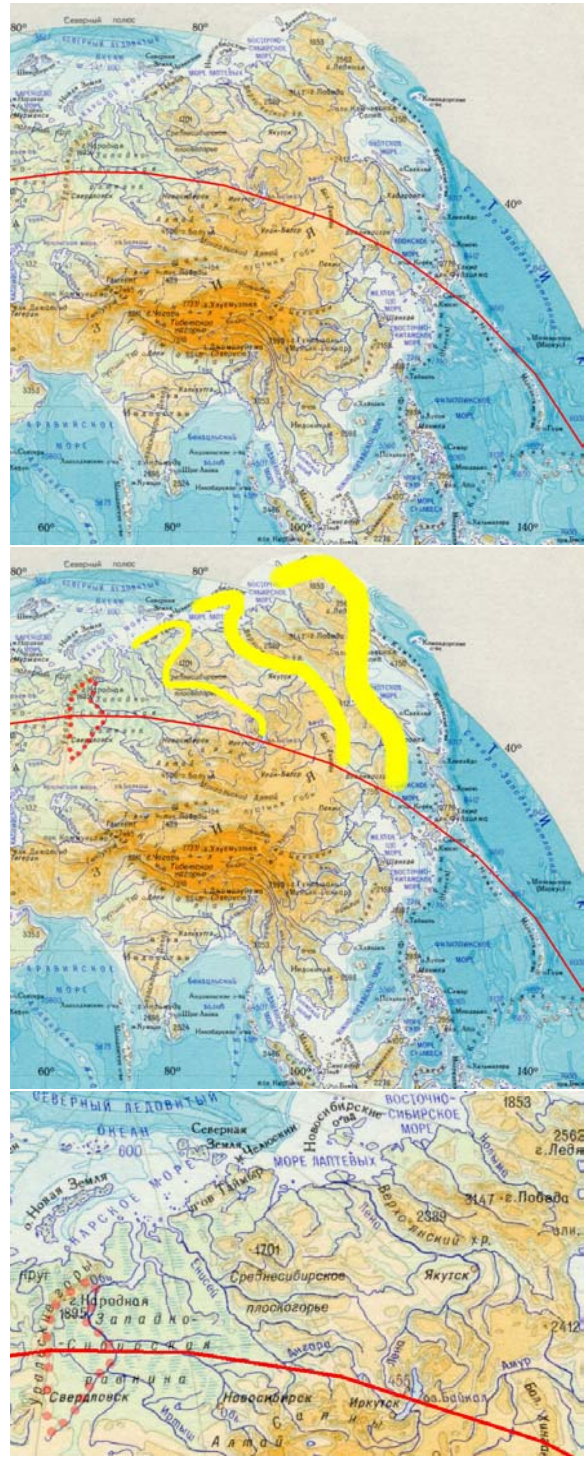

**Fig. 1.** Philippine fish (red, radius of the fish moving circle is $(\pi - 3)$ radians, see the "Moving basins" section) has a sequence of wave fronts (yellow). The straight line (white) is a direction of fish' moving, it connects the most Eastern point of the fish and the most deep point of the South America deepening (the Andes). So the fish could become supersonic. See the "Supersonic Australia", "Aitken basins", "Icelandiae" sections.

**Fig. 2.** The Kara Sea is a concave Chukotca, the Gulf of Ob is a Kamchatka, the Laptev Sea is an Alaska. Khanty-Mansi fish (red) is convex, it is a small copy of Philippine fish. Radius of the fish moving circle is $(\pi - 3)/4$ radians. Baikal Lake is an element of its foregoing shock wave, so the fish could become supersonic.

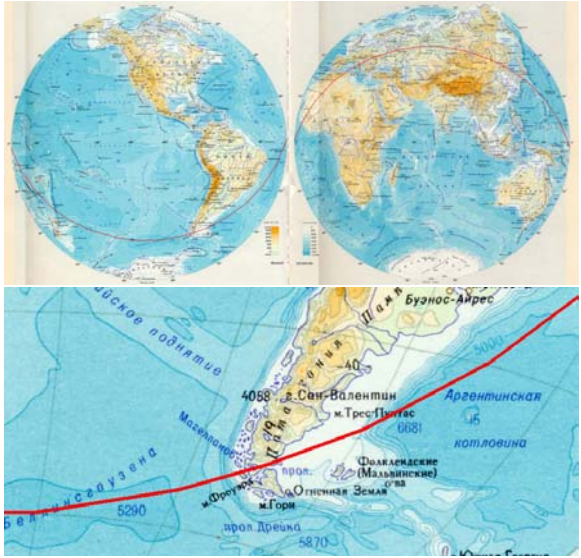

**Fig. 3.** The straight line (red) is a direction of moving for Khanty-Mansi fish. The line separates the triangle of Tierra del Fuego (it seems to be a shock wave triangle) from South America.

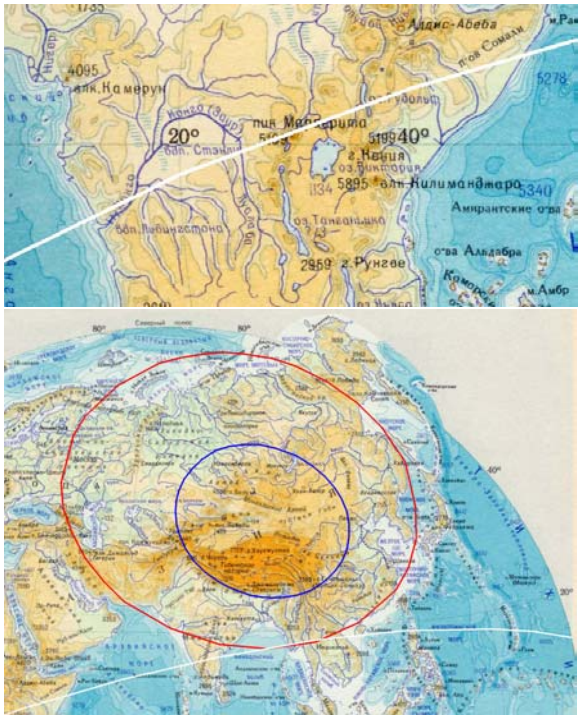

**Fig. 4.** The straight line (white) is a direction of moving for Philippine fish. **Upper:** Africa. The line goes through the center of Congo basin (exactly). See the "Africa inside Artemis" section. **Lower:** The line is tangent to the Asian circle. Radii of the circles are $2(\pi - 3)$ and $4(\pi - 3)$ radians. Chukotca and Indochinese Chukotca (see the "North America in Asia" section) are shock waves of Asian basin.

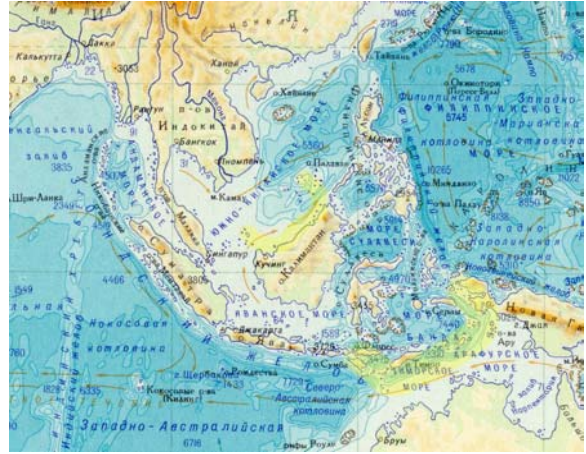

**Fig. 5.** Foregoing wave fronts of supersonic Indo-chinese Chukotca.

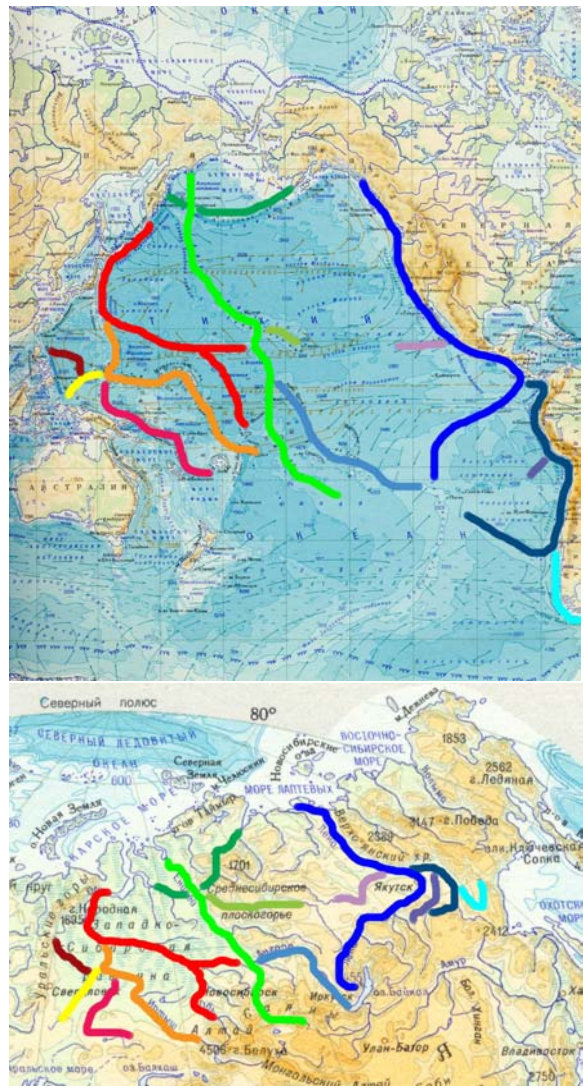

**Fig. 6.** Comparing shock waves.

**16. MOVING BASINS ON THE TEMPEL 1 COMET NUCLEUS.**

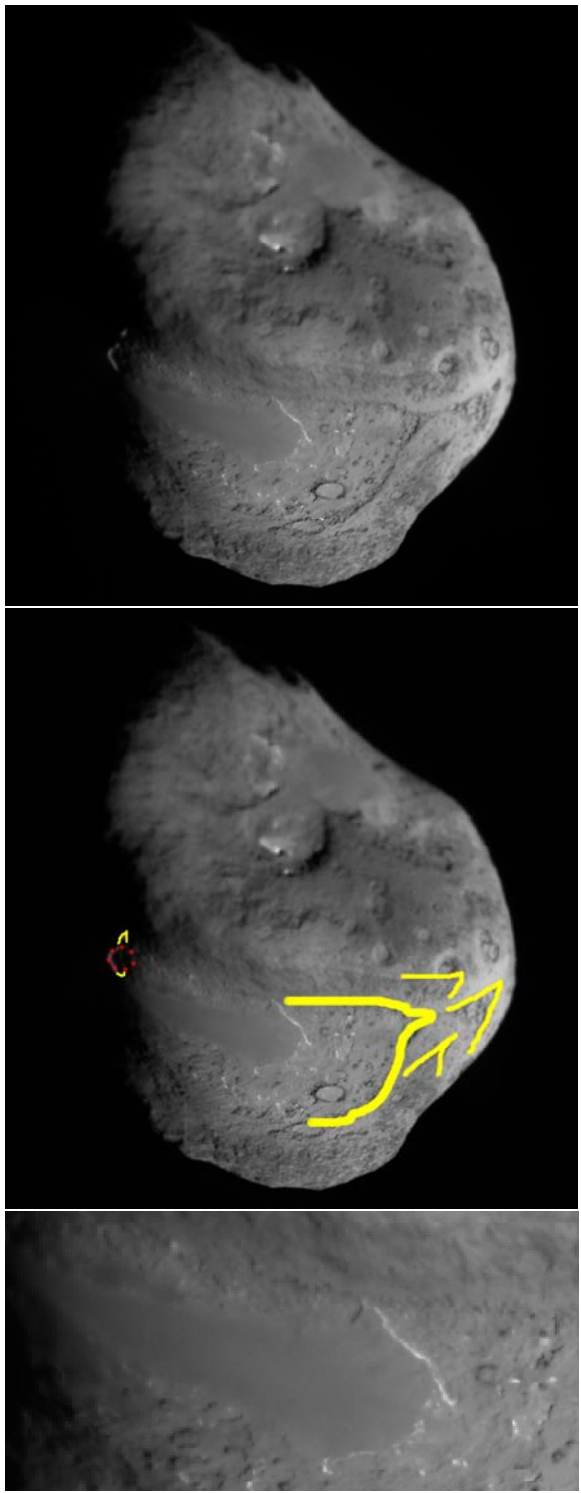

**Fig. 1.** PIA02142. Tempel 1 nucleus before the impact (the Deep Impact mission). Moving basin, slightly oscillating trace of the basin. Yellow fins behind the trace are shock waves. The trace is some kind of a continent, see the "Supersonic Australia" section.

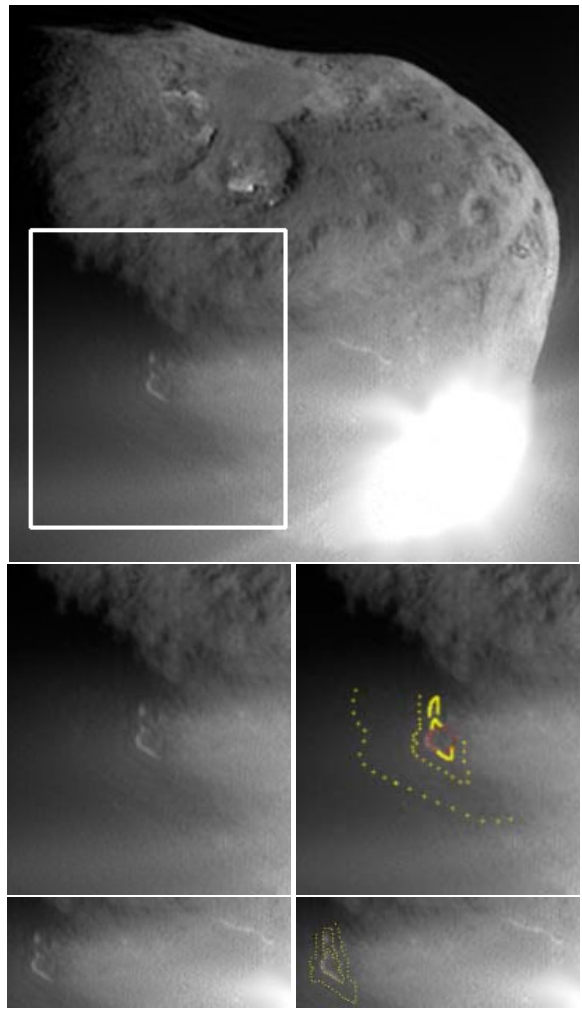

**Fig. 2.** PIA02137. Tempel 1 nucleus, 67 seconds after the impact. The moving basin has foregoing wave fronts and weakly oscillating trace, so it is supersonic (Fig. 4). The initial point of moving is near the impact site. Suppose the basin be running from the impact. Some analogous subconscious effects (for people, not for basins) are known for the author. Mathematics of such effects is nonlocal and projective, including parallels with Euler's projective arithmetic [5, 15]. See also [6, 16–18].

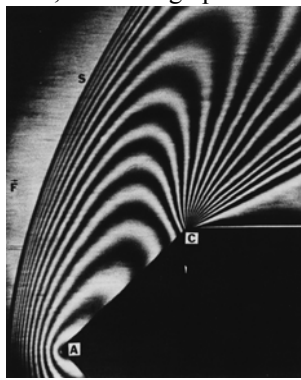

**Fig. 3.** [1, ph. 234]. Interferogram, lines of constant density, M = 2.5. Supersonic foregoing shock wave and subsonic area between it and the body. See Fig. 2.

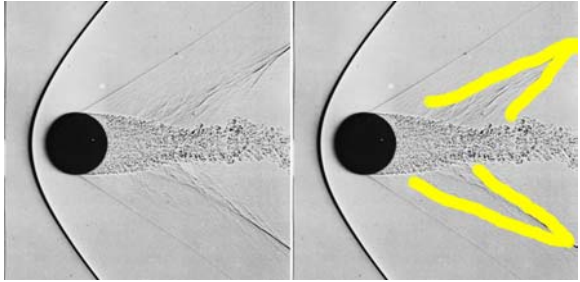

**Fig. 4.** [1, ph. 266]. Supersonic moving in air. M = 1.53. See Fig. 5, 6.

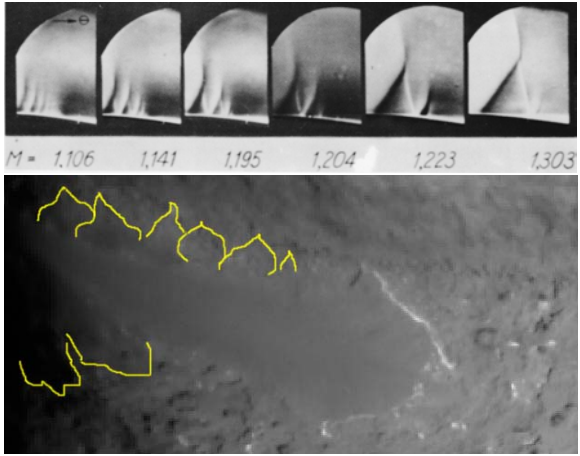

**Fig. 5. Upper:** [1, ph. 247]. Sequence of shock waves merges into one shock wave when local Mach is increasing. Moving is from right to left. **Lower:** Sequence of shock waves instead of a cluster of waves (gluing in one wave) at Fig. 4. Also see Fig. 6.

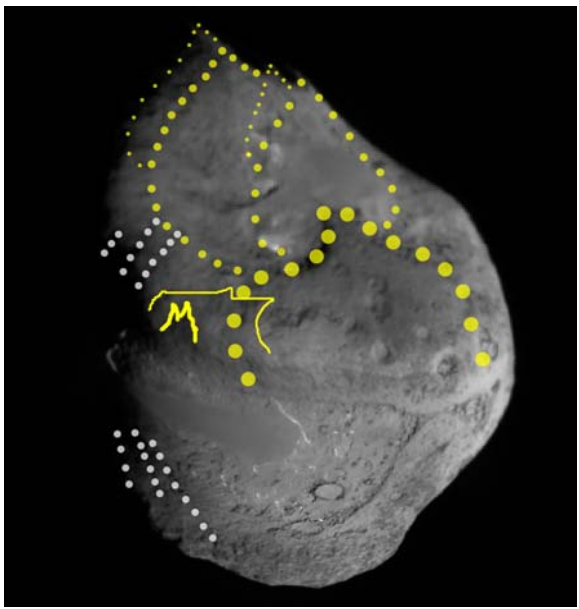

**Fig. 6.** PIA02142 (Fig. 1). Some big shock waves.

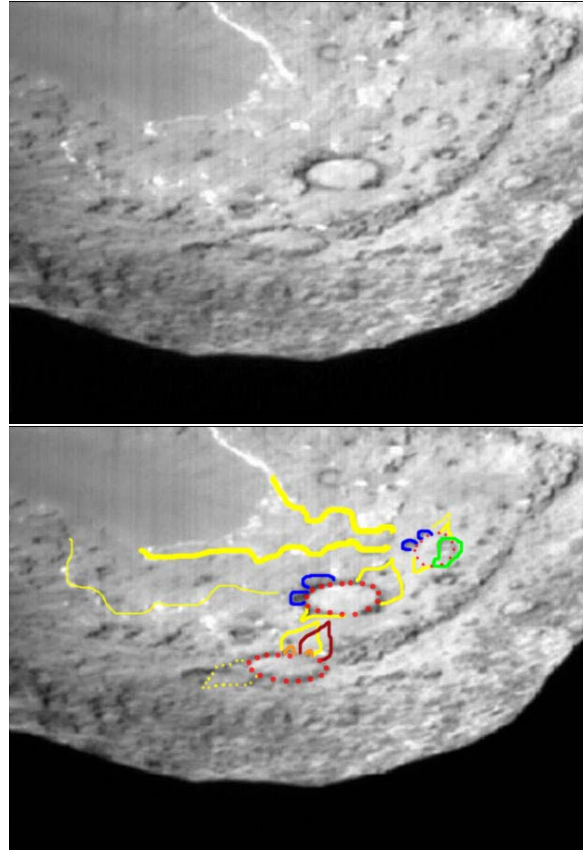

**Fig. 7.** PIA02135. Moving basins near the initial point of expanding trace.

**Comment:** By assumption, the basin is running from the impact (Fig. 2). This idea isn't too unbelievable. Probably there exist some another forms of life with another scale of time. One could try to recognize such forms by studying such effects. Also recall the concept of alive matter (it is obvious for a quarter of all people) [16]. Density of comet matter seems to be small, 'cause a comet probably consists of very porous ice. Speed of Earthen mountain pulsating glaciers (with usual ice) is some meters per day (up to 200 meters per day). Therefore our running basin doesn't need million years for avoiding the impact.

Gamov, a physicist, writes about his grandfather, a priest [19, p. 12]. There was a war in Sevastopol in 1855, and the grandfather waited transport during an artillery bombardment. Suddenly this man felt insistent physical need to relieve. When he turned a corner, some projectile burst just at the place of his previous staying. This case is typical, and it doesn't require any knowledge about future from a man or a basin.

## 17. DISCRETE FIELDS OF FORCE ON PLANET SURFACES.

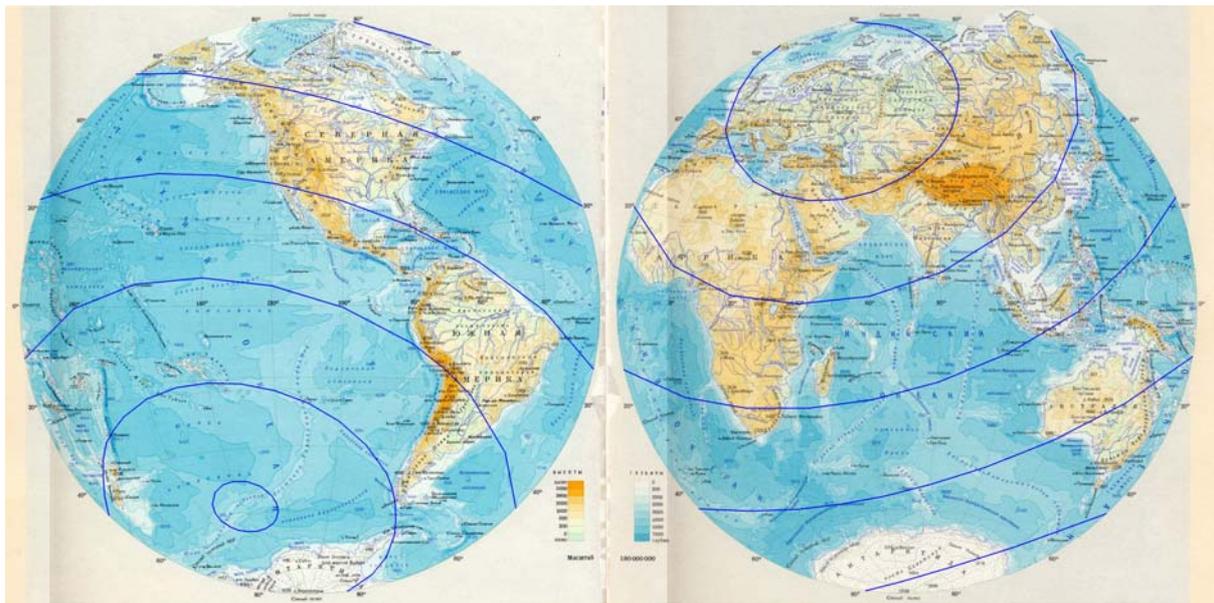

**Fig. 1.** Radii of the circles are 0.5, 1.0, 1.5, 2.0, 2.5, 3.0 radians, the center is Yaroslavl. There exists some global discrete field of force around Yaroslavl.

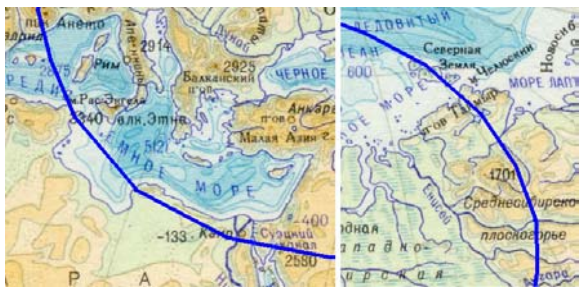

**Fig. 2.** Fragments of Fig. 1. Radius of the circle is 0.5 radian. **Left:** Geometry of the Mediterranean Sea correlates with the 0.5-circle. See also the "$(\pi - 3)$-basins" section, Fig. 21. **Right:** Positions of the Taimyr Peninsula and Taimyr Lake (shock waves) correlate with the 0.5-circle. See the "$(\pi - 3)$-basins" section, Fig. 22, the "Supersonic Australia" section, Fig. 10, and the "Chukotcae and Siberiae" section, Fig. 2.

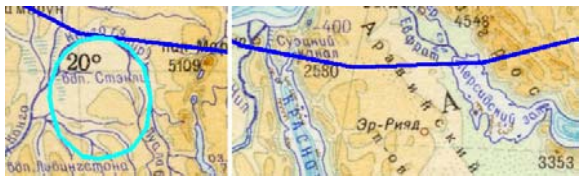

**Fig. 3.** Fragments of Fig. 1. **Left:** The 1-circle is tangent to Congo $(\pi - 3)/2$-basin. See Fig. 6, the "Africa inside Artemis" section, Fig. 6, and the "Chukotcae and Siberiae" section, Fig. 4. **Right:** Persian Gulf and Red Sea's gulfs (shock waves) are bounded by the 0.5-circle. See the "Icelandiae" section, Fig. 8.

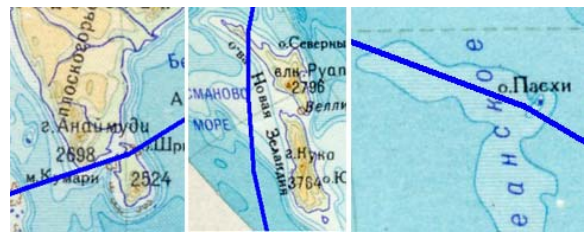

**Fig. 4.** Fragments of Fig. 1. **Left:** The 0.5-circle separates Ceylon from Hindustan. **Center:** The 2.5-circle near New Zealand. **Right:** The 2.5-circle near Easter Island (ostrov Paskhi), $\Delta = 40$ km. The submarine mountain range goes along the circle.

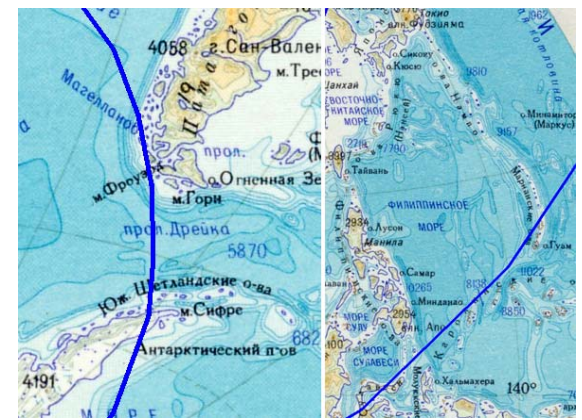

**Fig. 5.** Fragments of Fig. 1. **Left:** The 2.5-circle cuts the end of the Antarctidean shock wave. Compare with New Zealand (Fig. 4, center). See the "Antarctidae" section. **Right:** Philippine fish is bounded by the 1.5-circle. See the "Moving basins" section, Fig. 2, and the "Chukotcae and Siberiae" section, Fig. 1, 4..

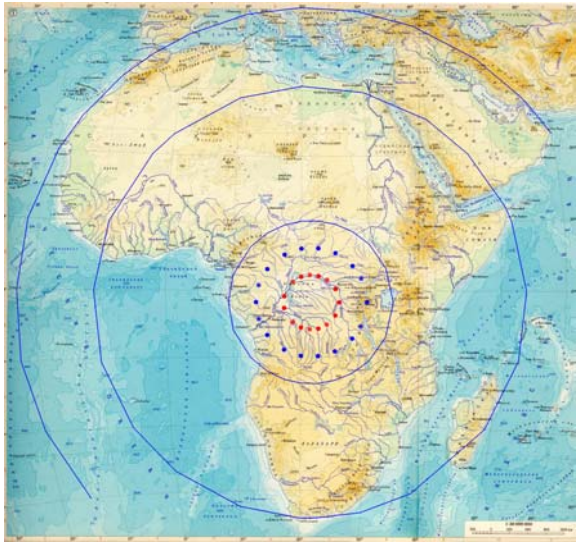

**Fig. 6** Circles around Congo basin. Radii of the circles are $n(\pi - 3)/2$ radians, $n = 1, 2, 3, 8, 11$. The center is $(1.76^0$ S, $21.46^0$ E). See Fig. 3.

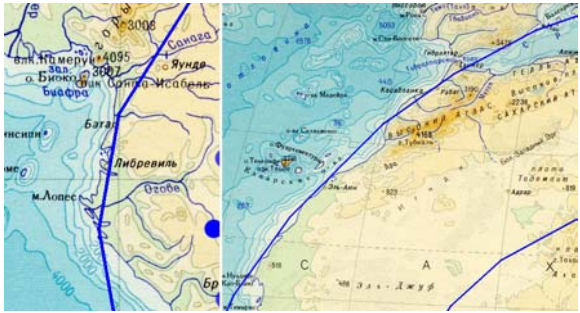

**Fig. 7.** Fragments of Fig. 6. **Left:** The Atlantic coast is bounded by the $3(\pi - 3)/2$-circle. **Right:** The Mediterranean Spain coast and the Atlantic coast are bounded by the $11(\pi - 3)/2$-circle.

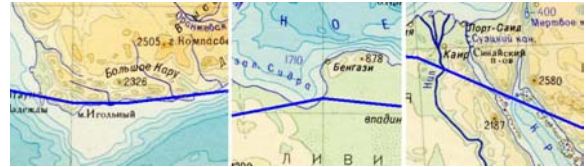

**Fig. 8.** Fragments of Fig. 6. **Left:** South Africa is bounded by the $8(\pi - 3)/2$-circle. **Center:** Mediterranean Sea's bay is bounded by the $8(\pi - 3)/2$-circle. **Right:** The Red Sea is bounded by the $8(\pi - 3)/2$-circle.

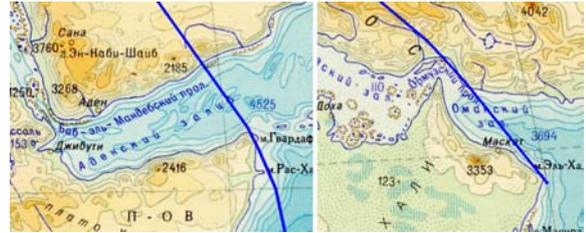

**Fig. 9.** Fragments of Fig. 6. **Left:** Somalia is bounded by the $8(\pi - 3)/2$-circle. **Right:** Arabia is bounded by the $11(\pi - 3)/2$-circle.

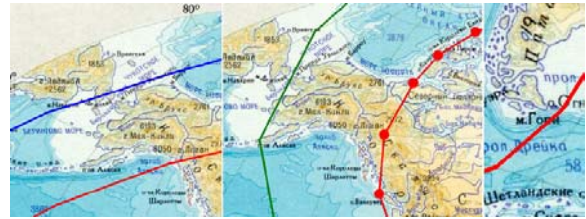

**Fig. 10.** Length of Chukotca and length of Alaska are $2(\pi - 3)$ radians. **Left:** Fragment of Fig. 11. **Center:** Circles around American basin. Radii of the circles are $2n(\pi - 3)$ radians, $n = 2, 3$; the center is $(48^0$ N, $77^0$ W). **Right:** Fragment of Fig. 11. Tierra del Fuego is bounded by the $20(\pi - 3)$-circle.

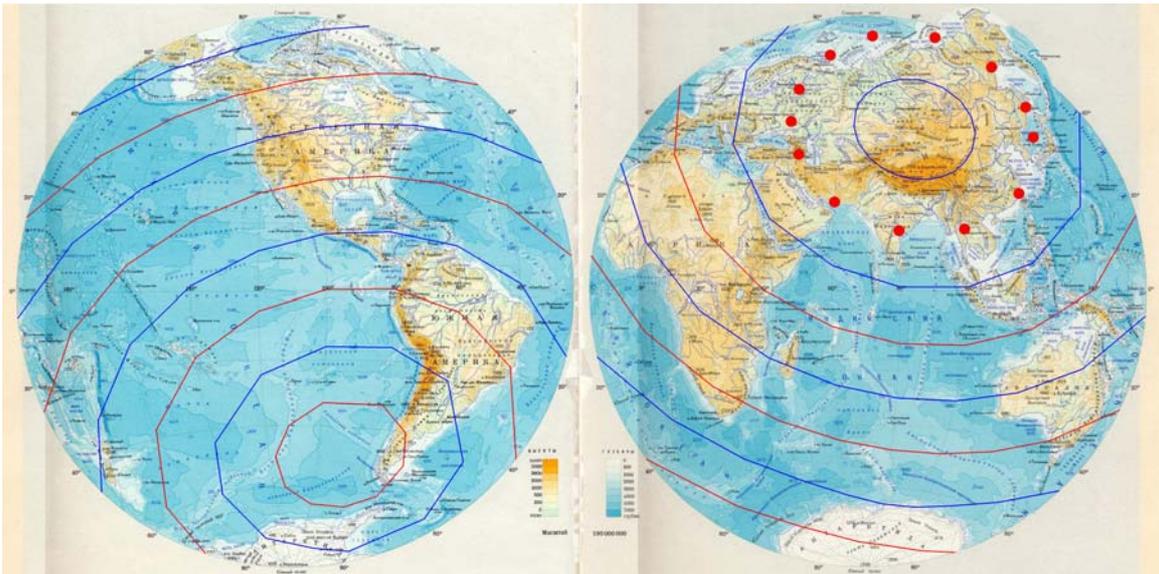

**Fig. 11.** Circles around Asian basin. Radii of the circles are $2n(\pi - 3)$ radians, $n = 1, 2, \ldots, 10$. The center is $(49^0$ N, $90^0$ E).

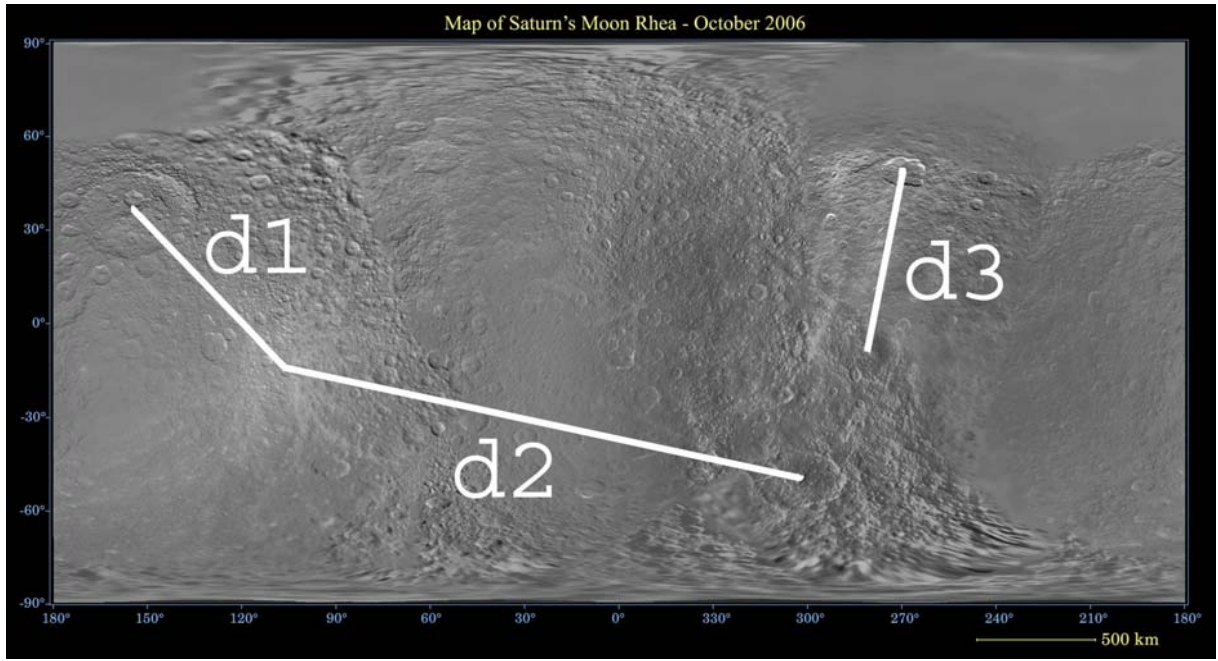

**Fig. 12.** Rhea, Saturn's satellite. PIA08343 (rectangular map). Distances between centers of basins. d1 = 1.07 ≈ 1.0 + (π − 3)/2 radians. See Fig. 3 left, Fig. 1: this is the distance from Yaroslavl to the center of Congo basin. d2 = 2.06 ≈ 2.0 + (π − 3)/2 radians. d3 = 1 radian.

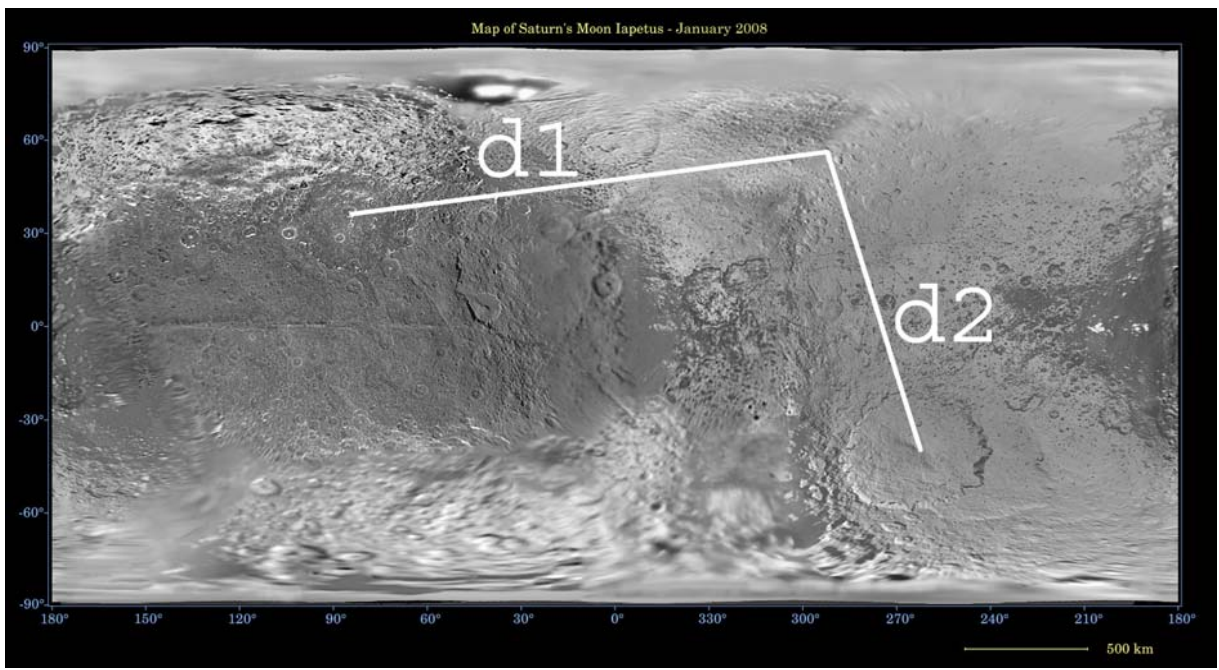

**Fig. 13.** Iapetus, Saturn's satellite. PIA08406 (rectangular map). Distances between centers of basins. d1 = 1.5, d2 = √3 radians.

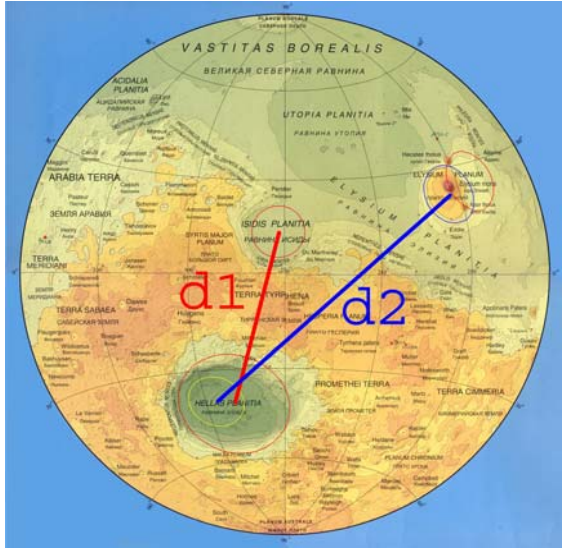

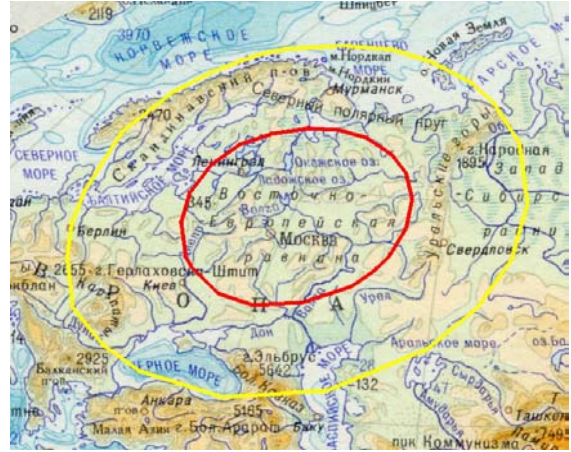

**Fig. 14.** Mars [7]. Distances between centers of basins. <span style="color:red">d1</span> = 1 radian, <span style="color:blue">d2</span> = $\sqrt{3}$ radians. See the "$(\pi - 3)$-basins" and "North America on Mars" sections. Embeddings of distances 1 km and $\sqrt{3}$ km were found by the author in [3, p. 64], [4, p. 44].

**Fig. 15.** Radii of the circles are $(\pi - 3)$ and $2(\pi - 3)$ radians. The center is Yaroslavl (57.641[0] N, 39.895[0] E). Curvature of the Urals–Novaya Zemlya shock wave is formed by the $2(\pi - 3)$-circle. See the "Moving basins" and "Moscow Antarctida in the Milky Way" sections.

## 18. INTERACTIONS OF DISCRETE FIELDS OF FORCE.

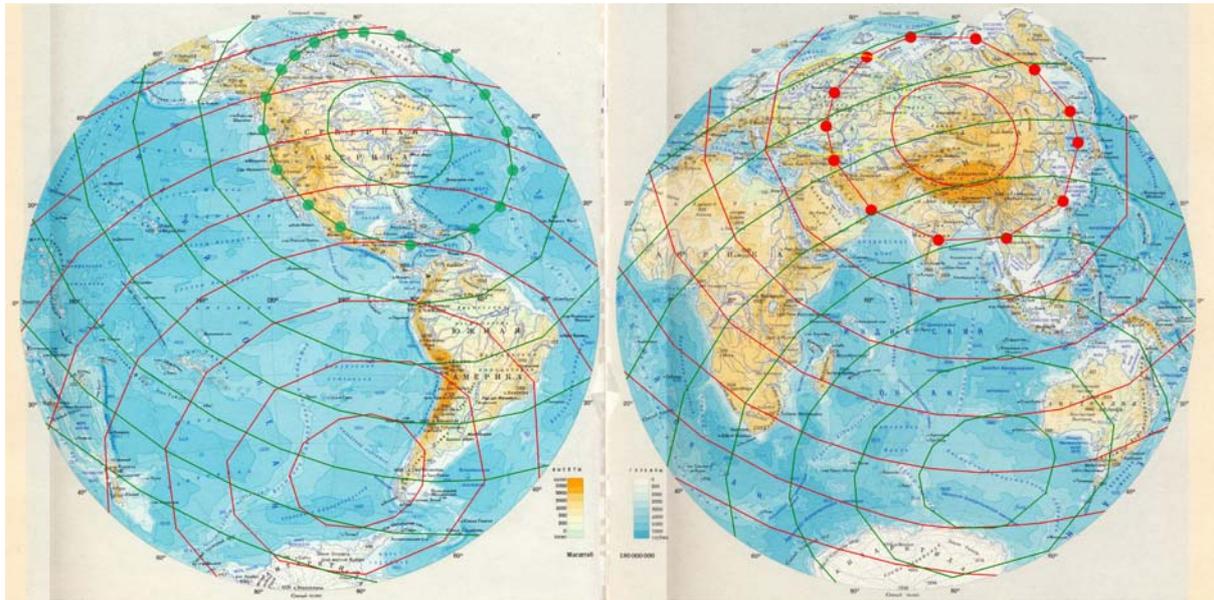

**Fig. 1.** Asian basin, centered at ($49^0$ N, $90^0$ E), and North American basin, centered at ($48^0$ N, $77^0$ W). Radii of circles are $2n(\pi - 3)$ radians, n = 1, 2, …, 10. For n = 11 there exists an (undrown) small circle, the radius is $r_0 = \pi - 22(\pi - 3)$ radians (169 km). The circles are centered at antipodal points too, radii are $r_0 + 2m(\pi - 3)$ radians, m = 1, 2, …, 10. The distance between Asian center and the antipodal point ($48^0$ S, $103^0$ E) of American center is $12(\pi - 3)$ radians (error is $\Delta = 32$ km). Circles of the two families are quasi-tangent. The difference (nonregularity of tangency) is $r_0$.

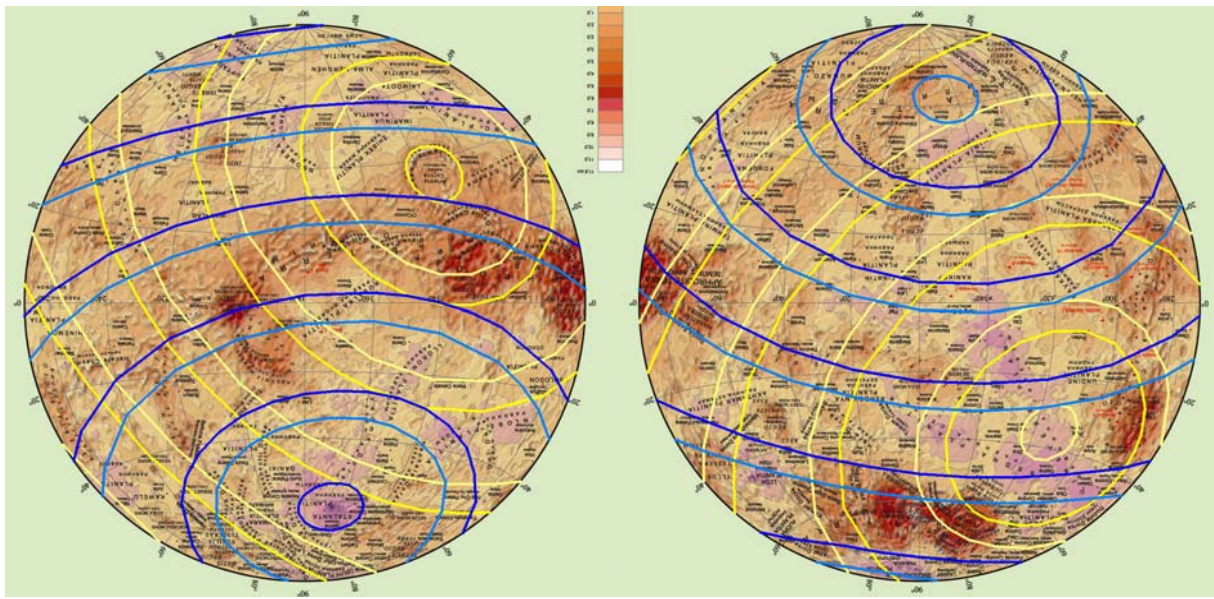

**Fig. 2.** Venus [9], $180^0$ rotated. North is down. Recall that rotation of Venus is contrary to rotation of the Earth. Circles are centered at Artemis Corona (upper left, 35.87$^0$ S, 134.09 E) and Atalanta Planitia (lower left, 62.4$^0$ N, 165.78$^0$ E), and at their antipodal points. Radii of circles are 0.5n, n = 1, 2, 3, 4, 5, 6 (light yellow for Artemis, yellow for Artemis antipode, light blue for Atalanta, blue for Atalanta antipode). For n = 6 radius is 3 radians, or $r_0 = (\pi - 3)$ radians for the same circle, centered at the antipodal point. Each family of circles is centered at the antipodal point, radii of circles are $r_0 + 0.5$m, m = 0, 1, 2, 3, 4, 5; $r_0 = (\pi - 3)$. The distance between Artemis center and Atalanta center is $12.5(\pi - 3)$ radians (error is $\Delta = 13$ km). Yellow and blue circles are quasi-tangent between Artemis and Atalanta, 'cause $12.5(\pi - 3) = 2(\pi - 3) + 1.5 - r_1$; $r_1 \approx 83$ km. The difference (nonregularity of tangency) is $r_1$. Light yellow and light blue circles are quasi-tangent between Artemis' and Atalanta's antipodes for the same reason. Note that the construction from Fig. 1 could be drawn here too, let radii be $k(\pi - 3)/2$, k = 1, 2, 3, …; see Congo basin at the "Discrete fields of force" section.

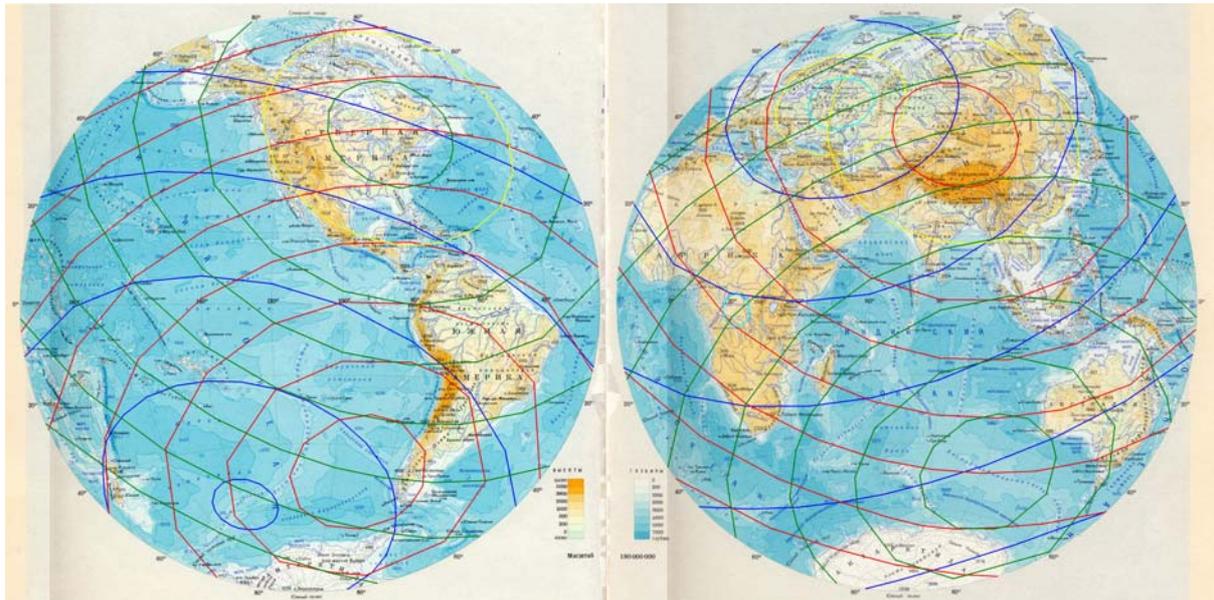

**Fig. 3.** Yaroslavl circles are added to Fig. 1; see the "Discrete fields of force" section. Blue circles are centered at Yaroslavl (57.641$^0$ N, 39.895$^0$ E); radii are 0.5n, n = 1, 2, 3, 4, 5, 6. Yellow circles are Yaroslavl $2(\pi - 3)$-basin, Asian $4(\pi - 3)$-basin, and North American $4(\pi - 3)$-basin. Cyan circles are Congo $(\pi - 3)/2$-basin, centered at (1.76$^0$ S, 21.46$^0$ E), and Yaroslavl $(\pi - 3)$-basin. One could see some situations of quasi-tangency of rational field (blue circles) and irrational fields (red and green circles).

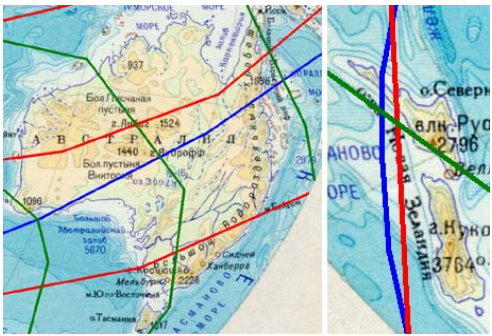

**Fig. 4. Left:** Blue and red circles are quasy-tangent at the center of Australia. **Right:** Quasi-tangency of blue and red circles at the 1-st shock wave (New Zealand) of supersonic Australian moving. See the "Supersonic Australia" section. Note that green circle is between two parts of New Zealand.

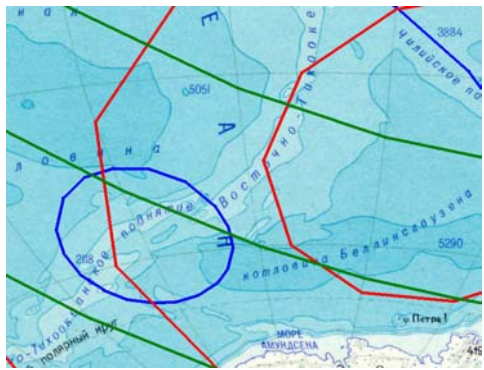

**Fig. 5.** Quasi-tangency of blue and red circles at the 2-nd shock wave of supersonic Australian moving.

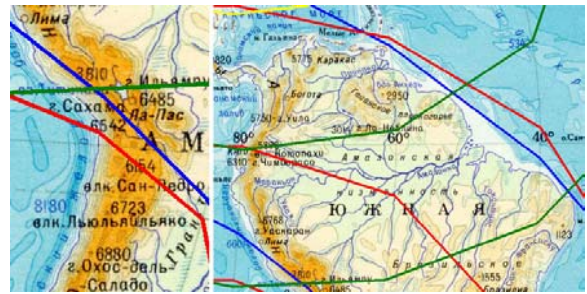

**Fig. 6. Left:** Quasi-tangency of blue and red circles at the 3-rd shock wave of supersonic Australian moving. **Right:** Quasi-tangency of blue and red circles at NE coast of South America. Thus all situations of quasi-tangency of blue and red circles are listed at Fig. 4–6.

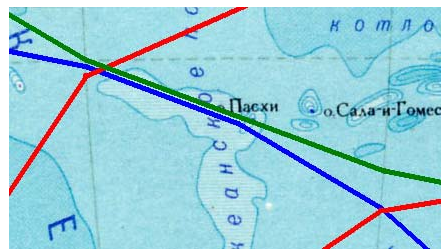

**Fig. 7.** Quasi-tangency of blue and green circles near Easter Island (ostrov Paskhi). The submarine mountain range goes along the circles. This singularity is analogous to Baikal Lake at the Pacific, see the "Chukotcae and Siberiae" section. The author doesn't know a key to another two situations of quasi-tangency.

## 19. VENUS: EXTERNAL GEOMETRY OF THE HOWE–DANILOVA–AGLAONICE CRATER TRIANGLE.

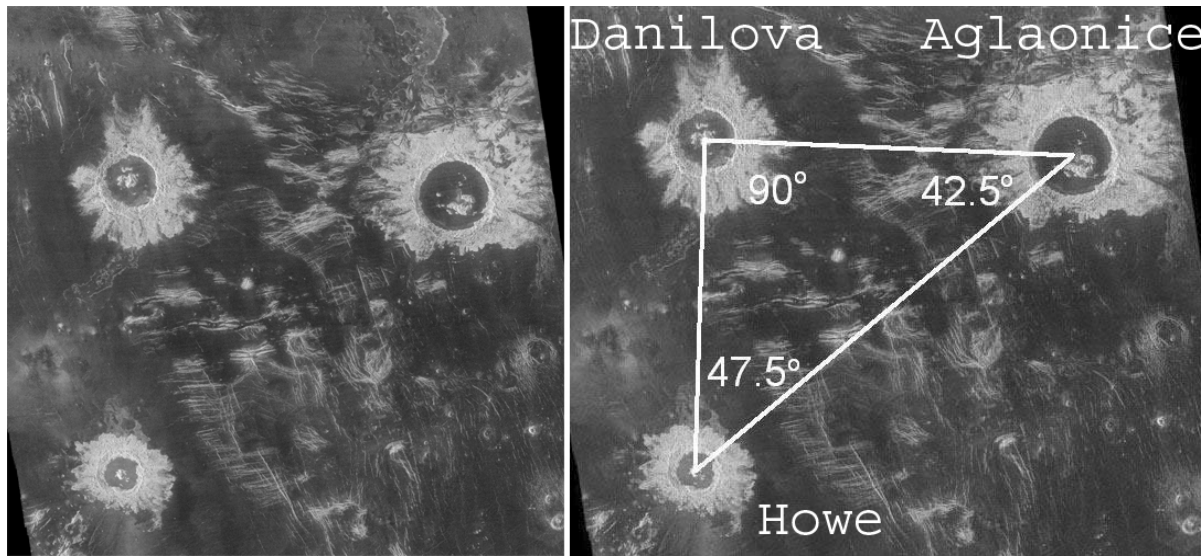

**Fig. 1.** Venus, the crater triangle under consideration. **Left:** PIA00086 (fragment). **Right:** By PIA00103: Howe (28.6 S, 337.1 E), diameter 37.3 km, Danilova (26.35 S, 337.25 E), diameter 47.6 km, Aglaonice (26.5 S, 340 E), diameter 62.7 km.

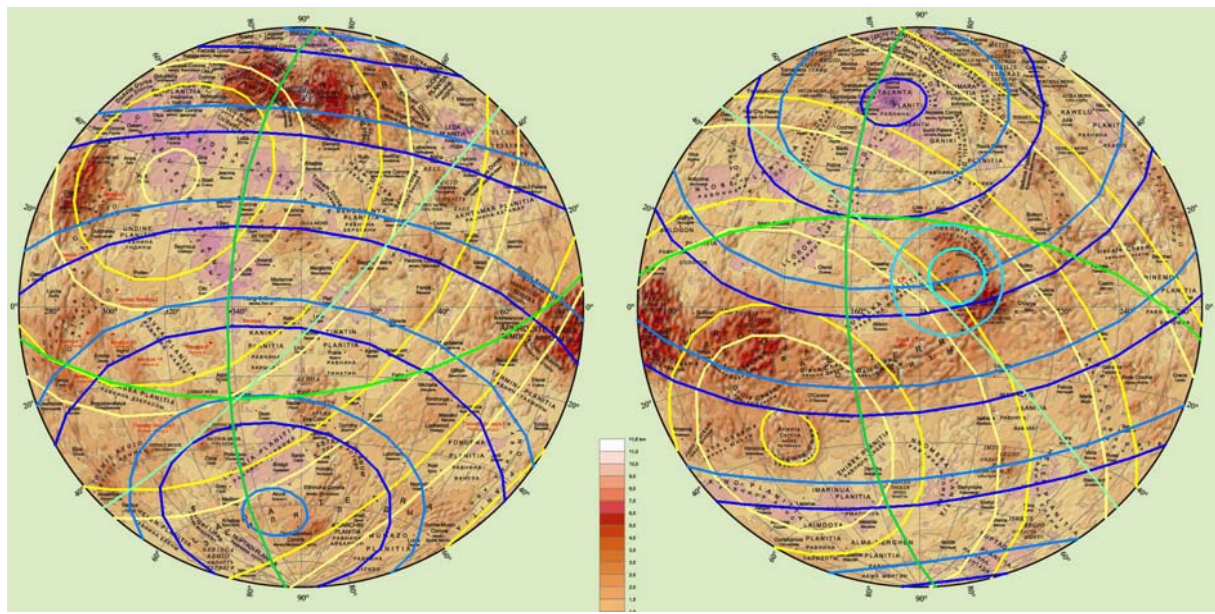

**Fig. 2.** Venus [9]. Direct lines of the triangle (prolonged sides, see Fig. 1) are green. Concentric circles are centered at Artemis Corona (35.87⁰ S, 134.09 E) and Atalanta Planitia (62.4⁰ N, 165.78⁰ E), and at their antipodal points. Radii of circles are 0.5n radians, n = 1, 2, 3, 4, 5, 6 (light yellow for Artemis, yellow for Artemis antipode, light blue for Atalanta, blue for Atalanta antipode). Cyan circle is centered at the center of the Sapas vortex (8.95⁰ N, 190.68⁰ E); radius is (π − 3) radians. Light-light blue circle is centered at Sapas mountain (7.97⁰ N, 187.90⁰ E); radius is 2(π − 3) radians.

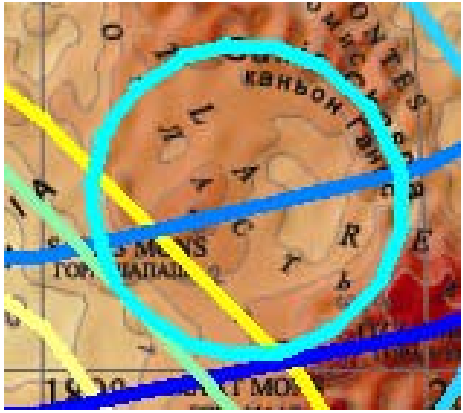

**Fig. 3.** Fragment of Fig. 2. **a)** The Howe–Aglaonice direct line (light green) is tangent to the $(\pi - 3)$-circle of the Sapas vortex (cyan). Error $\Delta = 11$ km is inside data precision. **b)** The Howe–Aglaonice direct line (light green) goes exactly between yellow and light yellow circles. Distance between the two circles is $(\pi - 3)$ radians, so the direct line is tangent to the circle with radius $1 + (\pi - 3)/2$ radians, centered at Artemis Corona. $\Delta = 18$ km is inside data precision. See the "Discrete fields of force" and "Interactions of discrete fields" sections.

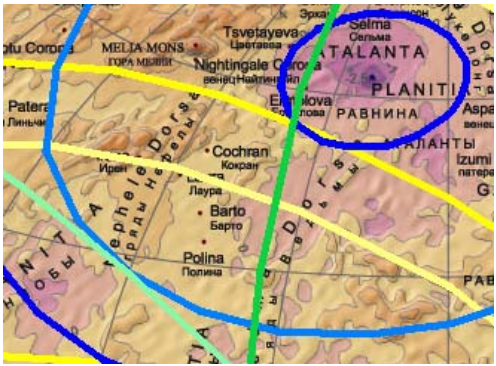

**Fig. 4.** Fragment of Fig. 2. The Howe–Aglaonice direct line (light green) is tangent to the 0.5-circle of Atalanta (light blue). $\Delta = 6$ km.

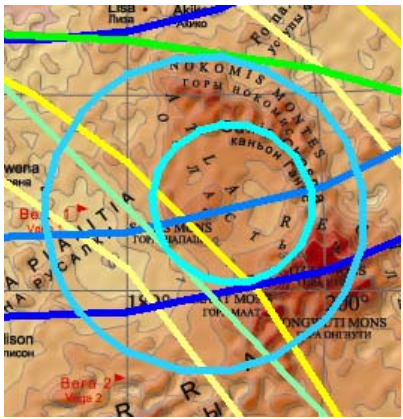

**Fig. 5.** Fragment of Fig. 2. The Danilova–Aglaonice direct line (bright green) is tangent to the $2(\pi - 3)$-circle of Sapas mountain (light-light blue). $\Delta = 0.2$ km.

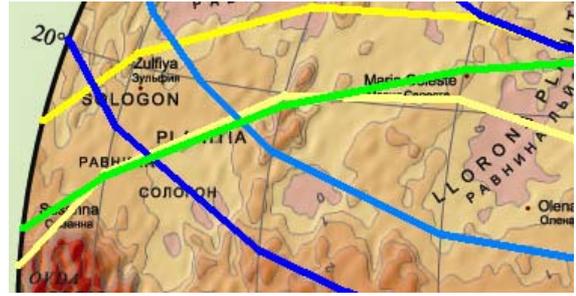

**Fig. 6.** Fragment of Fig. 2. The Danilova–Aglaonice direct line (bright green) is quasi-tangent to the 1-circle of Artemis (light yellow); $\Delta = 124$ km. More precise, this direct line is tangent to the circle with radius $0.98$ radian; $\Delta = 4$ km. The $0.98$ radian distance is embedded on Mars as the distance from the center of Isidis basin ($13.6^0$ N, $272.0^0$ W) to the summit of Elysium mons ($24.6^0$ N, $213.3^0$ W); $\Delta = 0.1$ km. So $0.98$ radian could become some important planetary constant.

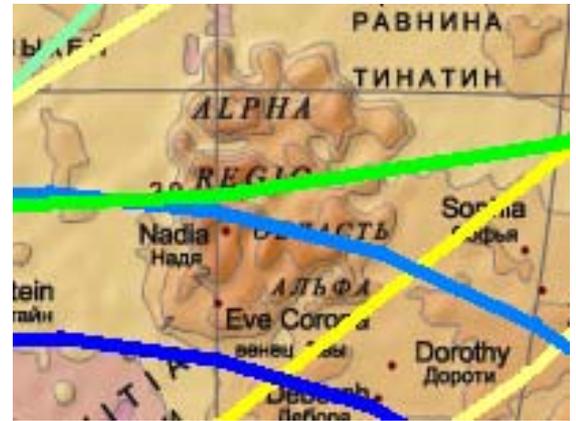

**Fig. 7.** Fragment of Fig. 2. The Danilova–Aglaonice direct line (bright green) goes through the highest summit of Alpha; $\Delta = 5$ km is inside data precision.

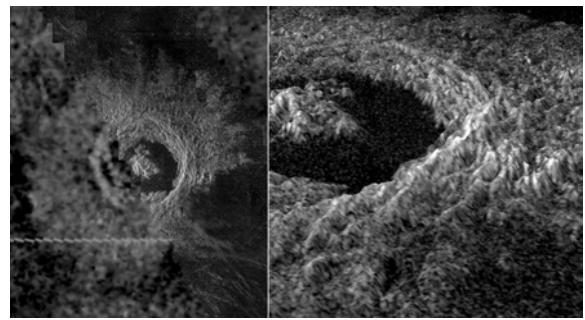

**Fig. 8.** PIA00210, PIA00209. Golubkina, a crater, centered at ($60.5^0$ N, $287.2^0$ E), diameter is 34 km (by PIA00209). The distance from this point to the Howe–Aglaonice direct line is 8 000 Venusian km; $\Delta = 0.3$ km. The distance from this point to the Danilova–Aglaonice direct line is 7 000 Venusian km; $\Delta = 0.8$ km. Recall that we consider distances at the exact sphere with the equator 40 000 Venusian kilometers ("arc" distances along the surface of the sphere), i.e. we study preimages of planets. See the "Central America" section.

## 20. $(\pi - n)$-BASINS.

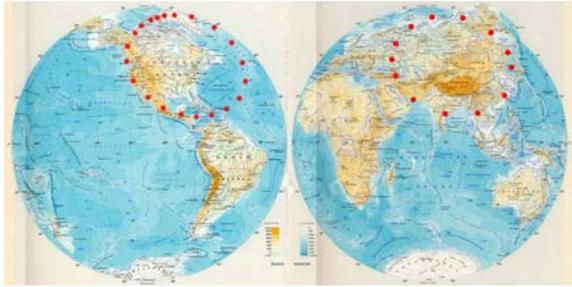

**Fig. 1.** North American basin and Asian basin. Diameters of the circles are $(\pi - 2)$ radians. Note that $(\pi - 2) = 8(\pi - 3) + \Delta$, here $\Delta = 22 - 7\pi \approx 56$ km. $(\pi - 2)$ radians is an important constant, diameter of a continental basin.

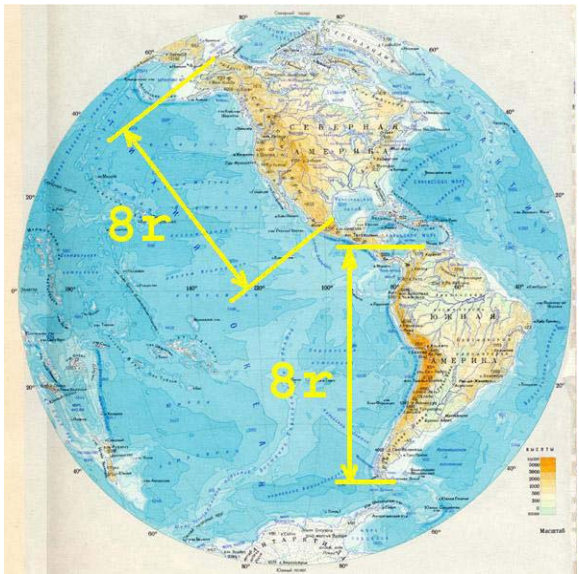

**Fig. 2.** Denote $r = (\pi - 3)$ radians, so $8r \approx (\pi - 2)$. $(\pi - 2)$ radians are embedded here as a continental diameter. South America is $\approx (\pi - 3)/2$ radians longer than $(\pi - 2)$.

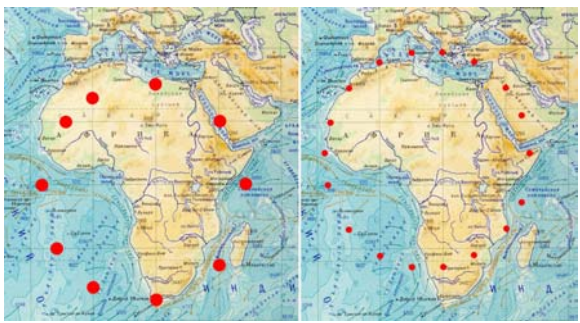

**Fig. 3.** Diameters of the circles are $(\pi - 2)$ radians. Diameter of Africa is nearly continental.

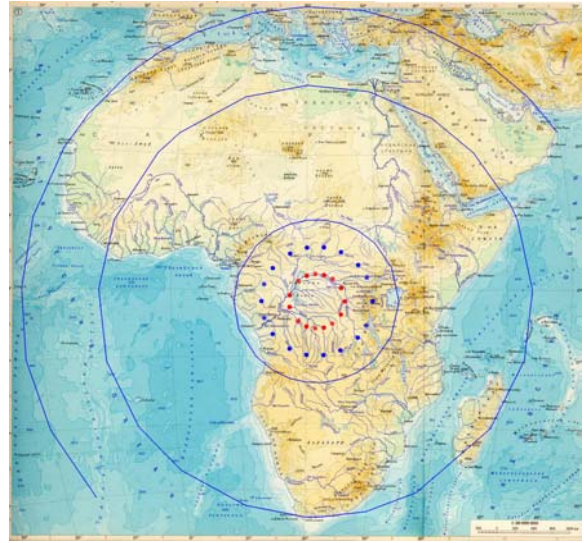

**Fig. 4.** Africa. Radii of the circles are $n(\pi - 3)/2$ radians, $n = 1, 2, 3, 8, 11$. The main continental basin of Africa ($n = 8$) has diameter $8(\pi - 3) \approx (\pi - 2)$ radians. This situation is similar to North America. See the "Discrete fields of force" section.

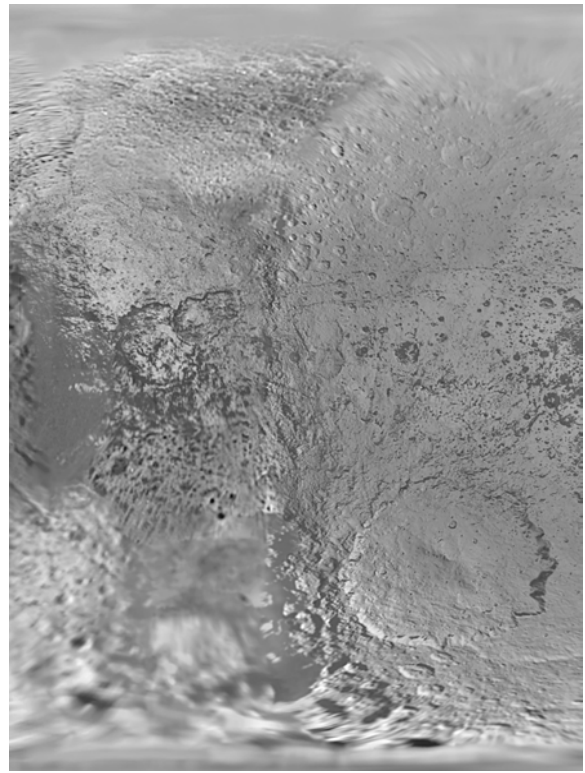

**Fig. 5.** Iapetus, Saturn's satellite. PIA08406 (a fragment of a rectangular map, latitude from $-90^0$ to $+90^0$). Big upper basin (if it really exists, 'cause the map could be conventional at polar areas) has continental diameter $(\pi - 2)$ radians.

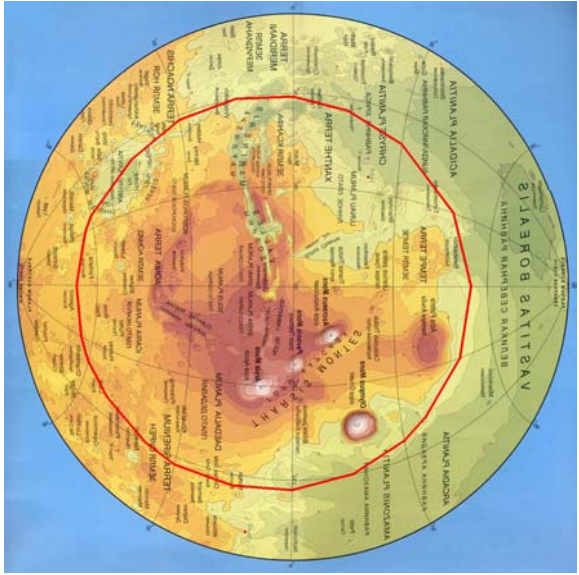

**Fig. 6.** North America on Mars [7], mirror image, rotated cw 90°. The circle is centered at (6.33° S, 92.2° W), radius is $(\pi - 1)/2$ radians. Thus North America on Mars has continental diameter $(\pi - 1)$ radians. See the "North America on Titan" and "North America on Mars" sections.

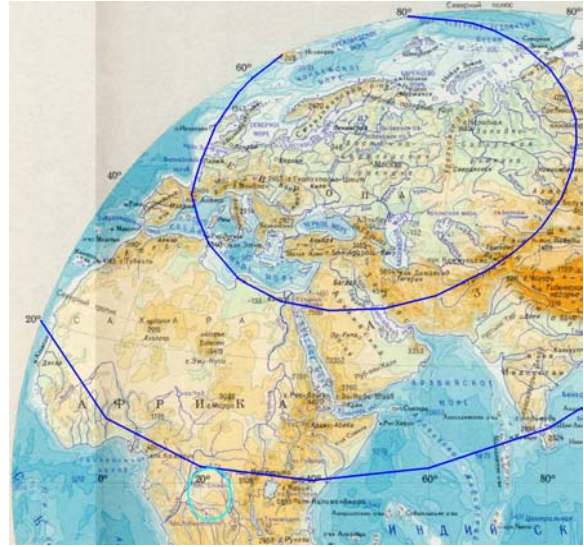

**Fig. 7.** Blue circles are centered at Yaroslavl, radii are 0.5 and 1.0 radians. The 1-circle is tangent to the Congo basin (cyan); its radius is $(\pi - 3)/2$ radians. See Fig. 4 and the "Discrete fields of force" section. Note that $1 + (\pi - 3)/2 = (\pi - 1)/2$ exactly. Thus the distance from Yaroslavl to the center of Congo basin is $(\pi - 1)/2$ radians. Also see the "External geometry of crater triangle" section, Fig. 3 b).

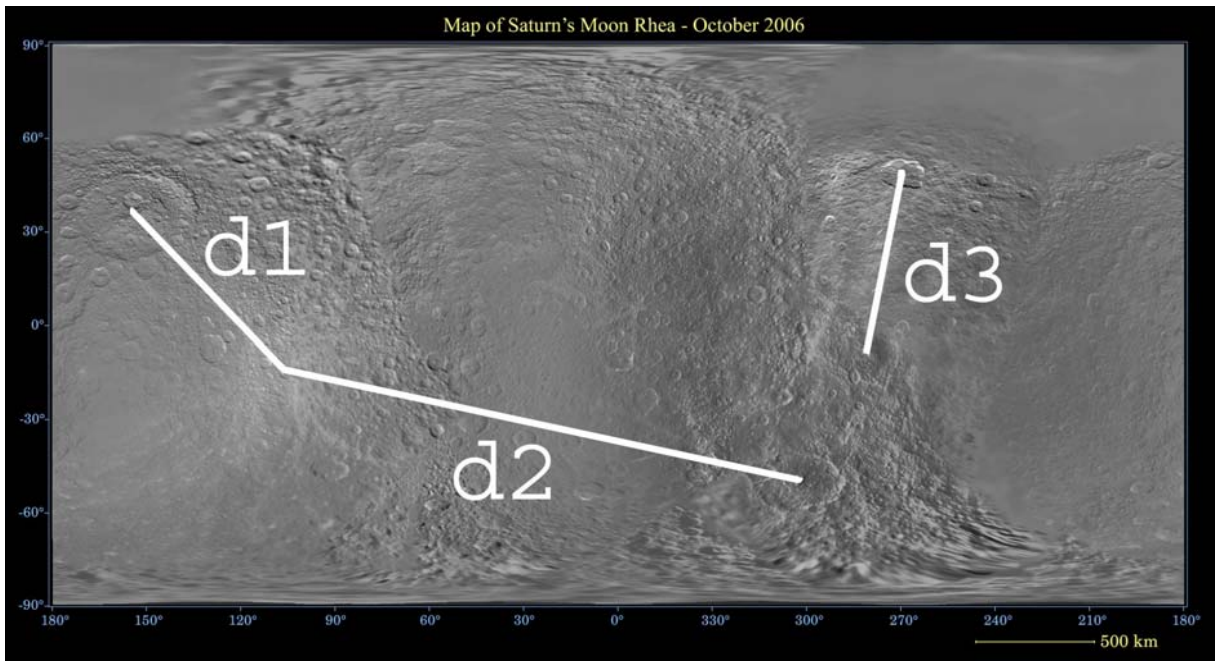

**Fig. 8.** Rhea, PIA08343. Distances between centers of basins. d1 = 1.07 ≈ $(\pi - 1)/2$ radians. d2 = 2.06 ≈ $1 + (\pi - 1)/2$ radians. d3 = 1 radian.

Conclusion: there exists some fundamental sequence of planetary continental constants. The initial element of this sequence is, obviously, $(\pi - 0)$.

Let radius of a continent be $(\pi - 0)$ radians. Then the continent is a whole planet. If $(\pi - 0)$ radians is a continental diameter, then the continent is a hemisphere. The rest of the planet is a hemisphere too.

Note that Mars is approximately divided into "continental" and "oceanic" hemispheres. See the map of Mars at the "$(\pi - 3)$-basins" section, Fig. 6.

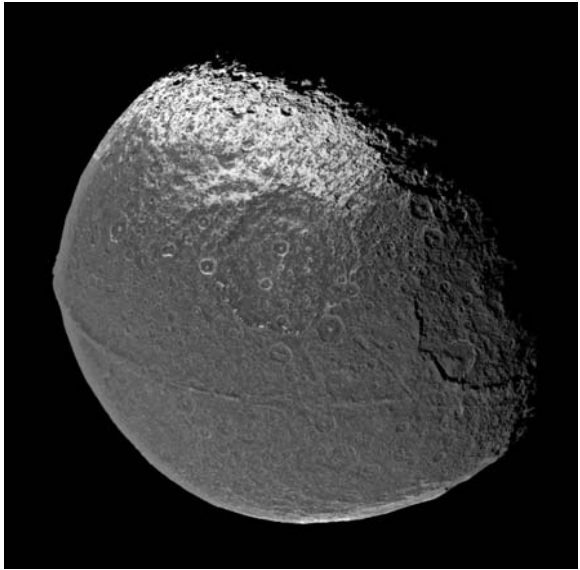

**Fig. 9.** PIA06166. Equatorial seam on Iapetus, Saturn's satellite. Hemispheres are basins with diameter $(\pi - 0)$ radians.

Consider $(\pi - n)$ series. Let *n* be a whole number. For n = −1 we have $(\pi + 1)$. A circle with radius $(\pi + 1)$ radians is (as a set of poins) a circle with radius 1 radian, centered at the antipodal point. Outstanding role of such circles is demonstrated at the "Discrete fields of force" section. It is wellknown in planetology that antipodal points of some basins have interesting relief, so professors say: "By terrifying impact of an asteroid the planet was punched throughout". See the "North America on Mars" section, Fig. 3.

For n = −2, −3, i.e. for $(\pi + 2)$, $(\pi + 3)$ we have the same situation as for n = −1. See the "Discrete fields of force" section. A circle with radius $(\pi + 3)$ radians is (as a set of poins) a circle with radius $(\pi - 3)$ radians, centered at the same point. Therefore, for n < 0 the sequence is ringed. For n = −4 a circle with radius $(\pi + 4)$ radians is (as a set of poins) a circle with radius $1 - (\pi - 3) = (4 - \pi)$ radians, centered at the same point.

Consider $(\pi - 4)$. $(\pi - 4) < 0$, it could be interpreted as contrary direction of going from the center. A circle with radius $(\pi - 4)$ radians is (as a set of poins) a circle with radius $|\pi - 4|$ radians, centered at the same point. $|\pi - 4| = 6(\pi - 3) + \Delta$, here $\Delta = 22 - 7\pi \approx 56$ km (for the Earth), as in the $(\pi - 2)$ case, see Fig. 1.

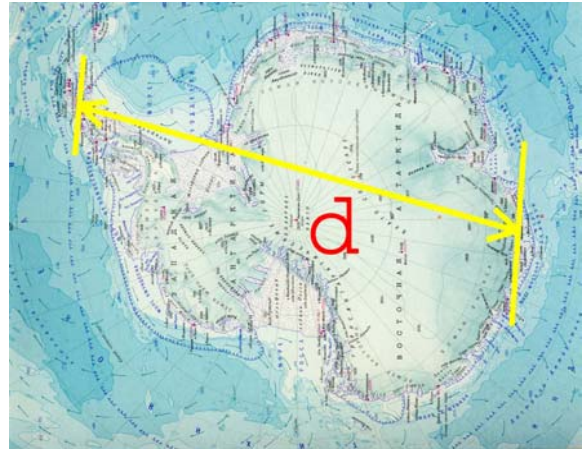

**Fig. 10.** Diameter of Antarctida is d $\approx |\pi - 4|$ radians. Antarctida is 200 km longer. So $(\pi - 4)$ radians is a continental diameter too.

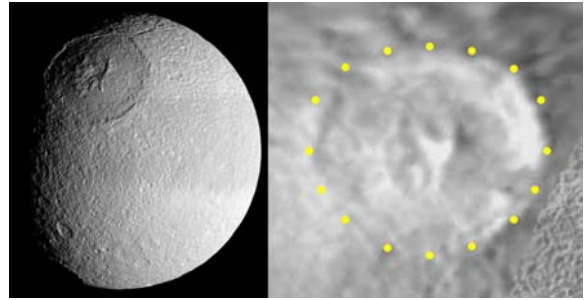

**Fig. 11.** Odissey, a big crater on Tethys, Saturn's satellite. Its bottom is a convex up surface with nearly the same radius of curvature as other planet surface. **Left:** PIA08400. **Right:** PIA07733 (fragment). Radius of the circle is $|\pi - 4|$ radians.

Thus it is reasonable to consider $(\pi - n)$ series at least for n = −4, −3, −2, −1, 0, 1, 2, 3, 4. Recall that for n = 0 we have a whole planet (if $\pi$ is basin's radius) or two hemispheres (if $\pi$ is basin's diameter).

## 21. BIPOLAR AMERICA AND ITS VORTEX STREET.

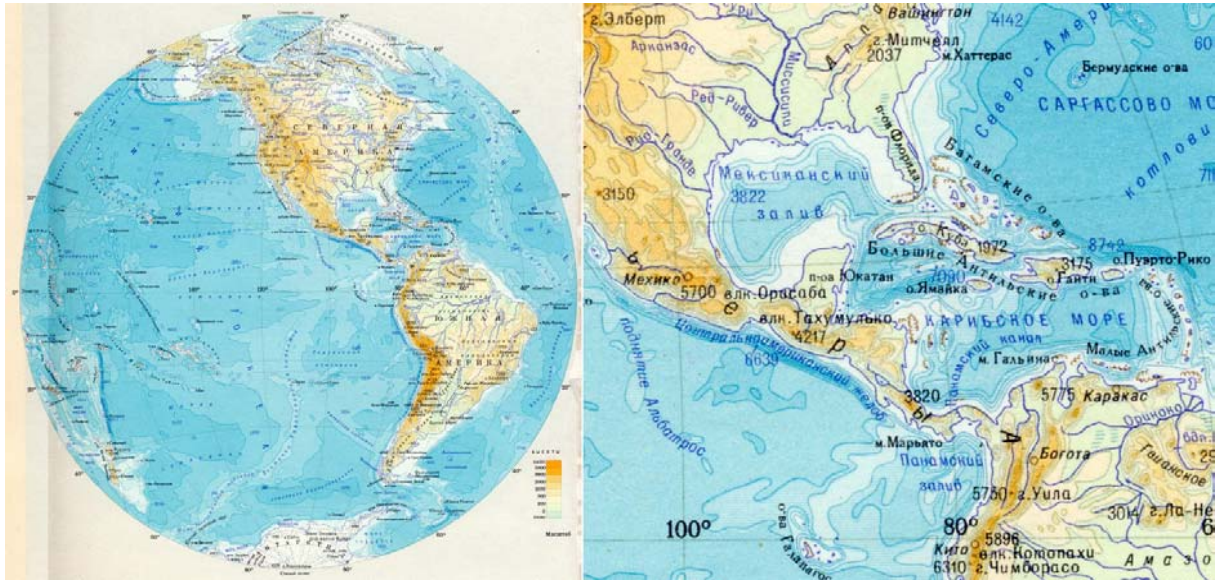

**Fig. 1. Left:** America is a bipole object. **Right:** There is a small bipole between two components of the main bipole. The direction of components of the small bipole is orthogonal to the direction of Central America.

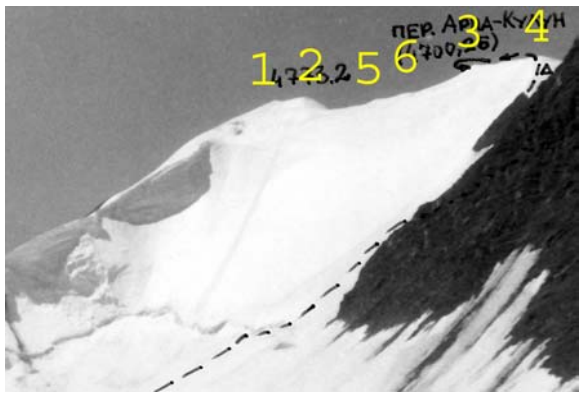

**Fig. 2.** 4773, the 2-nd peak of Ferganski range, South-Western Tian-Shan [3, Fig. 100]. 1, 2: the south summit (main), a bipole. 3, 4: the north summit, a bipole. 5, 6: a bipole between the two main summits. View from North–East, from Karakol glacier. Photo by Yu. N. Bratkov, 1994.

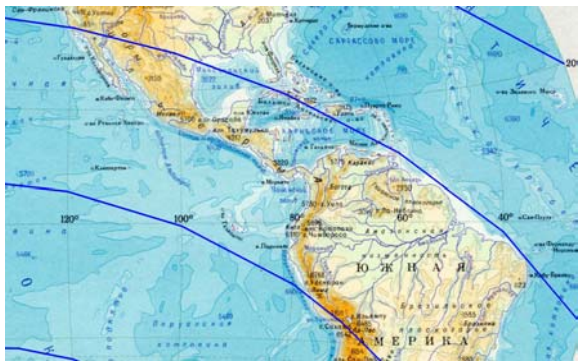

**Fig. 3.** Fragment of the "Discrete fields of force" section, Fig. 1. Central America goes along the two circles, centered at Yaroslavl. Radii of the circles are 1.5 and 2 radians.

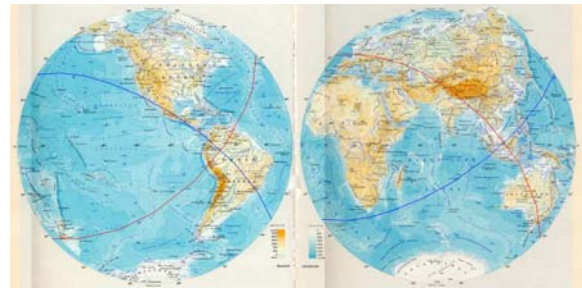

**Fig. 4.** Chomolungma and Chogory are two highest mountains on the Earth. Let them be two poles of the Earthen bipole. (The most obvious bipole structure is defined by North Pole and South Pole, but these poles doesn't give any direction on the surface of the planet.) Red straight line goes through the two mountains. Blue straight line is perpendicular to red, it goes through Orisaba volcano (the highest peak of Central America) [3, Fig. 142]. Central America goes along blue line. So the small bipole (Fig. 1 right) goes nearly along red line. More correctly, blue line defines some set of perpendiculars, including red line.

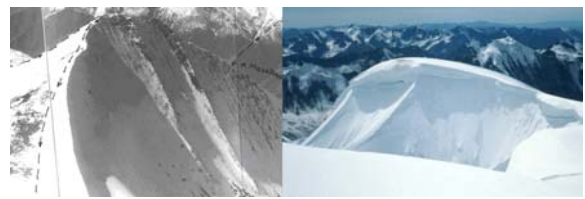

**Fig. 5.** Two mountain edges (a bipole) are directed to the East, or nearly in the direction of red line at Fig. 4, as far as two waves of Central America (Fig. 1 right). **Left:** The south summit of Otsolrak [3, Fig. 110]. View from the north (main) summit. Nukatl range, Dagestan, the Caucasus. Photo by Yu. N. Bratkov, 1994. **Right:** One of two heads of the summit of Nairamdal, the highest peak of Mongolian Altay [3, Fig. 122]. Photo by Yu. N. Bratkov, 1996.

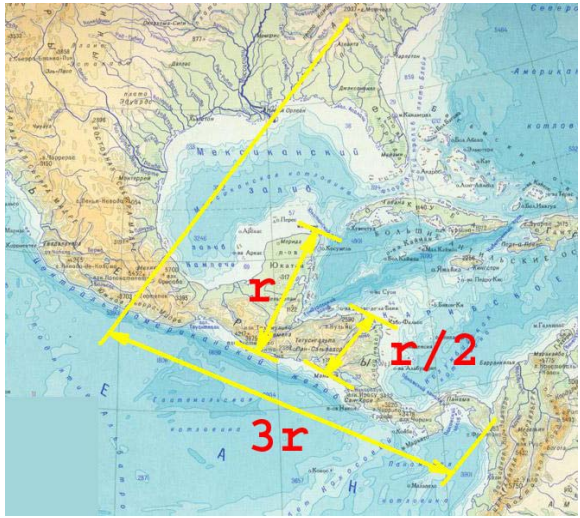

**Fig. 6.** Main sizes of Central America are given by eigenvalues. Here r = (π − 3) radians. 3r is the distance from South America to the prolongation of the straight line of the Appalachians.

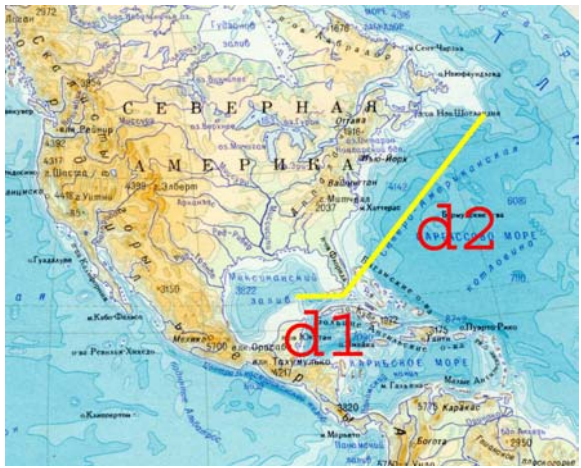

**Fig. 7.** Let the ends of Yucatan, Florida, Newfoundland be centers of vortices of some vortex street (Fig. 10) behind Central America. Suppose there are some more elements of this sequence. Let the coefficient of the sequence be k = d2/d1, then d2 = k·d1, d3 = k·d2, d4 = k·d3, etc. The direction of the street seems to be nearly red line at Fig. 4.

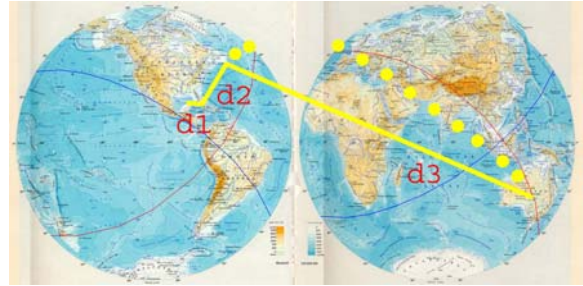

**Fig. 8.** See Fig. 4. d3 = k·d2. In the direction of the vortex sequence there is some obvious 4-th element. It is nearly the center of Australian 2(π − 3)-body. See the "Supersonic Australia" section. Yellow points are nearly the Florida–Newfoundland–Australia straight line.

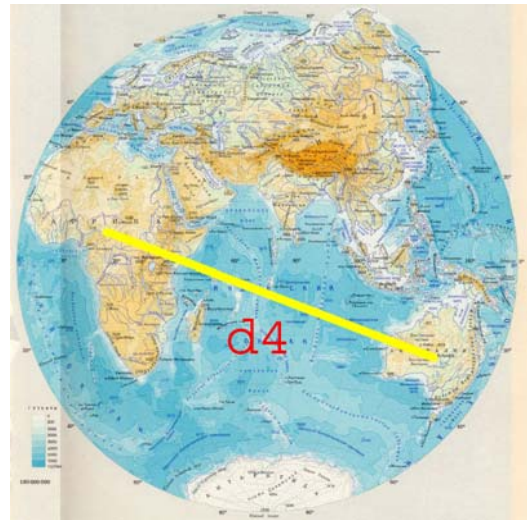

**Fig. 9.** d4 = k·d3. This distance is big. After twice going around the Earth we get the rest (d4 − 4π) and seek some appropriate object with this distance. It is the center of Africa. See the "Supersonic Australia" section, Fig. 1, 2. The center of the 4-th step is not the center of Congo basin, but nearly the center of the circle at the "(π − n)-basins" section, Fig. 3 (right). However, the direction from Australia to Africa is contrary to the direction of the previous sequence. One could explain it by Fig. 10. Next steps after d4 don't give interesting results.

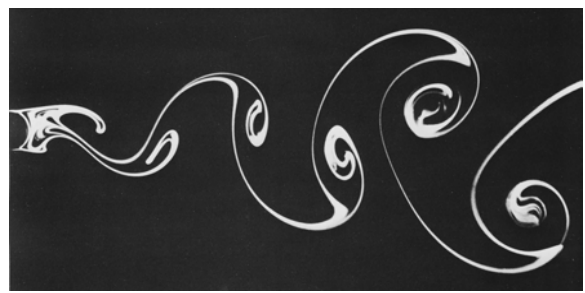

**Fig. 10.** [1, ph. 94], Re = 140. Oscillating trace in water behind streamlined cylinder. There is back direction of the main stream after each vortex. Recall that the 4-th step goes twice around the planet.